\documentclass[11pt, a4paper]{report}
\usepackage{graphicx}
\usepackage{latexsym}
\usepackage{nomencl}
\usepackage{amsmath}
\usepackage{amssymb}
\usepackage{bm}
\usepackage{multirow}
\usepackage{appendix}
\usepackage[T1]{fontenc}
\usepackage{courier}
\usepackage{sectsty}
\usepackage{indentfirst}
\usepackage[bf]{caption}
\RequirePackage{fancyhdr}

\usepackage[bookmarks=true, bookmarksnumbered=true, hypertexnames=true, breaklinks=true, linktocpage=true, linkbordercolor={0 0 1}, pdfborder={0 0 .5}]{hyperref}
\hypersetup{
pdfauthor = {Anna Simon},
pdftitle = {Correlated radiative electron capture in ion-atom collisions},
pdfsubject = {Ph.D. Dissertation},
pdfkeywords = {radiative double electron capture, RDEC, electron capture, REC, bremsstrahlung}}

\newcommand{\spacing}{1.5}
\renewcommand{\baselinestretch}{\spacing}

\newcommand{\captionfonts}{\small}

\chapterfont{\LARGE}
\chaptertitlefont{\LARGE}

\makeatletter  % Allow the use of @ in command names
\long\def\@makecaption#1#2{%
  \vskip\abovecaptionskip
  \sbox\@tempboxa{{\captionfonts #1: #2}}%
  \ifdim \wd\@tempboxa >\hsize
    {\captionfonts #1: #2\par}
  \else
    \hbox to\hsize{\hfil\box\@tempboxa\hfil}%
  \fi
  \vskip\belowcaptionskip}
\makeatother   % Cancel the effect of \makeatletter

\newcommand{\thesistitle}[6]{%
\thispagestyle{empty}
\begin{center}
 
\vspace*{2cm}
\textbf{\Large #3}
\vspace*{2cm}

\textbf{\Huge #1}
\vspace*{1cm}

\textbf{\LARGE \it #2}

\renewcommand{\baselinestretch}{\spacing}
\vspace{\fill}
\renewcommand{\baselinestretch}{1.0}%

\vspace*{1.in}

#4 \\

\vspace{\fill}
\begin{figure}[h]
\begin{center}
  \includegraphics[scale=.4]{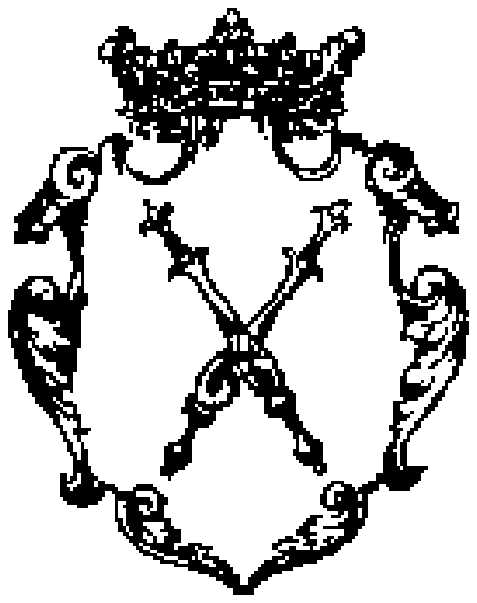}
\end{center}
\end{figure}

#5 \\

#6 \\
\end{center}
}

\newcommand{\munepsfig}[3][1.0]{%
	\begin{figure}[!tbp]
	\centering
	\vspace{10mm}
	\scalebox{#1}{\includegraphics{figures/#2.eps}}
	\renewcommand{\baselinestretch}{1.0}%
	\caption{#3}
	\renewcommand{\baselinestretch}{\spacing}%
	\label{fig:#2}
	\end{figure}
}

\usepackage{array}

\newcolumntype{x}[1]{%
>{\centering\hspace{0pt}}p{#1}}%

\newcommand{\muntab}[4]{%
	\begin{table}%[!htbp]
	\centering
	\small{
	\renewcommand{\baselinestretch}{1.0}%
	\caption{#3}
	\label{tab:#2}
	\vspace{2mm}
	\begin{tabular*}{.95\textwidth}{#1}
	#4
	\end{tabular*}
	}
	\vspace{5mm}
	\renewcommand{\baselinestretch}{\spacing}%
	\end{table}
}

\newcommand{\muneqn}[2]{%
	\begin{equation}
	\centering
	\label{eqn:#1}
	#2
	\end{equation}
}

\newcommand{\nomen}[2]{
\nomenclature{\parbox{5cm}{#1}}{#2}
}

\newcommand{\abbrev}[2]{
\nomenclature{\parbox{1.5cm}{#1}}{#2}
}

\begin{document}

\pagenumbering{roman}

\thesistitle
	{Correlated radiative electron capture in ion-atom collisions}
	{ }
	{Anna Simon}
	{Ph.D. Dissertation \\
	prepared under the supervision of \\
	Prof. Andrzej Warczak}
	{Marian Smoluchowski Institute of Physics\\
	Jagiellonian University }
	{Krak\'{o}w, April 2010}

\makenomenclature

\newpage

\begin{center}
\vspace*{2cm}
\textbf{\LARGE Abstract}
\end{center}

Radiative double electron capture (RDEC) is a one-step process where two free (or quasi-free) target electrons are captured into a bound state of the projectile, e.g. into an empty K-shell, and the energy excess is released as a single photon. This process can be treated as a time inverse of double photoionization. However, unlike in case of photoionization experiments, bare ions are used during RDEC observations. Thus, RDEC can be considered as the simplest, clean tool for investigation of the electron-electron interaction in the presence of electromagnetic fields generated during ion-atom collisions. 

Within this dissertation, the \(38\) MeV O$^{8+}$ + C experiment, conducted at Western Michigan University using the tandem Van de Graaff accelerator, is discussed and the first experimental evidence of the RDEC process is presented. The cross section obtained experimentally is compared to the latest theoretical calculations. 

\newpage

\begin{center}
\vspace*{2cm}
\textbf{\LARGE Abstrakt}
\end{center}

Skorelowany radiacyjny wychwyt dw\'{o}ch elektron\'{o}w (RDEC) jest procesem, podczas kt\'{o}re-go dwa swobodne (albo kwaziswobodne) elektrony tarczy wychwytywane s\k{a} do stanu zwi\k{a}zane-go pocisku (np. nieobsadzonej pow\l oki K), a r\'{o}\.{z}nica energii pomi\k{e}dzy ko\'{n}cowym a pocz\k{a}tko-wym stanem eletron\'{o}w emitowana jest w postaci pojedynczego fotonu. Proces ten mo\.{z}na traktowa\'{c} jako odwr\'{o}cenie w czasie podw\'{o}jnej fotojonizacji. Jednak\.{z}e, w przeciwie\'{n}stwie do eksperyment\'{o}w dedykowanych fotojonizacji, do obserwacji RDEC stosuje si\k{e} jony ca\l kowicie pozbawione elektron\'{o}w, co pozwala na wyeliminowanie t\l a pochodz\k{a}cego od elektron\'{o}w nie bior\k{a}cych bezpo\'{s}rednio udzia\l u w badanym procesie. RDEC mo\.{z}e by\'{c} wi\k{e}c traktowany jako najprostsze narz\k{e}dzie do badania oddzia\l ywania elektron-elektron w obecno\'{s}ci pola elektromagnetycznego generowanego podczas zderzenia.

Rozprawa ta po\'{s}wi\k{e}cona jest procesom atomowym zachodz\k{a}cym w zderzeniach O$^{8+}$ + C przy energii \(38\) MeV podczas eksperymentu przeprowadzonego przy u\.{z}yciu alceleratora Van de Graaffa w Western Michigan University. Przedstawione zosta\l o w niej pierwsze do\'{s}wiadczalne potwierdzenie procesu RDEC. Uzyskany eksperymentalnie przekr\'{o}j czynny zosta\l ~por\'{o}wnany z wynikami najnowszych przewidywa\'{n} teoretycznych.

\newpage
\begin{center}
\vspace*{2cm}
\textbf{\LARGE Acknowledgements}
\end{center}

First, I would like to give my sincere thanks to my supervisor, Professor Andrzej Warczak. He offered me advice, patiently supervised me and always guided me to the correct direction. I have learned much from him, without his help I would have never finished my dissertation successfully.

Special thanks are also given to Professor John A. Tanis. He is the one who invited me to Western Michigan University for my research during the 2008-2009 academic year. His help and encouragement made me feel confident enough to fulfill my desires and to overcome the difficulties I encountered. His understanding, encouragement and personal guidance have provided a good basis for my thesis. It is not sufficient to express my gratitude with such a few words. 

I am very grateful to Professor Bogus\l aw Kamys, the Head of the Nuclear Physics Division at the Institute of Physics, Jagiellonian University and Professor Paul Pancella, the Chair of Physics Department, Western Michigan University, for the financial support of my stay at WMU.

Additionally, I owe my most sincere gratitude to Dr. Asghar Kayani for his patience while teaching me how to operate the WMU tandem Van de Graaff accelerator and his willingness to immediately solve any beam related problems during my experiment.
My sincere thanks should also go to Rick Welch and Allan Kern for their help with the maintenance of the experimental setup.

I would also like to thank Janusz Kopczy\'{n}ski, Adam Malarz and Adam Mucha for their care and support during my work at Jagiellonian University.

I am also very grateful to Prof. Thomas St\"{o}hlker, the Head of Atomic Physics Group at GSI, Darmstadt, who frequently invited me to join his group's experiments during which I had a chance to learn the secrets of the experimental work of atomic physicists. I'm grateful to Dr. Angela Br\"{a}uning-Demian and Dr. Christophor Kozhuharov for inspiring conversations that led to many ideas implemented in this thesis.

I am very grateful to my friends, David Cassidy, Buddhika Dassanayake, Ma\l gorzata Makuch, Dagmara Rozp\k{e}dzik and Andrzej Pezarski, for always being there for me.

Last, but not least, I wanted to thank my parents for their support and encouragement.

\tableofcontents
\addtocontents{toc}{\protect\thispagestyle{fancy}}

\addcontentsline{toc}{chapter}{List of Tables}
\listoftables
\addtocontents{lot}{}%\protect\thispagestyle{fancy}}

\addcontentsline{toc}{chapter}{List of Figures}
\listoffigures
\addtocontents{lof}{\protect\thispagestyle{fancy}}

\addcontentsline{toc}{chapter}{List of Symbols and Abbreviations}
\thispagestyle{plain}

\printnomenclature

\abbrev{RR}{Radiative recombination}
\abbrev{REC}{Radiative electron capture}
\abbrev{RDEC}{Radiative double electron capture}
\abbrev{DREC}{Double radiative electron capture}
\abbrev{AB}{Atomic bremsstrahlung}
\abbrev{SEB}{Secondary electron bremsstrahlung}
\abbrev{NB}{Nucleus bremsstrahlung}
\abbrev{RECC}{Radiative electron capture to continuum}
\abbrev{QFEB}{Quasifree electron bremsstrahlung}
\abbrev{RI}{Radiative ionization}
\abbrev{NREC}{Nonradiative electron capture}

\abbrev{PWBA}{Plain wave Born approximation}
\abbrev{CBE}{Coulomb Born exchange}

\abbrev{\(\Omega\)}{Detector solid angle}
\abbrev{\(d\)}{Target thickness [particles/cm$^{2}$]}
\abbrev{\(d_t\)}{Target thickness [mm]}
\abbrev{\(b_w\)}{Beam diameter [mm]}
\abbrev{\(I\)}{Beam intensity [ions/s]}
\abbrev{\(\theta\)}{Observation angle in the laboratory frame}
\abbrev{\(N\)}{Number of counts}
\abbrev{\(b\)}{Number of background counts}
\abbrev{\(T\)}{Statistical variable of the $\chi^2$ test}
\abbrev{\(\Im(p_z)\)}{Compton profile}
\abbrev{\(q\)}{Projectile charge state}
\abbrev{\(T_m\)}{Maximum energy transfer during ion-atom collision (in the laboratory frame)}
\abbrev{\(v_e\)}{Electron velocity}
\abbrev{\(A\)}{Projectile mass}
\abbrev{\(A_t\)}{Target mass}
\abbrev{\(E\)}{Projectile kinetic energy [MeV/u]}
\abbrev{\(E_B\)}{Binding energy of an electron in the bound state of the projectile}
\abbrev{\(E_{Bt}\)}{Binding energy of an electron in the bound state of the target}
\abbrev{\(T_r\)}{Kinetic energy of the free target electron calculated in the projectile frame}
\abbrev{\(\sigma\)}{Cross section}
\abbrev{\(P_{REC}\)}{Probability that the photon is registered in the REC range of the x-ray spectrum}
\abbrev{\(P_{RDEC}\)}{Probability that the photon is registered in the RDEC range of the x-ray spectrum}
\abbrev{\(\hbar \omega\)}{Photon energy in the laboratory frame}
\nomen{\(\alpha=1/137\)}{Fine structure constant}
\nomen{\(\hbar=6.582\cdot10^{-16}\) eV$\cdot$s}{Planck constant}
\nomen{\(a_0=5.29 \cdot 10^{-11}\) m}{Bohr radius}
\nomen{\(c=2.997 \cdot 10^{8}\) m/s}{Speed of light}

\abbrev{\(Z\)}{Projectile atomic number}
\abbrev{\(Z_t\)}{Target atomic number}
\abbrev{\(\eta\)}{Electron momentum within the target bound state}
\abbrev{\(p\)}{Momentum}
\abbrev{\(\xi\)}{Dimensionless parameter describing collision velocity (adiabacity parameter)}
\abbrev{\(v\)}{Projectile velocity}
\abbrev{\(m_e\)}{Electron rest mass}
\abbrev{\(m_p\)}{Proton rest mass}
\abbrev{\(\nu\)}{Sommerfeld parameter for K-shell electron}

\pagenumbering{arabic}

\chapter{Introduction}
\label{chap:intro}
Since the first observation of the photoelectric effect by Hertz \cite{hertz} and its explanation by Einstein \cite{eins} the interaction between electrons and light has been of considerable attention. The fundamental process occurring due to this interaction is photoionization, where absorption of a photon of energy \(\hbar\omega\) results in the emission of an electron:
%----------------------
\muneqn{photoionization}{
A + \hbar \omega \rightarrow A^+ + e^-.
}
%----------------------
Simple photoionization experiments usually are restricted to neutral atoms, where the influence of the electrons, which do not participate in the process directly, cannot be neglected. This complicates comparison of the experimental results with theoretical predictions.

However, based on the principle of detailed balance \cite{landau,shev} the photoionization can be studied via the time reversed processes, i.e. radiative recombination (RR) and radiative electron capture (REC) \cite{ichi1,ichi2,eich1}. During these processes a free (RR) or loosely bound (REC) electron is captured to the bound state of the projectile and a photon with energy equal to the difference between the final and initial electron states is emitted. 
Unlike single photoionization of multielectron systems, REC has been investigated for bare ion-atom interactions \cite{stoe1,stoe2} and offers clean conditions for exploration of photoionization with only one electron, allowing for observation of pure photon-electron interactions.

During the last thirty years double photoionization has been of considerable interest \cite[and references therein]{dal}. As a photon typically interacts only with one electron, double photoionization is caused by the electron-electron interaction \cite{smit}. However, double photoionization studies have been performed mainly for low \(Z\) atoms, such as He \cite{he1,he2,he3}, Ne \cite{ne1,ne2,arne}, and Ar \cite{ar,arne}. This is due to the background contributions from other electrons for high \(Z\) systems, which make the subtle electron correlation effects difficult to observe. Fortunately, similar to single photoionization, double photoionization can be studied by means of the time inversed process -- RDEC, for which this background is absent.
Radiative double electron capture (RDEC) involves transfer of two target electrons into a bound state of the projectile with simultaneous emission of a single photon \cite{war, bed}. Since bare ions are used during the experiment, RDEC can be considered as the simplest, clean tool for investigation of the electron-electron interaction \cite{war} in the presence of electromagnetic fields generated during ion-atom collision.
Thus, investigation of the RDEC process can provide crucial information necessary for a proper description of electron correlations within atomic systems and provide data required to define the wave function of two correlated electrons in the projectile continuum.

During the last twenty years the RDEC process was addressed not only experimentally \cite{war, bed}, but also theoretically \cite{mir,yak1,yak2}. The calculations were found to be in disagreement with the experimental data \cite{bed} and verification of the RDEC process was not possible. The more recent calculations not only explained previous experimental results, but also suggested the choice of low energy mid-\(Z\) (\(Z\leq35\)) collision systems for observation of RDEC \cite{mikh1,mikh2, druk}. It is also noted that for these systems capture to an excited \(1s^12s^1\) state might significantly enhance the process and contribute to the observed x-ray spectra \cite{nef}.
These calculations provided the main motivation for yet another experiment dedicated to the RDEC process. To fully take advantage of the new calculations, two collision systems at two different accelerator complexes were chosen:
\begin{itemize}
 \item Xe$^{54+}$~+~C at \(20\)~MeV/u, to be performed at GSI, Darmstadt in Germany,
 \item O$^{8+}$~+~C at \(2.375\)~MeV/u, realized by means of the Van de Graaff accelerator at WMU, Kalamazoo, MI, USA.
\end{itemize}
So far the WMU experiment was carried out. During six months of the experiment preparations and data taking, \(43\) days of beam time were used. At the moment when this thesis is being written the GSI beam time is still pending.

Within this dissertation the O$^{8+}$~+~C at \(38\)~MeV experiment is discussed and the first experimental evidence of the RDEC process is presented. The cross section obtained experimentally is compared with the latest theoretical calculations \cite{mikh1,mikh2,nef, druk}.

This thesis begins with an introduction to the most important atomic processes that occur during ion-atom collisions. In Chapter~\ref{chap:processes} special attention is paid to the processes which add to the background for the x-ray spectrum registered during the experiment and formulae allowing for estimations of contributions of these processes are suggested.
Chapter~\ref{chap:rdec} addresses the RDEC process in detail. The history of the experimental approach and the theoretical calculations of the RDEC cross section are presented. Additionally, this chapter focuses on the recent theoretical calculations which were the main motivation for the experiment discussed in this dissertation.
The goal of the experiment was the observation of x rays emitted during collisions of bare oxygen ions with carbon atoms. The x-ray spectra were registered in coincidence with ongoing particles which underwent single or double charge exchange. The experimental setup which allowed for achieving this goal is presented in Chapter~\ref{chap:exp}. The operation principle of a Van de Graaff accelerator is explained and the construction of the target chamber, particle spectrometer and x-ray detector are described in detail.
Chapter~\ref{chap:anal} is dedicated to data analysis, with a particular focus on processes that may contribute to the x-ray spectrum within the RDEC region. Various approaches to estimation of the background and calculations of the cross section are discussed. In Chapter~\ref{chap:cross} the experimentally obtained RDEC cross section is compared with the theoretical value and the possible reasons for the obtained discrepancy are given.
In Chapter~\ref{chap:mc} results of the Monte Carlo simulations of the x-ray spectrum generated during the O$^{8+}$~+~C collisions are compared with the experimental results. 
Finally, in Chapter~\ref{chap:conclusions} suggestions for further investigations of the RDEC process are given, with indication of necessary improvements of the experimental setup.

\chapter{Atomic processes during ion-atom collisions at low energy}
\label{chap:processes}
Interaction between an incoming ion and a target atom may lead to many different atomic processes, such as:
\begin{itemize}
\item 
ionization, mainly of the target atom, as the electrons are usually less bound to a light target than to a partially ionized projectile,
\item
electron transfer from the target to the projectile,
\item 
excitation of both target and projectile states -- such states deexcite after the collision emitting characteristic x rays.
\end{itemize}

Within the following sections the most important processes that were considered competitive to RDEC for the presented experiment are discussed.

\section{Nonradiative electron capture (NREC)}

The Coulomb interaction between the projectile and the target electrons can lead to a process called Coulomb capture or nonradiative electron capture (NREC). Here, the energy difference between the initial and final state of the electron is converted into the kinetic energy of the collision partners.
The most convenient and frequently used scaling formula that estimates the cross section for nonradiative electron capture is the one given by Schlachter \cite{schla}. It is a semiempirical formula which allows for calculation of the \(\sigma_{NREC}\) as a function of the projectile energy for various projectiles with an accuracy better than \(30\%\).

The NREC process occurs mainly at the velocity matching condition \(v \approx v_e\), where \(v_e\) is the velocity of the captured electron, bound in the target atom. For \(v \gg v_e\) in the nonrelativistic approximation, as shown for example in \cite{eich}, the NREC cross section scales as:
%----------------------------------
\muneqn{nrec_approx}{
\sigma_{NREC} \sim \dfrac{Z_t^5 Z^5}{v^{12}}.
}
%----------------------------------
%Dividing this approximation by the one for the radiative electron capture (REC) (Eq.~\ref{eqn:rec_approx}) shows that nonradiative process dominates at low collision energy or for high-\(Z\) targets.
%, or when the inequality:
%----------------------------------
%\muneqn{nrec}{
%E_{proj} \leq 13 Z _{t} ^{10/7} 
%}
%----------------------------------
%is fulfilled.

\section{Radiative electron capture (REC)}
Radiative electron capture (REC) is one of the best known atomic processes observed in heavy ion-atom collisions. It was first observed in early seventies of the last century \cite{schno1,kie,schno2} and since that time has been intensively studied both experimentally \cite{kand,mok,spin,stoe,stoe1,stoe2,stoe3,stoe4,stoe5,stoe6,tan2,tan3} and theoretically \cite{eich1,eich2,hino,ichi1,ichi2,soh}. During this process capture of a single target electron is followed by a photon emission (Fig.~\ref{fig:rec}). The energy \(E_{REC}\) of the emitted photon fulfills the energy conservation rule for this process. Thus, it is given by:
%----------------------------------
\muneqn{Erec}{
E_{REC} = T_r + E_B - E_{Bt} + \overrightarrow{v} \overrightarrow{p},
}
%----------------------------------
where \(E_B\) and \(E_{Bt}\) are the binding energies of the projectile and target, respectively, \(\overrightarrow{v}\) is the projectile velocity and \(\overrightarrow{p}\) the momentum of the electron in the bound state of the target. \(T_r=({m_e}/{m_p})E\) is the kinetic energy of the quasifree target electron calculated in the projectile's frame of reference.
%----------------------------------

\munepsfig[.75]{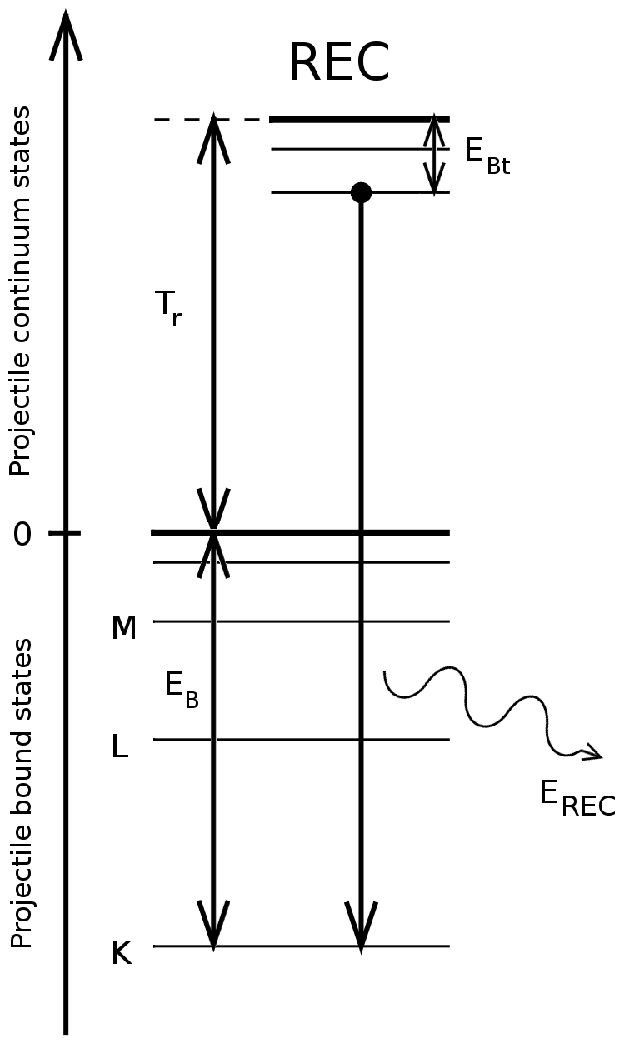}{Radiative electron capture (REC). A target electron is captured into the projectile bound state and the energy excess is emitted as a single photon.}
%----------------------------------

The REC line observed during experiments is much broader than the characteristic x-ray lines, as can be observed in Fig.~\ref{fig:rec_swiat}, which is due to the velocity distribution of target electrons. This distribution is described by Compton profile \(\Im(p_z)\) \cite{big}, which gives the probability of finding an electron with a given momentum projection \(p_z\), where (for ion-atom collisions) the \(z\)-axis is defined by the projectile velocity. The Compton profile depends on the target atomic number \(Z_t\) and its width increases with increasing \(Z_t\). Moreover, the width depends on the binding energy and is smaller for loosely bound electrons, than for a tightly bound \(1s\) electron as shown in Fig.~\ref{fig:compton}. 
%As it was shown by Tawara \cite{taw} the width of the REC line can be calculated as:
%----------------------------------
%\muneqn{E_FWHM}{
%\Delta E_{FWHM} = 2.04 \left( E E_{Bt}\right) ^{1/2}.
%}
%----------------------------------

%----------------------------------
\munepsfig[.45]{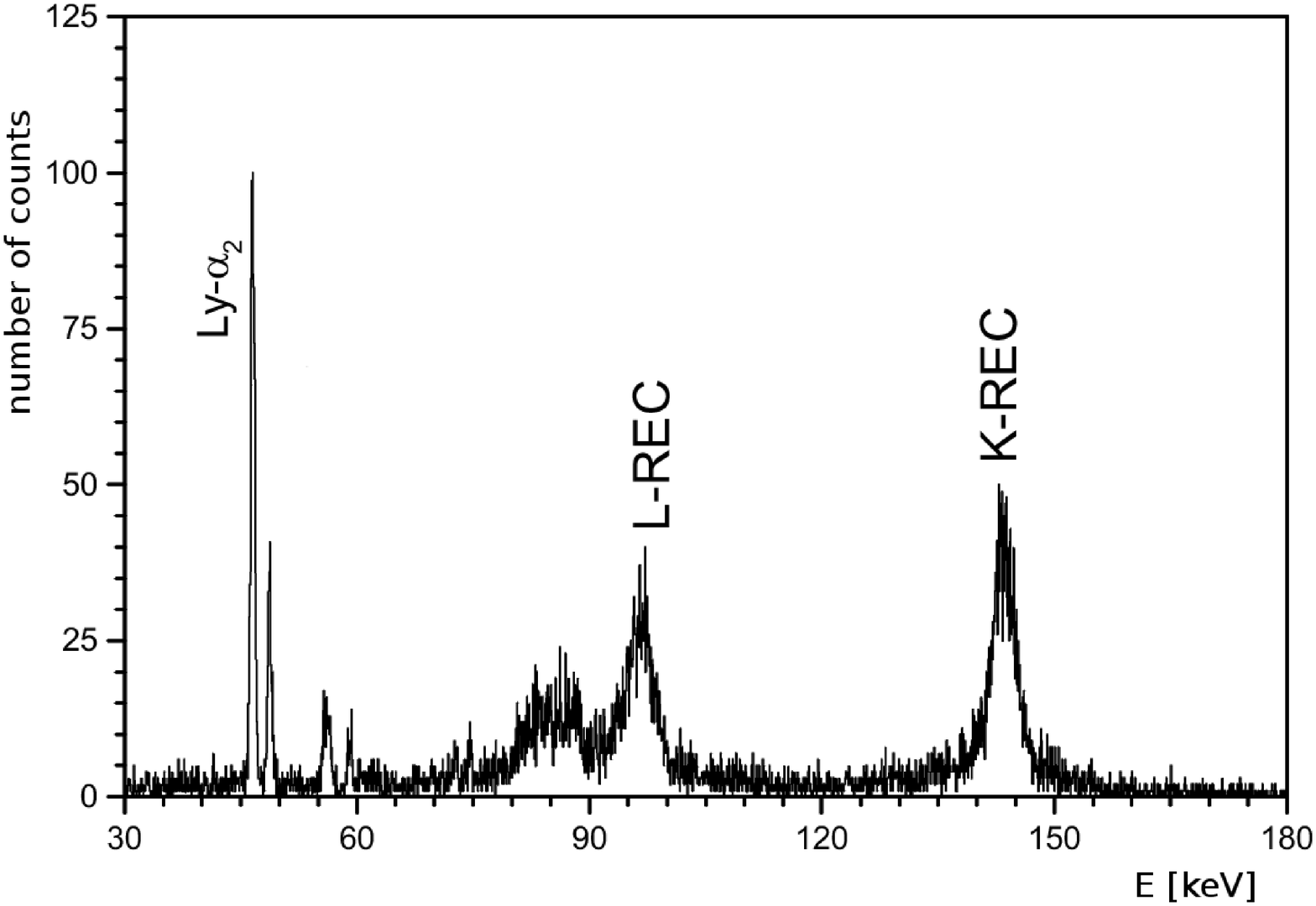}{Example of the x-ray spectrum registered in coincidence with single electron capture during U$^{92+}$~+~N$_{2}$ collisions at $309.7$~MeV/u \cite{sw2}.}
%----------------------------------
%----------------------------------
\munepsfig[.75]{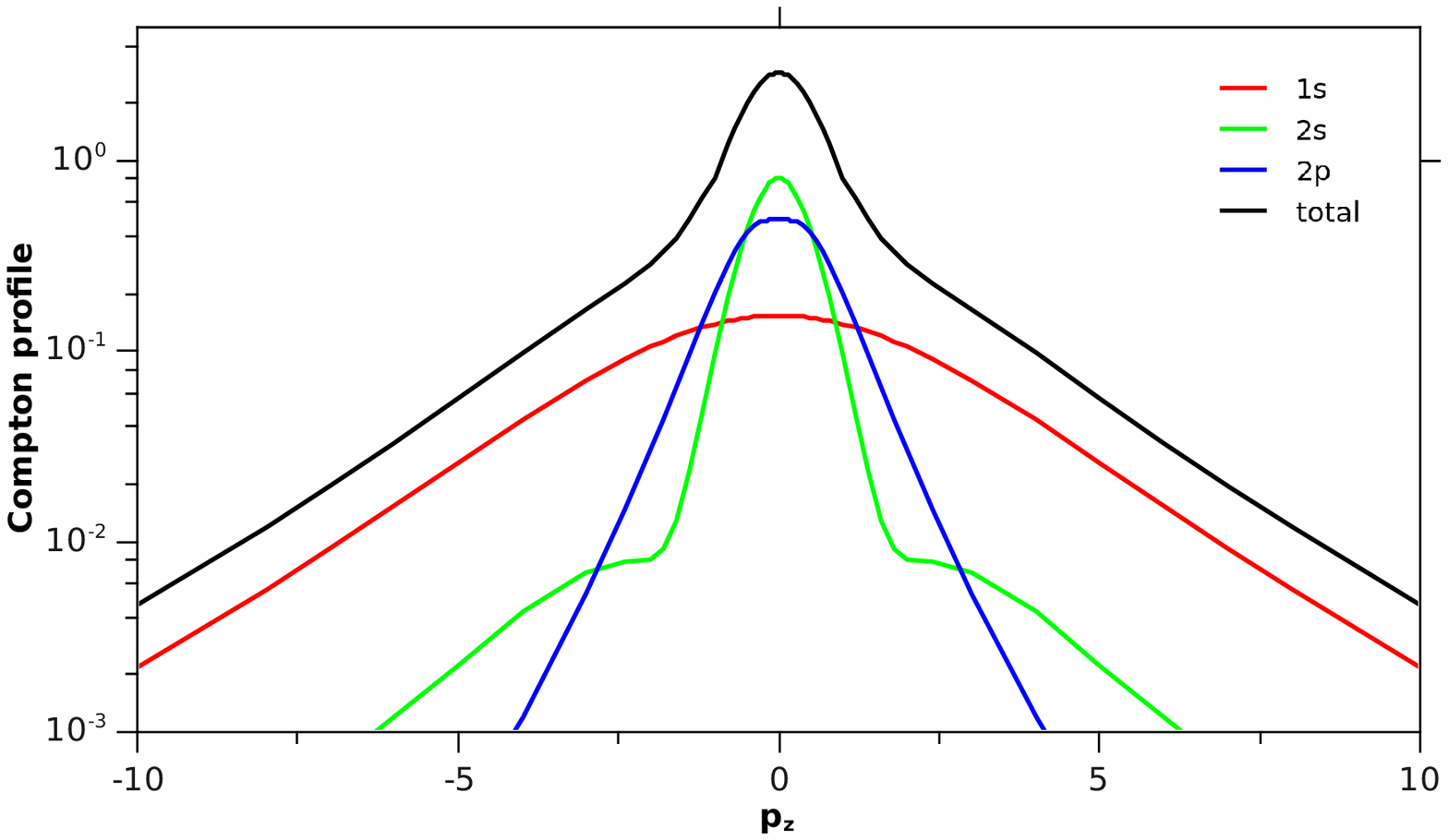}{Compton profile of electrons in carbon atom. It can be noticed that the structure of the $1s$ line is much broader than that for $n=2$ states \cite{big}.}
%----------------------------------

When the binding energy of the target electron is much smaller than \(T_r\), the captured electron can be treated as quasifree. This means that REC can be described as capture of a free electron (radiative recombination - RR), which is the time inverse of photoionization. As the cross section for single photoionization can be calculated from the well known formula given by Stobbe \cite{stob}, one can use it to calculate the REC cross section via the principle of detailed balance.

Principle of detailed balance describes the relation between the cross sections for direct (\(\sigma_{i\rightarrow f}\)) and time inverse (\(\sigma_{f\rightarrow i}\)) processes \cite{landau,shev}:
%----------------------------------
\muneqn{detailed}{
g_i p_i^2 \sigma_{i\rightarrow f}(p_i) = g_f p_f^2 \sigma_{f\rightarrow i}(p_f),
}
%----------------------------------
where \(g\) is the number of possible states given by angular momentum and spin combinations and \(p\) is the momentum of the particle in the center of mass system describing the size of a phase space accessible for the initial (\(i\)) and final (\(f\)) states.

Based on Eq.~\ref{eqn:detailed} and the Stobbe formula for the photoionization cross section, the cross section for REC to the projectile K-shell during collision of a bare ion with a hydrogen target can be expressed in the form:
%----------------------------------
\muneqn{stobbe}{
\sigma_{REC}=9.16\left(\dfrac{\nu^3}{1+\nu^2}\right)^2 \dfrac{\exp(-4\nu \cot^{-1}(1/\nu))}{1-\exp(-2\pi\nu)}\cdot10^{-21} [cm^2],
}
%----------------------------------
where \(\nu ={Z_t e^2}/{\hbar v}\) is the Sommerfeld parameter of the target K-shell electron and \(v\) is the projectile velocity. Thus, for fast collisions, the REC cross section scales with energy as:
%----------------------------------
\muneqn{rec_approx}{
\sigma_{REC} \sim \dfrac{Z_t Z^5}{v^{5}}.
}
%----------------------------------
When this result is compared with Eq.~\ref{eqn:nrec_approx}, one should notice that the radiative electron capture dominates for high energy collisions with light targets.

The angular distribution of the REC photons is given by the angular differential REC cross section calculated within the dipole approximation \cite{schno1,kie}:
%----------------------------------
\muneqn{diff}{
\dfrac{d\sigma_{REC}}{d\Omega} = \dfrac{3}{8\pi} \sigma_{REC} \sin^2 \vartheta.
}
%----------------------------------
Finally, the double differential cross section \({d^2\sigma_{REC}}/{d\Omega dE_\gamma}\) can be expressed as:
%----------------------------------
\muneqn{double_diff}{
\dfrac{d^2\sigma_{REC}}{d\Omega dE_\gamma}=\dfrac{1}{v} \dfrac{d\sigma_{REC}}{d\Omega}\Big{|}_{p=p_0+p_z}\Im(p_z),
}
%----------------------------------
where \(\Im(p_z)\) is the Compton profile of the target electrons. This formula describes the shape of the REC line registered within the x-ray spectrum at a given observation angle.

\section{Bremsstrahlung}
When a charged particle penetrates a gaseous or solid target a continuous x-ray spectrum is emitted. This spectrum is a result of bremsstrahlung processes occurring in the target material, when a charged particle is accelerated (or decelerated) in the Coulomb field of the target components. A schematic explanation of this process for an electron in a field of an ion is presented in Fig.~\ref{fig:bremsstrahlung}. 
Bremsstrahlung was first observed by R\"{o}ntgen in 1895 \cite{rentgen,rentgen2} and since that time has been intensively studied \cite{ishi,ishi1,mir,chu,jak,lud}.
%----------------------------------
\munepsfig[1.4]{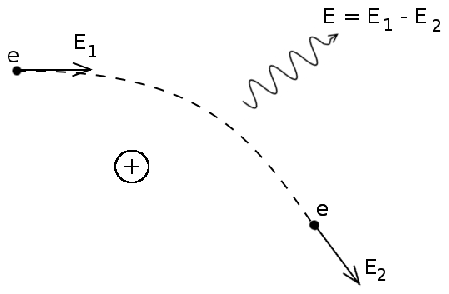}{Example of the bremsstrahlung process for an electron in the electromagnetic field of an ion.}
%----------------------------------

During ion-atom collisions both the ion and ejected electrons may undergo bremsstrahlung processes.
However, the total power radiated via bremsstrahlung is proportional to \(\gamma^4\) (when \(d\overrightarrow{v}/dt \perp \overrightarrow{v}\)) or \(\gamma^6\) (when \(d\overrightarrow{v}/dt \parallel \overrightarrow{v}\)) \cite{gryf}. Since \(E = \gamma mc^2\), where \(m\) is the rest mass of the moving particle, the total radiated power is proportional to \(1/m^4\) or \(1/m^6\), respectively. The above means that electrons lose energy via the bremsstrahlung process much more rapidly than heavier charged particles. This is why electron bremsstrahlung dominates over the ion-related processes.

Quasifree electron bremsstrahlung (QFEB), secondary electron bremsstrahlung (SEB), atomic bremsstrahlung (AB) and nucleus-nucleus bremsstrahlung (NB) dominate among various bremsstrahlung processes that can occur during ion-atom collision. These processes were taken into account during data analysis and are more thoroughly discussed in the following sections.

\subsection{Electron bremsstrahlung}
%----------------------------------
\munepsfig[.75]{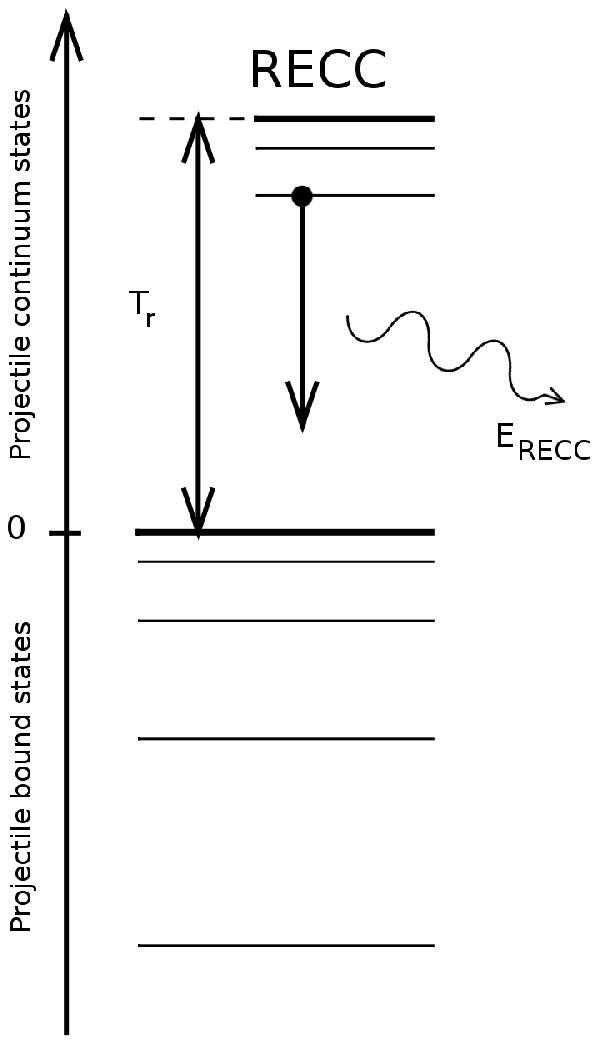}{Radiative electron capture to continuum (RECC). Target electron is captured into a continuum state of the projectile and a photon is emitted.}
%----------------------------------
Radiative electron capture to continuum (RECC), sometimes referred to as quasi-free electron bremsstrahlung (QFEB), is a process where the target electron is captured to the projectile continuum, which means it becomes a free electron. Energy conservation in this process is fulfilled by a photon emission (Fig.~\ref{fig:recc}). 

The maximum kinetic energy (\(T_r\)) of the involved electron, calculated in the projectile frame assuming \(T_r \gg E_{Bt}\), is given by:
%----------------------------------
\muneqn{Tr}{
T_r = \dfrac{1}{2} m_e v^2 = \dfrac{m_e}{m_p} E,
}
%----------------------------------
where \(v\) is the velocity of the incoming ion in the laboratory frame (equal to the velocity of the captured electron in the projectile reference frame). \(T_r\) is the maximum energy (in the projectile frame of reference) of the photon emitted during the RECC process. As the maximum energy of the emitted photons is well defined, the spectrum of the emitted x-rays will have an edge at this value. This edge was observed, for example, during collisions of carbon ions with Be- and C-targets \cite{bed2}, as shown in Fig.~\ref{fig:edge}.

%----------------------------------
\munepsfig[.45]{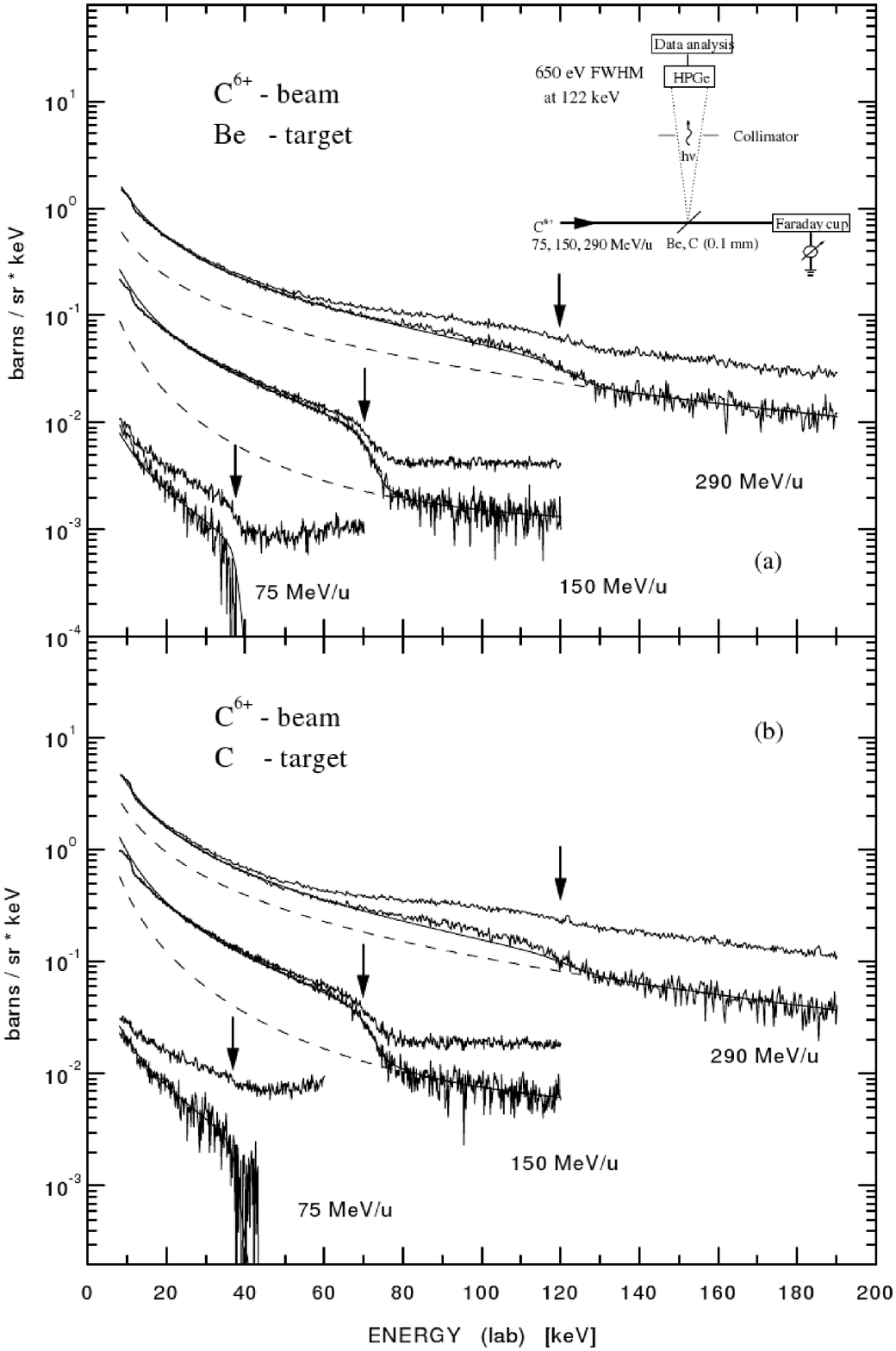}{Spectra observed during the experiment of \cite{bed2} for: (a) Be-target, (b) C-target. For each energy-target combination two spectra are displayedwith the upper spectrum showing the raw data and the lower spectrum with the results after background subtraction. For presentation purposes spectra were multiplied by factors; (a) Be-target:
$1/20$ for $75$~MeV/u, $10$ for $290$~MeV/u; (b) C-target: $1/8$ for $75$~MeV/u, $10$ for $150$~MeV/u, $50$ for $290$~MeV/u. Dashed line: SEB contribution, solid line: RECC (relativistic approximation) + K-REC + SEB. Arrows show the RECC-edge energy $T_r$ transformed to the laboratory frame. Inset in (a) represents the experimental setup.
}
%----------------------------------%----------------------------------
\munepsfig[2]{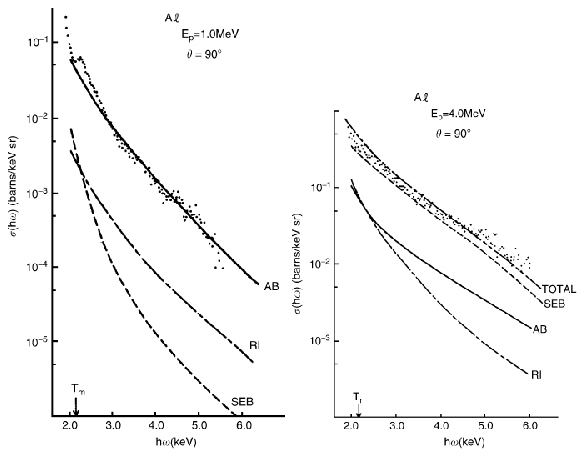}{Example of the contribution of various bremsstrahlung processes to the continuous x-ray spectrum during pl~+~Al collisions at $1$ and $4$~MeV \cite{ishi1}. It can be noticed that SEB becomes a dominating process at higher beam energy.}
%----------------------------------

Ejected target electrons may scatter in the Coulomb field of other target nuclei, producing additional bremsstrahlung. This process is referred to as secondary electron bremsstrahlung (SEB).
In this case the maximum energy (\(T_m\)) of the emitted photons is equal to the maximum transfer of the kinetic energy during ion-electron collisions, given by:
%----------------------------------
\muneqn{Tm}{
T_m = 4 T_r = 4 \dfrac{m_e}{m_p} E.
}
%----------------------------------
Thus, similar to RECC, SEB spectrum has an edge at the photon energy of \(T_m\) \cite{ishi}.

%-----------------------------------
\munepsfig[.7]{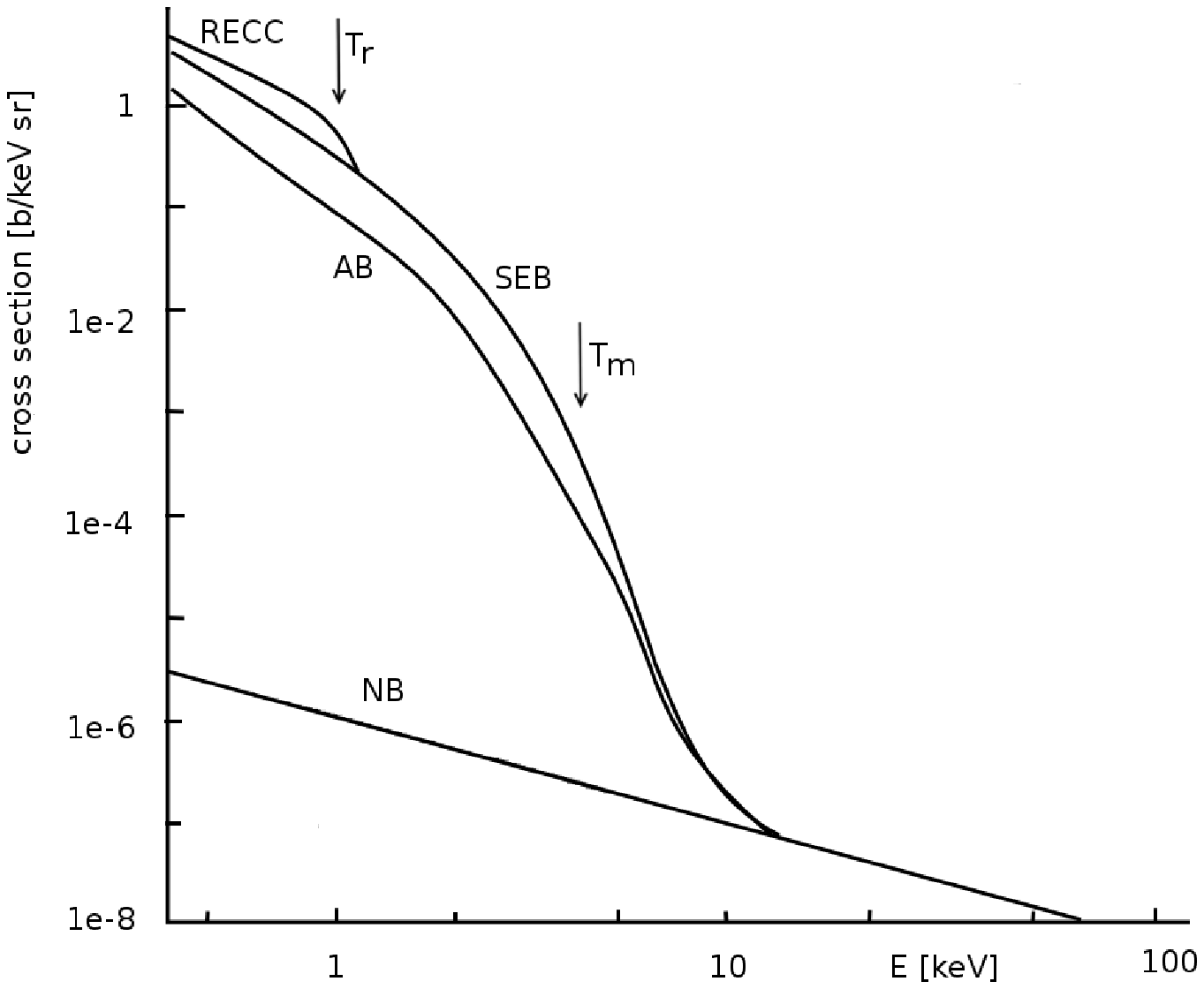}{Bremsstrahlung processes observed during p~+~C collisions at $2$~MeV. Plot based on Fig.~3 in \cite{ishi1}.}
%-----------------------------------

During the atomic bremsstrahlung process (AB), the projectile excites a target electron to a target continuum state. This electron can be recaptured by a target atom with simultaneous emission of x rays. The electron can also lose only part of its energy but remain free, in which case this process is referred to as radiative ionization (RI).

It was shown in \cite{ishi1} that the relative contribution of the above processes strongly varies with projectile energy. The theoretical description of bremsstrahlung cross sections given in \cite{ishi1} is in agreement with experimental data. Comparison of experimental data for p~+~Al collisions at \(1\) and \(4\)~MeV with theoretical calculations are presented in Fig.~\ref{fig:ishii}.
Simple scaling formulae describing double differential bremsstrahlung cross sections were proposed in \cite{ishi1}:
%----------------------------------
\muneqn{scalingRECC}{
\dfrac{(\hbar \omega)^2}{Z^2} \dfrac{d^2 \sigma_{RECC}}{d(\hbar \omega) d\Omega} = Z_t f(\dfrac{m_e}{m_p}\dfrac{E}{\hbar \omega}),
}
\muneqn{scalingAB}{
\dfrac{(\hbar \omega)^3}{Z^2} \dfrac{d^2 \sigma_{AB}}{d(\hbar \omega) d\Omega} = f(\dfrac{a_0 \omega}{v_p}),
}
\muneqn{scalingSEB}{
\dfrac{(\hbar \omega)^2}{Z^2} \dfrac{d^2 \sigma_{SEB}}{d(\hbar \omega) d\Omega} = Z_t^2 f(\dfrac{m_e}{m_p}\dfrac{E}{\hbar \omega}).
}
%----------------------------------
where \(\hbar\omega\) denotes photon energy, \(a_0\) is the Bohr radius and \(f\) is a universal function discussed extensively in \cite{ishi1}.
The bremsstrahlung processes for protons interacting with various targets at a wide range of energies were thoroughly studied for example by Folkmann \cite{folk1,folk2}. By means of the above formulae, the bremsstrahlung contribution to the experimental data can be estimated from the proton data.
 %This allows for the above formulae tested in many experiments to estimate the bremsstrahlung contribution in the experiment discussed in this thesis.

\subsection{Nucleus-nucleus bremsstrahlung (NB)}
\label{section:drec}
Nucleus-nucleus bremsstrahlung is a consequence of the projectile scattering in the Cou-lomb field of the target nuclei. The emitted x-ray spectrum extends up to the projectile energy.
The differential cross section for NB process can be calculated from the formula given by Mokler \cite{sel}:

%-----------------------------------
\muneqn{nnb}{
\dfrac{d\sigma_{NB}}{d(\hbar \omega)}=C \dfrac{Z^2 Z_t^2}{(\hbar \omega) E} A \left(\dfrac{Z}{A} - \dfrac{Z_t}{A_t}\right)^2,
}
\muneqn{c}{
C=\ln \left(\dfrac{\left( 1+\sqrt{1-x}\right)^2}{x} \right)\cdot4.3\cdot10^{-28} [cm^2],
}
\muneqn{x}{
x=\dfrac{A+A_t}{A_t}\dfrac{(\hbar \omega)}{E},
}
%-----------------------------------
where \(A\), \(A_t\) are projectile and target masses in atomic units, respectively.

Fig.~\ref{fig:ishii_background} shows the contribution of RECC, SEB, AB and NB to the x-ray spectra obtained during collisions of \(2.0\)~MeV protons with carbon. Upper limits of RECC and SEB x-ray spectra can be observed at \(T_r\) and \(T_m\), respectively. The NB cross section is significantly lower than those of the electron bremsstrahlung processes. Thus, NB plays a significant role only within the x-ray spectrum range above \(T_m\).

\section{Multielectron capture processes, noncorrelated double radiative electron capture (DREC)}
\label{section:rdec}

During a single ion-atom collision capture of more than one target electron to the projectile bound state is possible.
The simplest example of a noncorrelated capture of two electrons is double radiative electron capture (DREC) for which the capture of two electrons is accompanied by the emission of two independent REC photons (Fig.~\ref{fig:drec_rdec}). During this process the captured electrons do not interact with each other and the capture of each of them can be treated as a separate process.

Radiative noncorrelated double capture was theoretically addressed by Meyerhof \cite{mey}. In this paper the author calculated the REC cross section as an integral of the process probability given as a function of impact parameter.
The single electron capture cross section can be calculated as an integral of the probability \(P(b)\) of an electron capture given as a function of the impact parameter \(b\):
%----------------------------------
\muneqn{single_capture}{
\sigma_{REC}=\int_0^\infty db 2 \pi b P(b).
}
%----------------------------------
In case of REC, \(P(b)\) is given by:
%----------------------------------
\muneqn{mey_REC}{
P(b)=\sigma_{REC}(Z_t=1)\int_{-\infty}^\infty dz\rho(R),
}
%----------------------------------
\(R\) being the projectile to target atom distance and \(\rho\) the electron density. As the electron density is normalized: 
%----------------------------------
\muneqn{normalized}{
\int_0^\infty db 2 \pi b \int_{-\infty}^\infty dz\rho(R) = Z_t,
}
%----------------------------------
the REC cross section for a multielectron target is simply given by \(Z_t \sigma_{REC}(Z_t=1)\).

%----------------------------------
\munepsfig[.75]{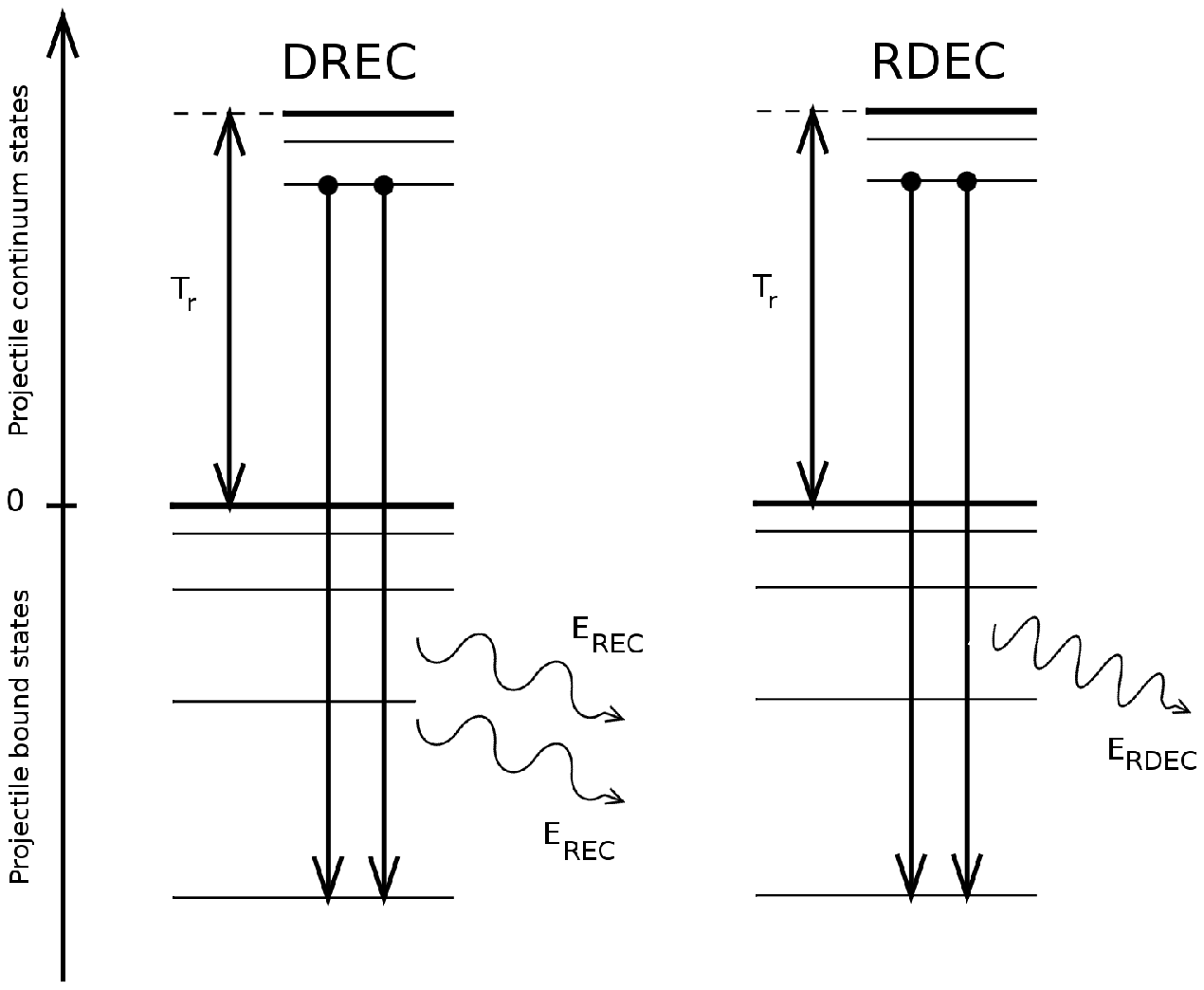}{Comparison of the DREC and RDEC processes.}
%---------------------------------- 
The same method was applied to noncorrelated double radiative capture.
If \(P_0(b)\) is the probability of a single electron capture into the bare ion,  and \(P_1(b)\) is the corresponding probability of electron capture into the H-like ion, the cross section for noncorrelated double electron capture can be expressed as \cite{mey}:
%----------------------------------
\muneqn{double_capture}{
\sigma_{DREC}=\int_0^\infty db 2 \pi b P_0(b) P_1(b).
}
%----------------------------------
For double electron capture (DREC) one obtains the cross section \cite{mey}:
\muneqn{mey_DREC}{
\sigma_{DREC}=0.13Z_t^{1/2}\sigma_{REC}^2(Z_t)a_0^{-2}.
}
The above formula was verified experimentally by Bednarz \cite{bed}.

When the captured electrons interact with each other during the collision, the process is referred to as correlated capture. Radiative double electron capture (RDEC) is the basic example of a correlated process and can be treated as a time inverse of double photoionization. Thus, due to the electron-electron interaction of the two captured electrons only one photon is emitted and its energy is about two times greater than that of a single REC photon (Fig.~\ref{fig:drec_rdec}). The RDEC process is discussed in detail in Chapter~\ref{chap:rdec}.

%----------------------------------
\munepsfig[.5]{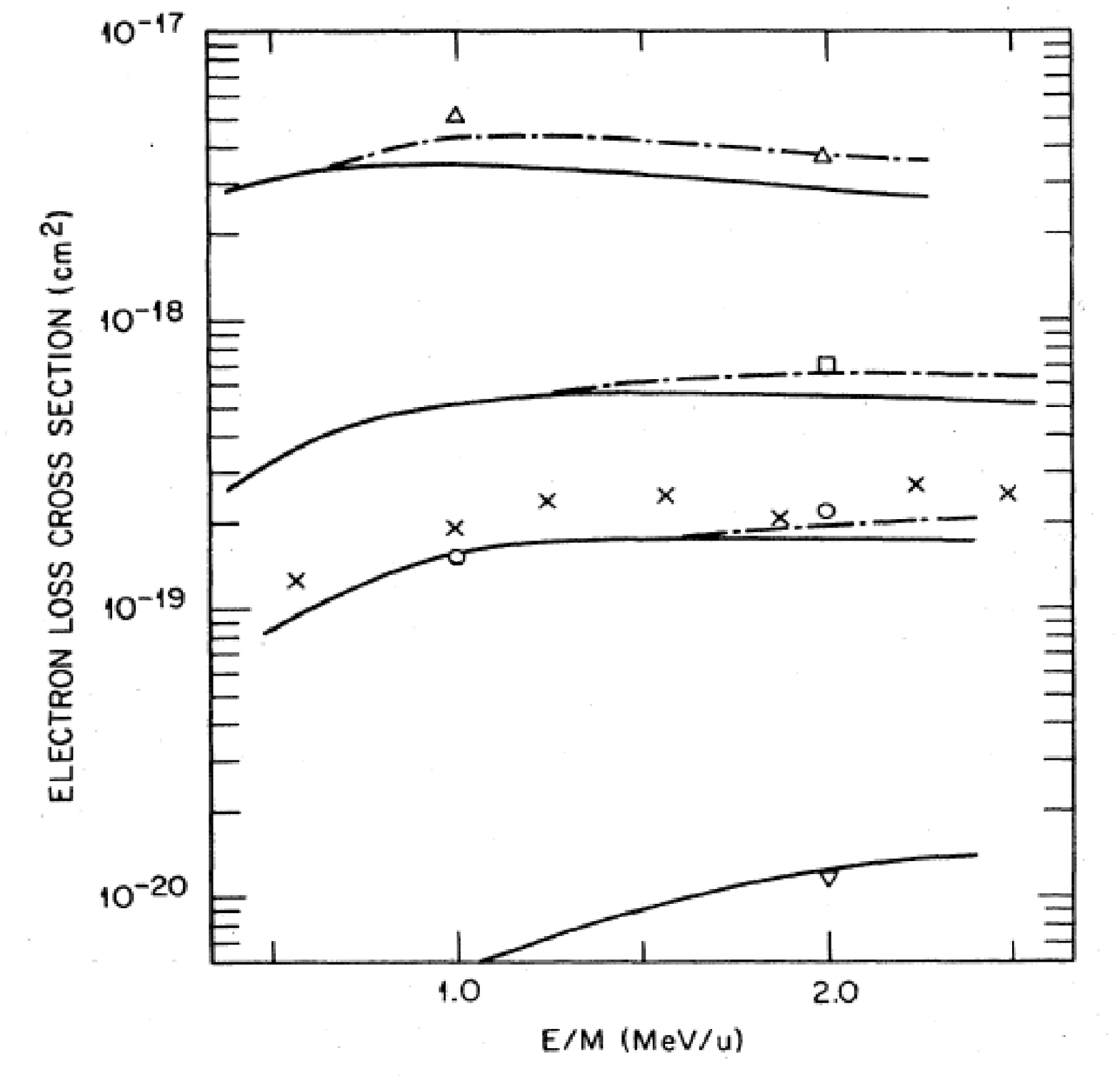}{Single electron loss cross section as given in \cite{hipp}. Solid line: PWBA calculations for He$^{2+}$ impact, dot-dashed line: includes contribution from free electron impact in CBE approximation. Symbols: $\vartriangle$ -- Si$^{8+}$, $\square$ -- O$^{6+}$, $\triangledown$ -- Si$^{13+}$, $\circ$ and $\times$ -- O$^{7+}$.}
%----------------------------------

\section{Projectile ionization -- electron loss}
\label{section:loss}
There is a variety of terminology used in the literature when reffering to electron detachment, which leads to confusion. Here, a nomenclature from \cite{bom,tan,hipp} is applied. The term ionization is used when the electron is detached from the target atom, while the removal of the electron from the projectile bound state is reffered to as electron loss.
Consequently, electron loss is the process where an electron is removed from the projectile and remains free afterwards:
%----------------------------------
\muneqn{loss}{
A^{q+} + A_t \rightarrow A^{(q+x)+} + A_t + xe^-,
}
%----------------------------------
where \(x\) is the number of electrons lost by the projectile (\(q+x \leq Z\)). Electron loss processes have been extensively studied during the late seventies and eighties for various elements and charge states within energy range up to $10$~MeV/u \cite{grah1,grah2,ols}.

Boman et al. \cite{bom} developed a simple scaling formulae for electron loss cross section. The single electron loss cross section for oxygen ions at \(1\)~MeV/u can be estimated as: 
%----------------------------------
%\\
\begin{itemize}
\item 
for \(q=5:\) \muneqn{loss_5}{\sigma_1^5=(3.27 \cdot 10^{-18}) Z_t^{0.98} [cm^2],}
\item 
for \(q=6:\) \muneqn{loss_6}{\sigma_1^6=(8.83 \cdot 10^{-19}) Z_t^{0.78} [cm^2],}
\item 
for \(q=7:\) \muneqn{loss_7}{\sigma_1^7=(2.22 \cdot 10^{-19}) Z_t^{0.33} [cm^2],}
\end{itemize}
%----------------------------------
where \(q\) denotes the initial charge state of the ion.
It has been also checked by the authors that in case of Si$^{8+}$~+~He collisions at \(1.0\)~MeV/u the ratio of single to double electron loss cross sections \({\sigma_1^8/}{\sigma_2^8} \approx 40\). Thus, it can be assumed that the double electron loss process can be neglected for the case of the more tightly bound K-shell electrons in O$^{6+}$.
As can be observed in Fig~\ref{fig:loss} the single electron loss cross section does not change significantly when the beam energy is increased from \(1\) to \(2\)~MeV/u. Thus, the scaling formulae given by Eqs~\ref{eqn:loss_5}-\ref{eqn:loss_7} can be used to estimate the cross sections within this energy range.

\chapter{Radiative double electron capture (RDEC)}
\label{chap:rdec}
\thispagestyle{fancy}
Radiative double electron capture (RDEC) is a one-step process for which two free (or quasifree) target electrons are captured into bound states of the projectile, e.g. into an empty K-shell, and the energy excess is released as a single photon (Fig.~\ref{fig:drec_rdec}). This process has to be compared with a two-step double radiative electron capture (DREC) during which two electrons are captured independently and two single REC photons are emitted.

While for the DREC process both electrons can be treated separately (see Section~\ref{section:drec}), in case of the RDEC one has to go beyond the independent electron model. Here, due to the electron-electron interaction, transitions of two target electrons into the projectile bound states occur with an emission of one photon with an energy about two times greater than that of a single DREC photon.

In general, captured electrons may originate from two different orbitals in the target and arrive finally at different final states in the projectile. Thus, the energy of the RDEC photon can be expressed as:
%----------------------------------
\muneqn{Erdec}{
E_{RDEC} \approx 2T_r + E_B^{(1)} +E_B^{(2)} - E_{Bt}^{(1)} - E_{Bt}^{(2)} + \overrightarrow{v}\overrightarrow{p}^{(1)} + \overrightarrow{v}\overrightarrow{p}^{(2)},
}
%----------------------------------
where the indices \((1)\) and \((2)\) correspond to each of the captured electrons. The width of the peak is about twice as large as that of the REC line. Roughly, it is determined by the sum of Compton profiles of the two active electrons \cite{mir}.

\section{Initial experiments dedicated to RDEC}

The first experiment dedicated to RDEC was performed at GSI in 1994 with \(11.4\)~MeV/u Ar$^{18+}$ ions from the UNILAC impinging upon a carbon foil. A detailed description of this experiment is given in \cite{war}. A typical spectrum obtained during that experiment is presented in Fig.~\ref{fig:argon}. As shown in this figure, no significant line structure related to the RDEC process was observed. However, the number of counts collected in the expected RDEC energy window provided an upper limit of the RDEC cross section of about \(5.2\)~mb.
%-----------------------------------
\munepsfig[.6]{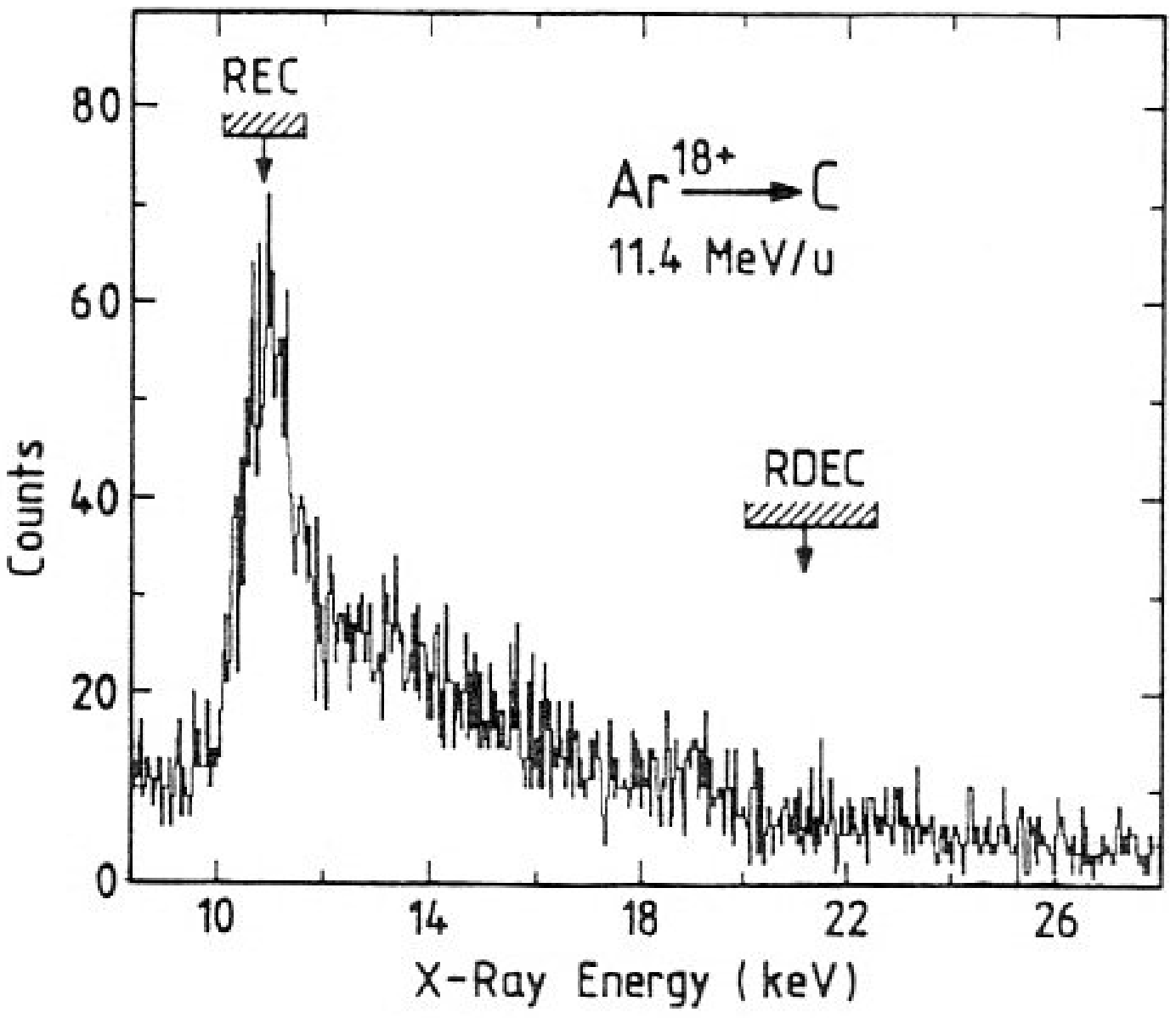}{Typical x-ray spectrum obtained during argon experiment \cite{war}.}
%-----------------------------------
A rough theoretical estimate of the cross section ratio \(\sigma_{RDEC}/\sigma_{REC}\) was also suggested, based on the principle of detailed balance and considering REC as a time reversal of photoionization. 
The REC and RDEC cross sections can be written as:
%-----------------------------------
\muneqn{rec_ph}{
\sigma_{REC}(\beta)=Z_t\left( \dfrac{\hbar \omega}{\gamma \beta m c^2}\right)^2 \sigma_{PI}(\hbar \omega),
}
\muneqn{rdec_ph}{
\sigma_{RDEC}(\beta)=F Z_t (Z_t-1) \left( \dfrac{\hbar \omega'}{2\gamma \beta m c^2}\right)^2 \sigma_{DPI}(\hbar \omega'),
}
%-----------------------------------
where \(\sigma_{PI}\) and \(\sigma_{DPI}\) are the cross sections for single and double photoionization, respectively. The factor \(F\) (\(F\leq 1\)) describes the phase space fraction of double photoionization accessible for the RDEC process.
Thus, the \(\sigma_{RDEC}/\sigma_{REC}\) ratio can be expressed in terms of single and double photoionization cross sections as \cite{war}:
%-----------------------------------
\muneqn{ratio}{
R=\dfrac{\sigma_{RDEC}}{\sigma_{REC}}=F(Z_t-1)\left(\dfrac{\omega'}{2\omega}\right)^2\left(\dfrac{\sigma_{DPI}(\hbar \omega')}{\sigma_{PI}(\hbar \omega)}\right),
}
%-----------------------------------
or, as in case of RDEC \(\hbar \omega' \approx 2 \hbar \omega\):
%-----------------------------------
\muneqn{ratio2}{
R=\dfrac{\sigma_{RDEC}}{\sigma_{REC}}=F(Z_t-1) \dfrac{\sigma_{PI}(2\hbar \omega)}{\sigma_{PI}(\hbar \omega)} \dfrac{\sigma_{DPI}(2\hbar \omega)}{\sigma_{PI}(2 \hbar \omega)}.
}
%-----------------------------------
where the values of the expresssions: \(\sigma_{PI}(2\hbar \omega)/\sigma_{PI}(\hbar \omega)\) and \(\sigma_{DPI}(2\hbar \omega)/\sigma_{PI}(2 \hbar \omega)\) can be easily estimated from \cite{fano,amu2}. The experiment \cite{war} provided an upper limit for \(R\) of \(3.1\cdot10^{-6}\).

%-----------------------------------
\munepsfig[.6]{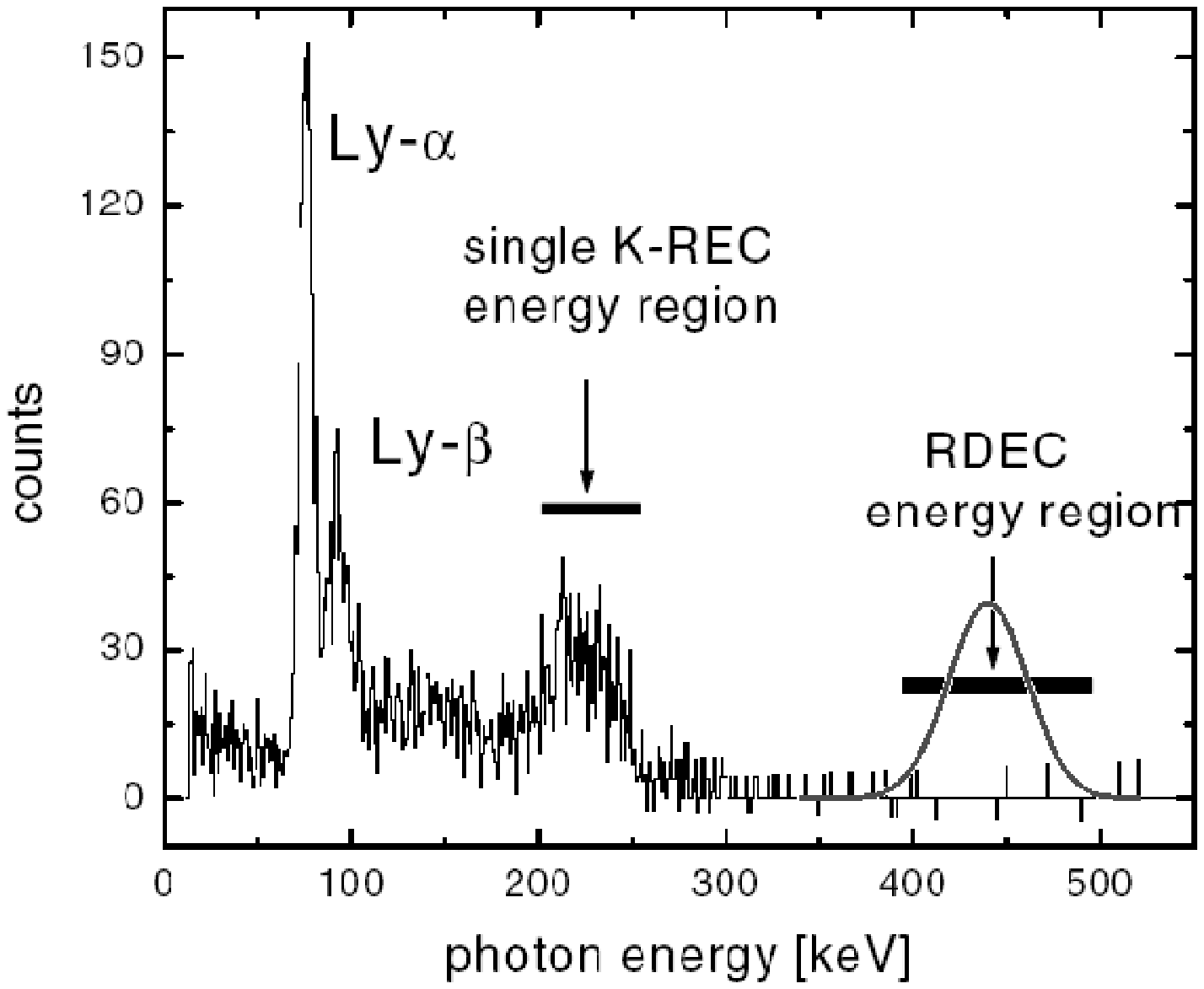}{Typical experimental x-ray spectrum obtained for uranium ions \cite{bed}. The Gaussian solid line shows the expected RDEC peak, which should be observed according to Yakhontov et al. \cite{yak1,yak2}. }
%-----------------------------------

This experiment stimulated theoretical treatment of the RDEC process \cite{yak1,yak2}. In these papers the authors presented nonrelativistic calculations of the RDEC process adapted to the kinematics and the energy range of the Ar$^{18+}$~+~C experiment. The calculations gave, for this particular collision system (Ar$^{18+}$~+~C at \(11.4\)~MeV/u), the RDEC to REC cross section ratio \(R\) of \(3.6\cdot10^{-6}\), which is close to the experimental upper limit.

Moreover, these calculations predicted a strong enhancement of RDEC during heavy ion-atom collisions at relativistic energies \cite{yak2}. These calculations were tested during the second experiment dedicated to RDEC. Here, bare uranium ions at an energy of \(297\)~MeV/u collided with an Ar target at the ESR storage ring of the GSI facility \cite{bed}. This experiment showed that for the collision system under consideration the RDEC cross section is certainly at least three orders of magnitude smaller than the theoretical prediction \cite{yak1,yak2}. Fig.~\ref{fig:uranium_gaus} shows a spectrum obtained during the experiment. Again, no significant line structure which could be assigned to the RDEC process was observed. The Gaussian line in Fig.~\ref{fig:uranium_gaus} shown within the RDEC region of the spectrum represents the shape of the RDEC line which should be observed in the spectrum, if the theoretical calculations \cite{yak1,yak2} were reliable. This experiment also provided only an upper limit for the RDEC cross section value of about \(10\) mb. 

\section{Recent theoretical approach to RDEC}

In order to explain the disagreement between the uranium experiment \cite{bed} and the theoretical treatment of RDEC \cite{yak1,yak2}, a new theoretical approach for the correlated double electron capture into the K-shell of bare ions was proposed \cite{mikh1,mikh2,nef}. Here, a brief description of this RDEC treatment is given with the notation used in the original papers. Indices \((1)\) and \((2)\) correspond to REC and RDEC, respectively, and natural units \((\hbar=c=1)\) are used throughout the text.

All the electrons involved in the process were considered as nonrelativistic and the energy \(\omega\) of the emitted photon was limited by \(I_{2K}\leq \omega \ll m\), where \(I_{2K}\) is the threshold energy for double photoionization of the projectile K-shell and \(m\) is the electron mass. In such case the Coulomb parameter (\(\alpha Z\), \(\alpha\) denotes the fine structure constant) is small (\(\alpha Z\ll 1\)) and perturbation theory with respect to the electron-electron interaction can be used.

In the reference frame of the incident ion the probability \(dW\) for double electron capture into the K-shell of bare ion with the emission of a single photon per unit time is given by \cite{mikh1}:
%-----------------------------------
\muneqn{dW_prob}{
dW=\dfrac{2\pi}{V^2}|A|^2 \dfrac{d \overrightarrow{k}}{(2\pi)^3}\delta(2E_P - \omega - I_{2K}),
}
%-----------------------------------
where \(E_P\) is the one-electron energy within the initial continuum state, \(\omega=|\overrightarrow{k}|=k\) is the energy of the emitted photon and \(I_{2K}=2I\) with \(I=\eta^2/2m\), the Coulomb potential for single ionization and \(\eta=m\alpha Z\) the characteristic momentum of the K-shell electron, and \(V\) is a normalization factor. Summation over all polarizations of the emitted photon is assumed in Eq.~\ref{eqn:dW_prob} and the delta function ensures energy conservation. The amplitude \(A\) was obtained from that for the double K-shell photoionization. A detailed description of this approach is given in \cite{mikh1}.

Dividing Eq.~\ref{eqn:dW_prob} by the current flux of the incident target electrons \(j=v/V\), where \(v=p/m\) is the absolute value of the initial velocity of the incident electrons before collision with the ion, one obtains the effective differential cross section:
%-----------------------------------
\muneqn{d_sigma}{
d\sigma^{(2)}=2\pi\dfrac{\omega^2}{vV}|A|^2\dfrac{d\Omega_k}{(2\pi)^3},
}
%-----------------------------------
which defines the angular distribution of the RDEC photons emitted into an element of a solid angle \(d\Omega_k\).

%-----------------------------------
\munepsfig[.4]{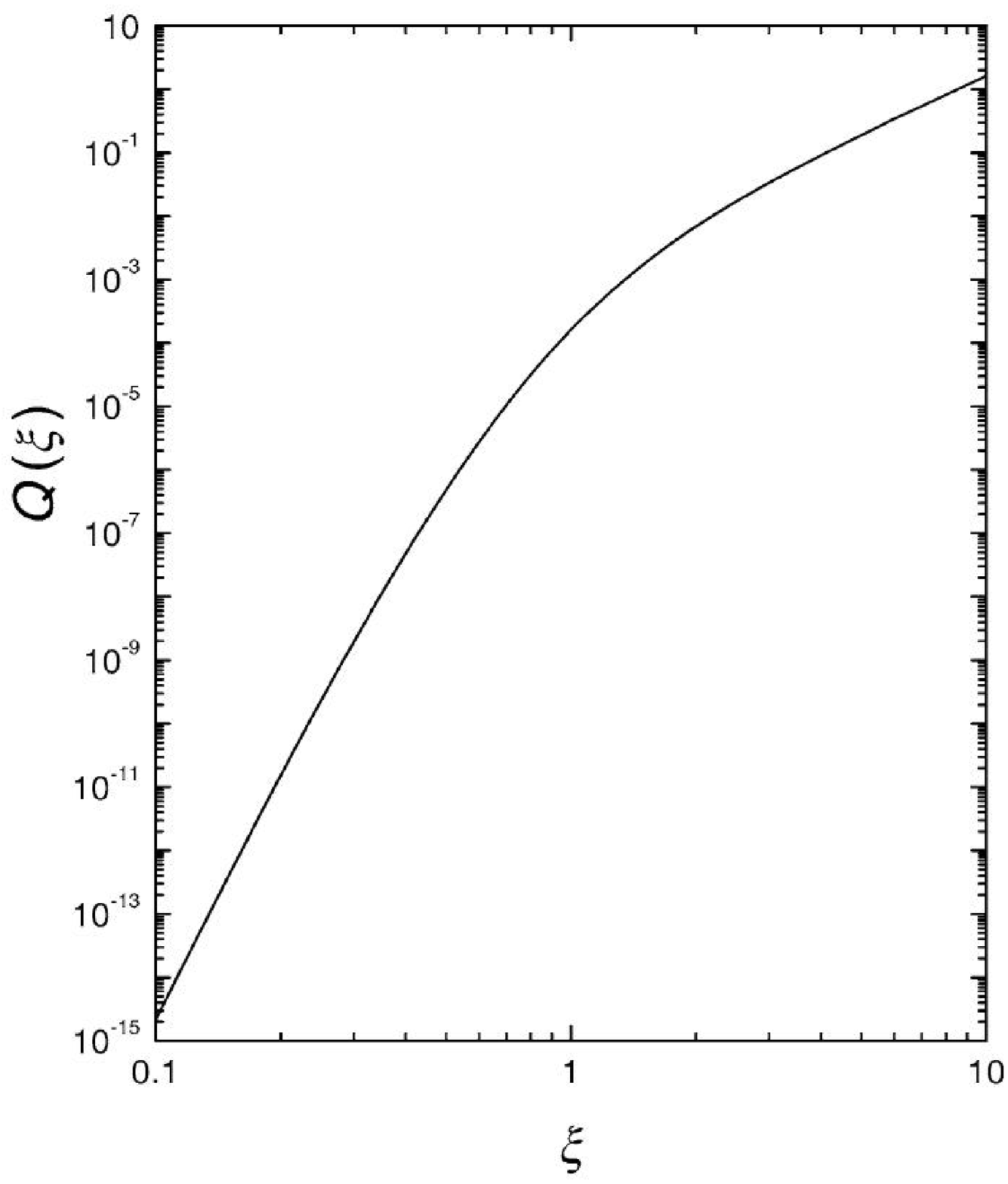}{Universal function $Q$ calculated as a function of the dimensionless variable $\xi$ \cite{mikh1}.}
%-----------------------------------

For the energy regime assumed in these calculations, it was possible to calculate the total cross sections within the electric dipole approximation. For collisions of heavy ions with light target atoms the total cross section for radiative double electron capture (RDEC) into the K-shell of the ion is given by:
%-----------------------------------
\muneqn{sigma_RDEC}{
\sigma^{(2)}=\dfrac{2^{19} Z_t^3}{3 \pi Z^5}Q(\xi),
}
%-----------------------------------
where \(\xi=\eta/p\) is a dimensionless parameter, \(\sigma_0 = \alpha^3 a_0^2\) and \(a_0\) denotes the Bohr radius. \(Q\) is a universal function of \(\xi\), which can be obtained by numerical integration (Fig.~\ref{fig:Q}). \(\xi \sim 1\) corresponds to the near-threshold domain, where the K-shell photoeffect reaches its maximum. For slow collisions \((\xi \gg 1)\) the RDEC cross section increases, while in case of fast collisions it decreases rapidly. Moreover, it has to be pointed out that the RDEC cross section rapidly drops with the projectile atomic number (\(\sim Z^{-5}\)) and increases significantly for low energy collisions.

%-----------------------------------
\munepsfig[.4]{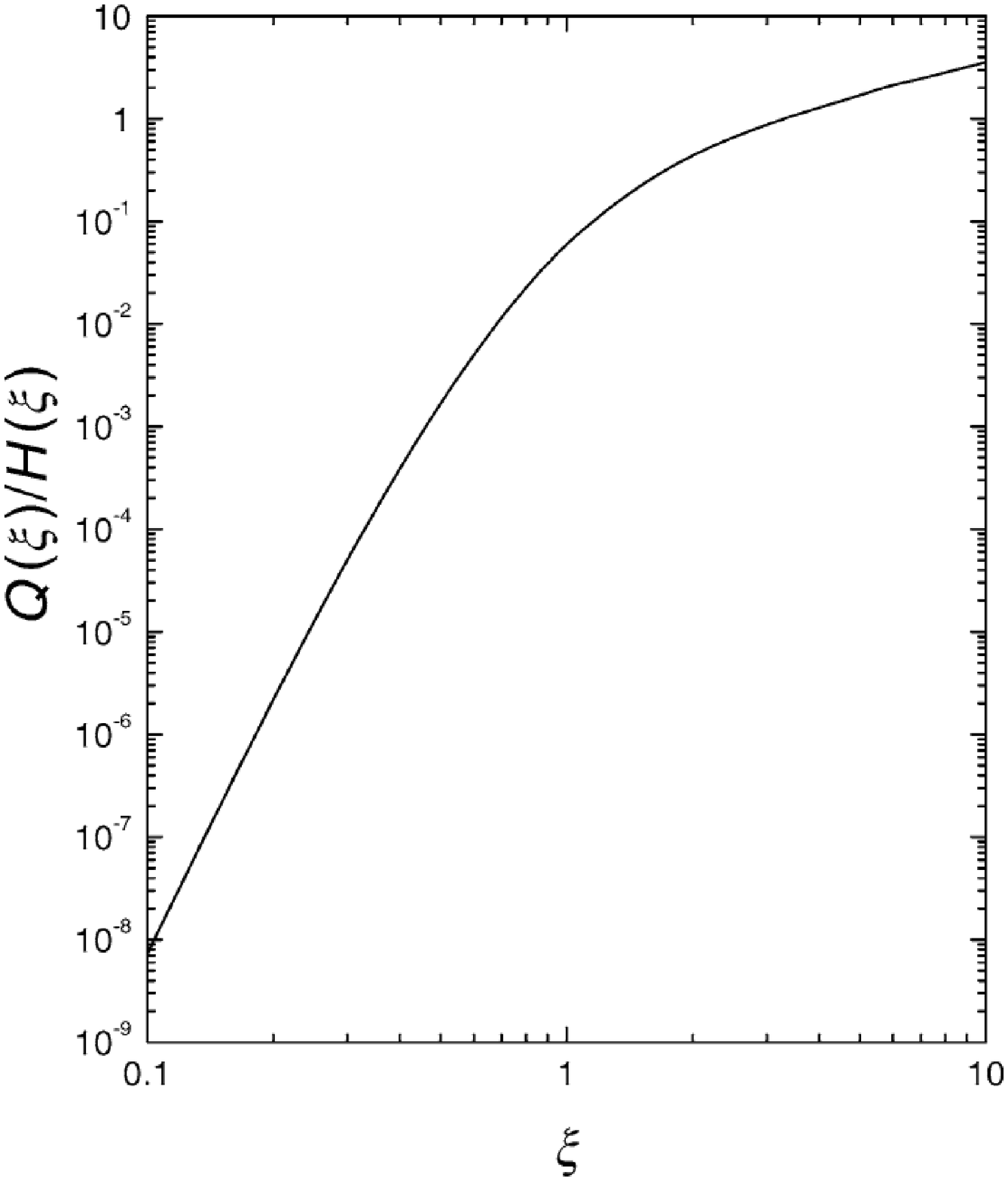}{Universal quantity $Q/H$ calculated as a function of the dimensionless variable $\xi$ \cite{mikh1}.}
%-----------------------------------

Another value calculated in \cite{mikh1} is the cross section ratio (\(R = \sigma^{(2)} / \sigma^{(1)}\)). The REC cross section can be expressed in terms of the photoionization cross section (\(\sigma_{PI}\)) via the principle of detailed balance. As \(\sigma_{PI}\) is known analytically (Stobbe formula \cite{stob}), for radiative electron capture into the K-shell of the projectile one obtains:
%-----------------------------------
\muneqn{sigma_REC}{
\sigma^{(1)}=\dfrac{2^{10}}{3}\pi^2 \sigma_0 Z_t H(\xi),
}
%-----------------------------------
%-----------------------------------
\muneqn{H}{
H(\xi)=\dfrac{\xi^2}{\varepsilon_\gamma^2}\dfrac{\exp(-4 \xi \cot^{-1}\xi)}{1 - \exp(- 2\pi \xi )},
}
%-----------------------------------
where \(\varepsilon_\gamma=\omega/I=2(1+\xi^{-2})\) is the dimensionless photon energy. Then the ratio \(R\) is given by:
 %-----------------------------------
\muneqn{R}{
R=\dfrac{2^9 Z_t^2 Q(\xi)}{\pi^3 Z^5 H(\xi)}.
}
%-----------------------------------
The function \(Q(\xi)/H(\xi)\) is presented in Fig.~\ref{fig:Q_H}.

These calculations are in disagreement with the previous relativistic approach \cite{yak2}, which was not able to explain the existing experimental data \cite{bed}. As shown in \cite{mikh1}, the enhancement of the wave function for the relativistic systems was calculated incorrectly by Yakhontov \cite{yak2} and even the corrected value, which is \(3\) orders of magnitude smaller \cite{mikh1}, was not confirmed by the experiment \cite{bed}. Therefore, the enhancement of the RDEC cross section for relativistic systems \cite{yak2} seems to be absent. This is similar to the behavior of the cross section for the REC process, where the cross section decreases when the projectile energy increases.

However, it has to be emphasized that the current estimate \cite{mikh1} of \(\sigma^{(2)}\) gives values closer to the experimentally obtained upper limits for both the nonrelativistic case (Ar$^{18+}$~+~C, \cite{war}) and the relativistic one (U$^{92+}$~+~Ar, \cite{bed}) (see Table~\ref{tab:comparison}), which suggests that \cite{mikh1} is so far the most reliable theoretical description of RDEC.
%-----------------------------------
\muntab{@{\extracolsep{\fill}}lcccccc}{comparison}{Comparison of experimentally obtained RDEC cross sections \cite{war,bed} and the calculated values given in \cite{yak2} and \cite{mikh1}.}
{\hline
\multirow{2}{*}{\(Z\)} &\multirow{2}{*}{\(E\) [MeV/u] }  & \multirow{2}{*}{\(\xi\) } & \multirow{2}{*}{ \(Z_t\) } & \multicolumn{3}{c}{$\sigma$$^{(2)}$ [mb]}\\
\cline{5-7}
 	& 	& 	& 	& Ref. \cite{mikh1} 	& Ref. \cite{yak2} 	& experiment\\
\hline
18 	& 11.4 	& 0.840 & 6 	& 3.2 			& 1.85 			& $\leq$5.2 \cite{war} \\
92 	& 297 	& 0.841	& 18 	& 2.5$\cdot$10$^{-2}$ 	& 5000 			& $\leq$10 \cite{bed} \\
\hline}
%-----------------------------------
%-----------------------------------
\munepsfig[.35]{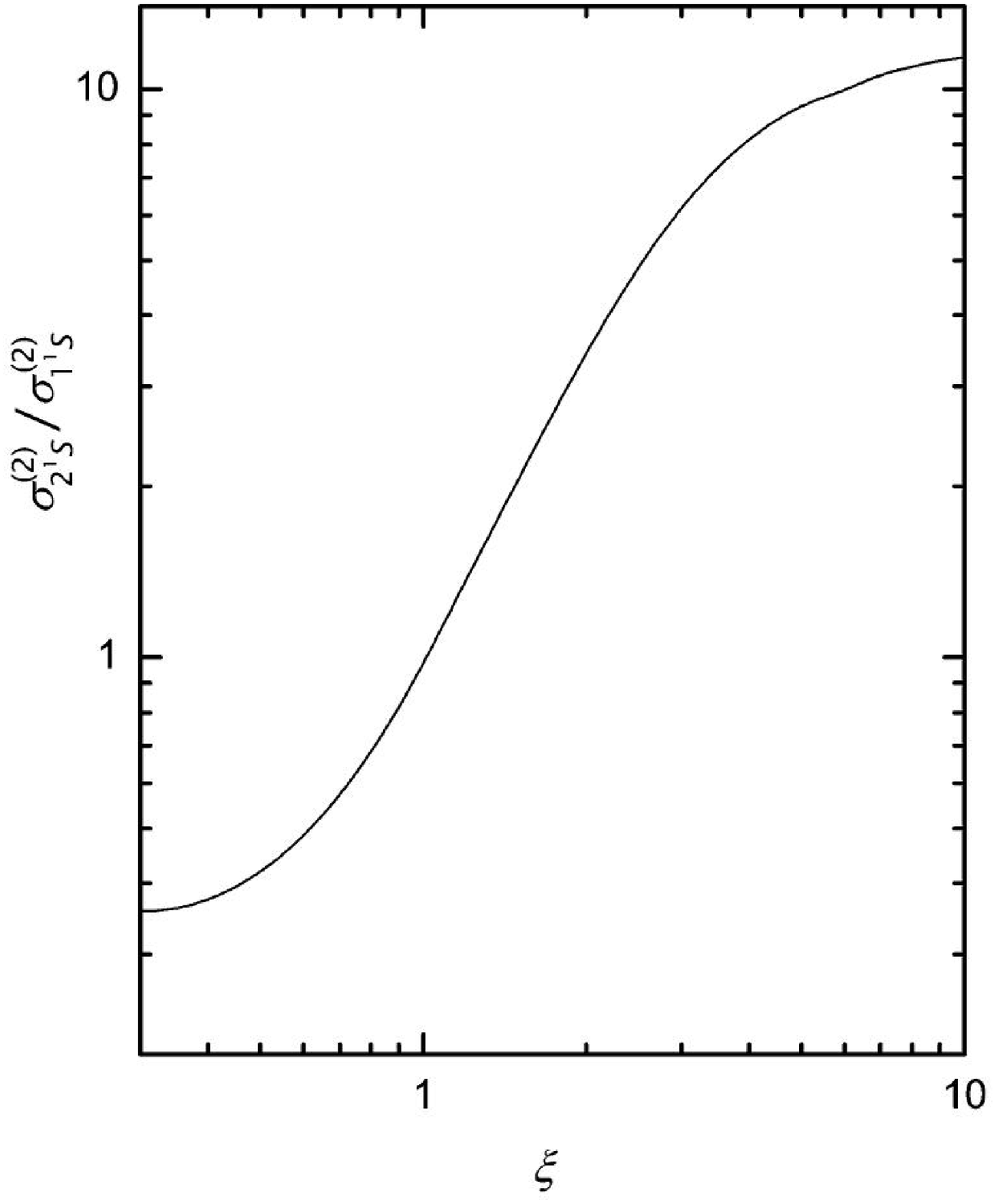}{The ratio of the RDEC cross sections to the excited ($\sigma^{(2)}_{2^1S}$) and ground ($\sigma^{(2)}_{1^1S}$) projectile states as a function of adiabacity parameter $\xi$ \cite{nef}.}
%-----------------------------------

In contradiction to predictions given in \cite{yak2}, the new calculations show that the RDEC cross section strongly depends on the target atomic number and electron density. One can expect much larger values of \(\sigma^{(2)}\) in case of slow collisions of multicharged ions with a solid state target with low atomic number \(Z_t\) \cite{mikh1}. As the orbital velocity of the target valence electrons is much smaller than that of the projectile, they can be considered as quasifree in the projectile's frame of reference. In this reference frame these electrons appear as an electron beam with velocity \(v\) and concentration \(n_e=\kappa \rho_t N_A / M_t\), where \(\kappa\) is the number of valence electrons, \(N_A\) is Avogadro's number and \(\rho_t\) and \(M_t\) are the density and molar mass of the target, respectively. Hence, by substituting \(V=n_e^{-1}\) in Eq.~\ref{eqn:d_sigma} \(\sigma^{(2)}\)can be expressed as:
 %-----------------------------------
\muneqn{RDEC_solid}{
\sigma^{(2)}=(n_e a_0^3)\dfrac{2^{19} \sigma_0}{3 Z^5}Q(\xi).
}
%-----------------------------------

In addition, the correlated double electron capture into the \(1s2s\) state increases the cross section for the RDEC process \cite{nef}. As shown in Fig.~\ref{fig:sigma}, the ratio of the cross section for RDEC to the \(1s2s\) state, \(\sigma_{2^1 S}^{(2)}\), to the cross section for RDEC to the \(1s^2\) ground state, \(\sigma_{1^1 S}^{(2)}\), is strongly dependent on the \(\xi\) value. As can be seen from Fig.~\ref{fig:sigma}, for \(\xi \gg 1\) (i.e. slow collisions) the cross section for electron capture to the \(1s2s\) state can greatly exceed that for \(1s^2\) state capture.

Recently, the calculations of Nefiodov and Mikhailov were continued by Drukarev \cite{druk}, who again addressed the high energy nonrelativistic limit (\(\xi \gg 1\)) of the RDEC process. As previously found, a strong energy dependence of the cross section was shown and the RDEC probability was compared with the one for noncorrelated capture. Obtained values of the REC, RDEC and DREC cross sections for Ar$^{18+}$~+~C and Mg$^{12+}$~+~C for various projectile energies are given in Table~\ref{tab:drukarev}.
%-----------------------------------
\muntab{@{\extracolsep{\fill}}lcccccccc}{drukarev}{The REC ($\sigma^{(1)}$), RDEC ($\sigma^{(2,\gamma)}$) and DREC ($\sigma^{(2,2\gamma)}$) cross sections and their ratios as given in \cite{druk}.}
{\hline
$Z$ &$\xi$ &$E$ [MeV/u] &$Z_t$ &$\sigma^{(1)}$ [kb] &$\sigma^{(2,2\gamma)}$ [mb] &$\sigma^{(2,\gamma)}$ [mb] &$\sigma^{(2,\gamma)}$/$\sigma^{(1)}$ &$\sigma^{(2,\gamma)}$/$\sigma^{(2,2\gamma)}$\\
\hline
\multirow{3}{*}{18}  &0.84 &11.4 &\multirow{3}{*}{6}  &0.36 &1.5 &3.2 &8.9$\cdot$10$^{-6}$ &2.1\\
		     &0.20 &646 &		      &1.5$\cdot$10$^{-3}$ &2.6$\cdot$10$^{-5}$ &1.0$\cdot$10$^{-6}$ &6.7$\cdot$10$^{-10}$ &3.2$\cdot$10$^{-2}$\\
		     &0.10 &804 &		      &6.4$\cdot$10$^{-5}$ &4.7$\cdot$10$^{-8}$ &1.6$\cdot$10$^{-10}$ &4.0$\cdot$10$^{-12}$ &3.4$\cdot$10$^{-3}$\\
\multirow{3}{*}{12}  &0.84 &5.1 &\multirow{3}{*}{6} &0.36 &1.5 &24 &6.7$\cdot$10$^{-5}$ &16\\
		     &0.20 &287 &		    &1.5$\cdot$10$^{-3}$ &2.6$\cdot$10$^{-5}$ &7.6$\cdot$10$^{-6}$ &5.1$\cdot10^{-9}$ &0.29\\
		     &0.10 &357 &		    &6.4$\cdot$10$^{-5}$ &4.7$\cdot$10$^{-8}$ &1.2$\cdot$10$^{-9}$ &1.9$\cdot10^{-11}$ &2.6$\cdot$10$^{-2}$\\
\hline}
%-----------------------------------

This theory \cite{mikh1,mikh2,nef,druk} suggests that the best systems for observation of the RDEC process are low energy collisions of mid-\(Z\) ions with light solid targets. This theory was a motivation for the next RDEC experiment and a reason for the choice of the conditions of the experiment presented in this thesis.

\chapter{Experimental setup at Western Michigan University}
\label{chap:exp}
\section{Van de Graaff accelerator}
The Van de Graaff accelerator is an electrostatic generator which uses a moving belt to accumulate very high, electrostatically stable voltage on a hollow metal sphere \cite{gra2}.
This type of generator was developed by Robert J. Van de Graaff at Princeton University. The first model was demonstrated in October 1929 and in 1931 a version able to produce a potential difference of 1~MV was described \cite{gra}. 

A simple Van de Graaff generator is presented in Fig.~\ref{fig:graaff}. A belt of dielectric material runs over two rollers, one of which is surrounded by a hollow metal sphere the high voltage terminal. 
Two electrodes, an upper and a lower one, are placed next to each roller. The upper electrode is connected to the sphere, while a high DC potential (with respect to the ground potential) is applied to the lower one, a positive potential in the example.

Due to the strong electric field the air around the lower electrode is ionized and the positive ions are repelled from the electrode and accumulated on the belt. Then they are transported towards the upper electrode which collects the charges from the belt and transports them onto the spherical collecting electrode. The potential of the HV electrode increases until the speed of its charging equals the speed of discharging. The maximum potential obtained on the HV electrode depends on the radius of the sphere and insulating properties of the gases surrounding it. SF$_6$ or a mixture of N$_2$ and CO$_2$ under a pressure even up to \(20\) bar are usually used \cite{hin}. The value of the terminal voltage in Van de Graaff accelerators may reach up to \(15\)-\(20\)~MV \cite{edw,shev}. 

If a source of positive ions is placed close to the high voltage terminal, as in Fig.~\ref{fig:graaff}, the ions are repelled by the positive charge of the terminal electrode and thus accelerated towards the ground potential.
The final kinetic energy of the ions depends on their charge state \(q\) and is proportional to the terminal voltage \(V_{terminal}\):
%----------------------------------
\muneqn{acc}{E=qV_{terminal}.}
%----------------------------------

%----------------------------------
\munepsfig[.7]{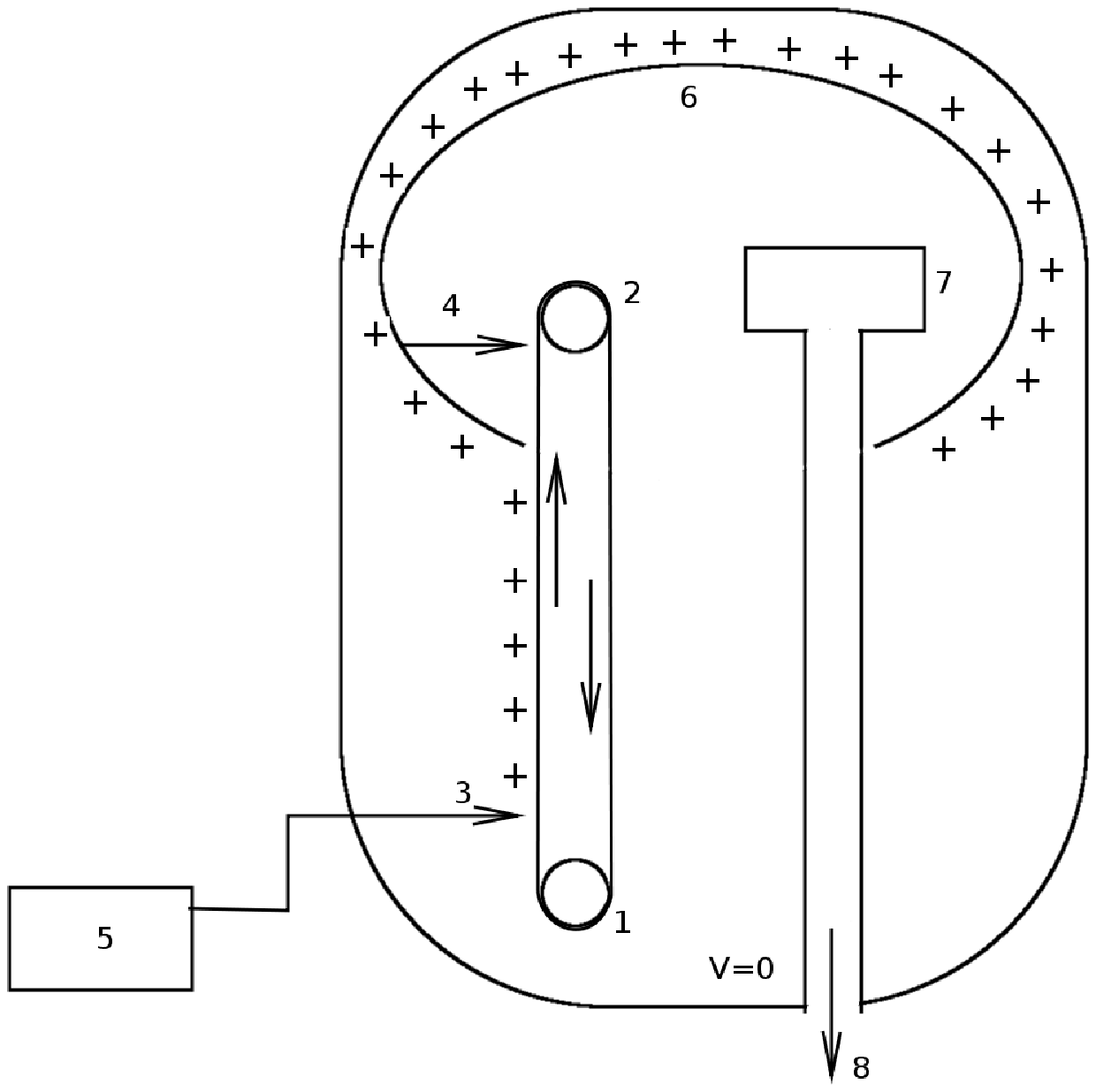}{
Schematic view of a classical Van de Graaff accelerator:
(1) lower roller,
(2) upper roller,
(3) charging electrode,
(4) electrode collecting positive charge,
(5) voltage generator,
(6) spherical electrode (high voltage terminal),
(7) ion source,
(8) extracted ion beam.
}
%----------------------------------

In modern ion accelerators with a Van de Graaff generator, electrodes located at entry and exit of the vacuum tube are grounded and the high-voltage terminal is located at the middle of the tube, as shown in Fig.~\ref{fig:tandem} \cite{hin,wied}. A source of negative ions is placed at the entrance of the tube and produced ions, usually singly charged, are accelerated within the tube towards the high-voltage terminal, where two or more electrons are removed from each ion as it passes through a stripping foil. The charge state of the ion changes from negative to positive and the ion is repelled from the terminal and accelerated towards the grounded exit of the tube. Compared to Van de Graaff accelerators of the ordinary type, by means of tandem Van de Graaff accelerators higher particle energies can be obtained since the potential difference is used for the acceleration twice. Thus, in this case the final kinetic energy can be calculated as:
%----------------------------------
\muneqn{acc_t}{E=(q+1)V_{terminal},}
%----------------------------------
where \(q\) is the ion charge state after passing through the stripping foil.
%----------------------------------
\munepsfig[.8]{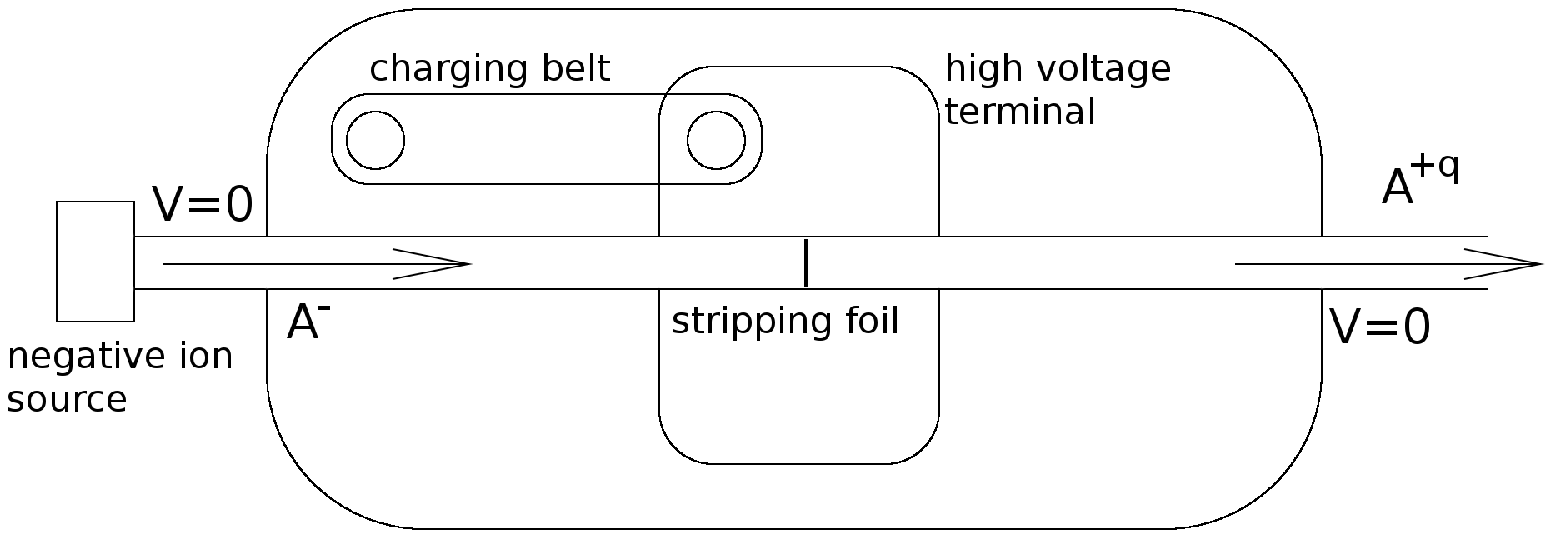}{Schematic view of a tandem Van de Graaff accelerator:
A negative ion entering the accelerator (A$^-$) is accelerated by the high terminal voltage. Some of its electrons are removed while the ion passes through the stripping foil. The positive ion (A$^{+q}$) is repelled by the high voltage terminal, thus causing additional acceleration.}
%----------------------------------

The experiment described in this dissertation was performed at Western Michigan University using \(6\)~MV tandem Van de Graaff accelerator. The WMU accelerator was built by the High Voltage Engineering Corporation, the company founded by Robert Van de Graaff. % It was installed in 1969 when the Physics Department building was being built \cite{WMU}.
The construction of the accelerator allows for obtaining stable beams of bare ions of all elements up to $^9$F with a total kinetic energy up to \(40\)~MeV.
%----------------------------------
\munepsfig[0.5]{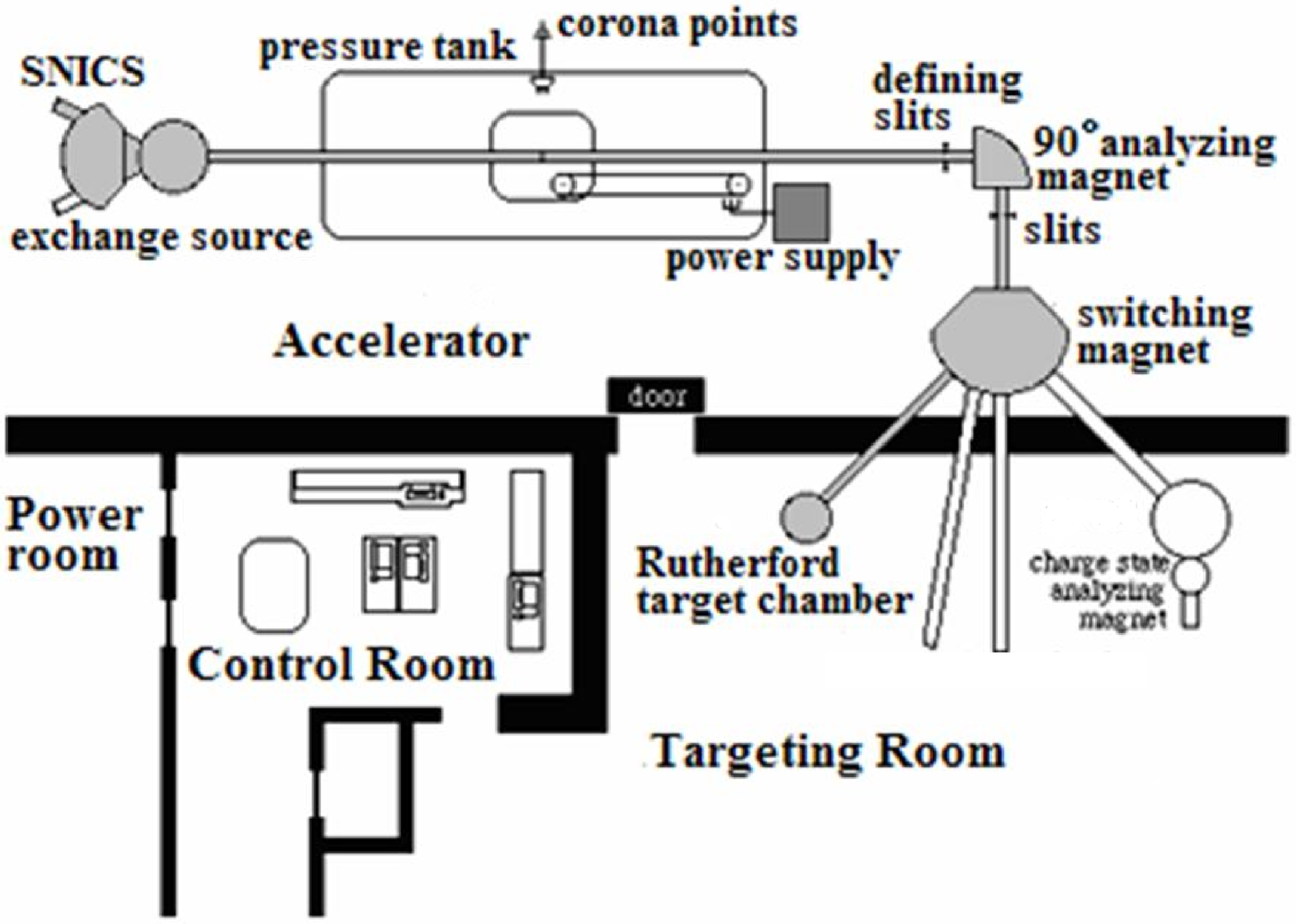}{Schematic view of the WMU van de Graaff accelerator facility \cite{WMU}.}
%----------------------------------

\section{Beam line setup at Western Michigan University}

A schematic view of the WMU accelerator beam line is presented in Fig.~\ref{fig:accelerator_map}. The accelerated beam passes through a \(90^{\circ}\) analyzing magnet which allows for choosing the appropriate ion charge state. At this point, the final energy of the beam is defined accordingly to Eq.~\ref{eqn:acc_t}. Then the beam passes through a post stripper followed by a switching magnet which directs the desired charge state towards the experimental area.
%----------------------------------
\munepsfig[.968]{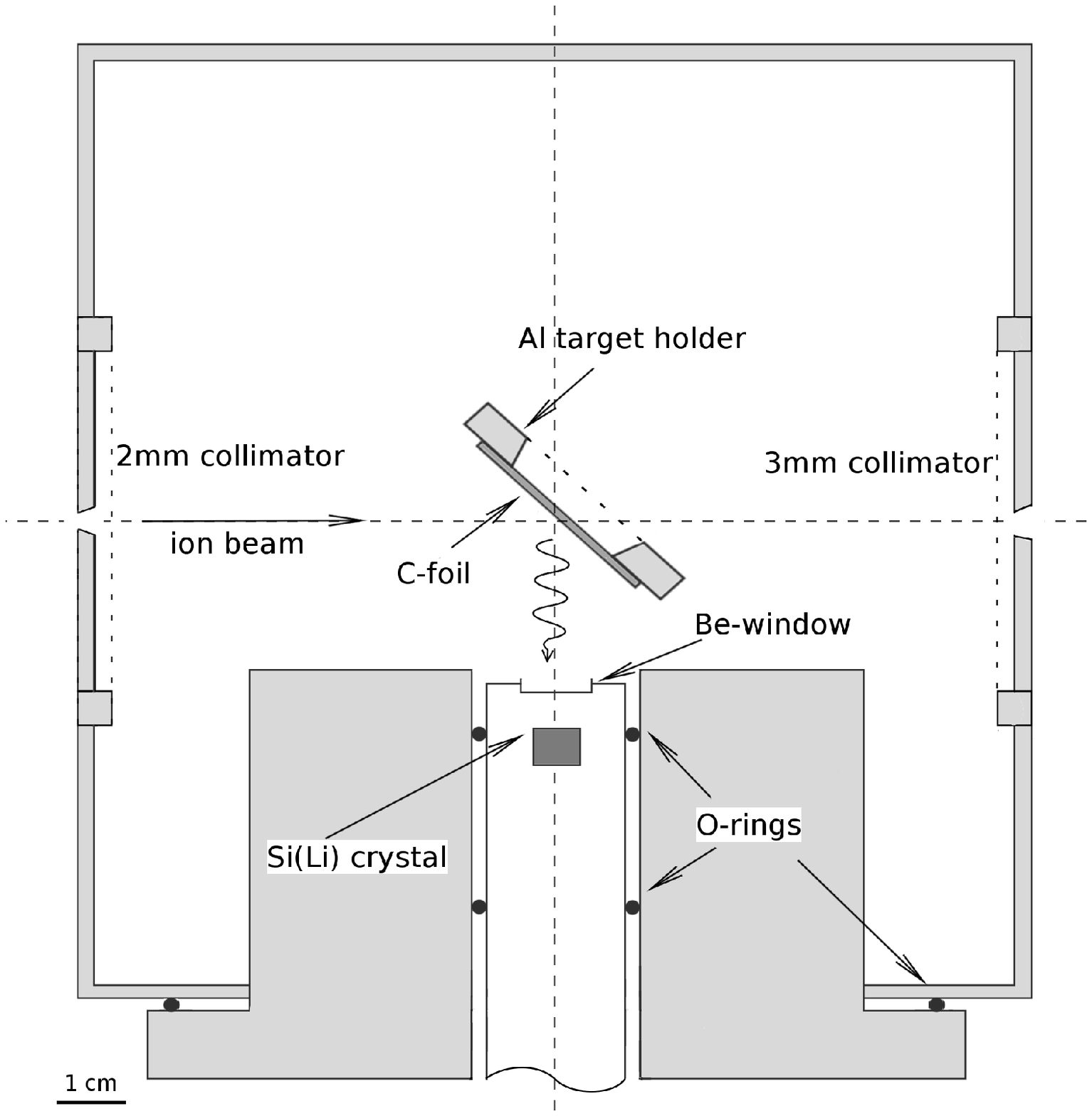}{The experimental target chamber in 1:1 scale. }
%----------------------------------
For the presented experiment a beam of O$^{6+}$ was extracted from the accelerator operating at the terminal voltage of \(5.43\)~MV, which produced a beam of energy equal to \(38\)~MeV. Then the beam traversed through a \(20\) $\mu$g/cm$^2$ carbon stripper foil, where bare and H-like oxygen ions were produced. Simply, by changing the magnetic field of the switching magnet one could choose the necessary charge state.
When a proton beam was accelerated, the post-stripper was removed from the beam line.

The experimental beam line farthest left, when looking along beam direction, was used during this experiment. There, an exclusively designed chamber for a solid target was placed, which not only allowed for mounting up to four films but also provided a simple mechanism for target rotation. This was necessary for optimization of the target position during the experiment. During data acquisition the target film was positioned at \(45^\circ\) to the beam direction, facing the x-ray detector as shown in Fig.~\ref{fig:uklad_exp}. This setup ensured a direct detection of emitted photons, as they did not traverse through the foil, so the unnecessary energy loss was avoided. It also allowed for usage of the whole active area of the x-ray detector, which was not covered by the aluminum frame of the target holder. The target foils used during the experiment were a few $\mu$g/cm$^2$ thick which corresponds to an areal density on the order of \(10^{17}\) particles/cm$^2$.

The target chamber was designed in a way that minimizes the distance between detector window and target center. The total crystal-target distance achieved was about \(25\)~mm, which gives a detection solid angle of \(\Delta\Omega=0.044(1)\)~sr.

%----------------------------------
\munepsfig[0.40]{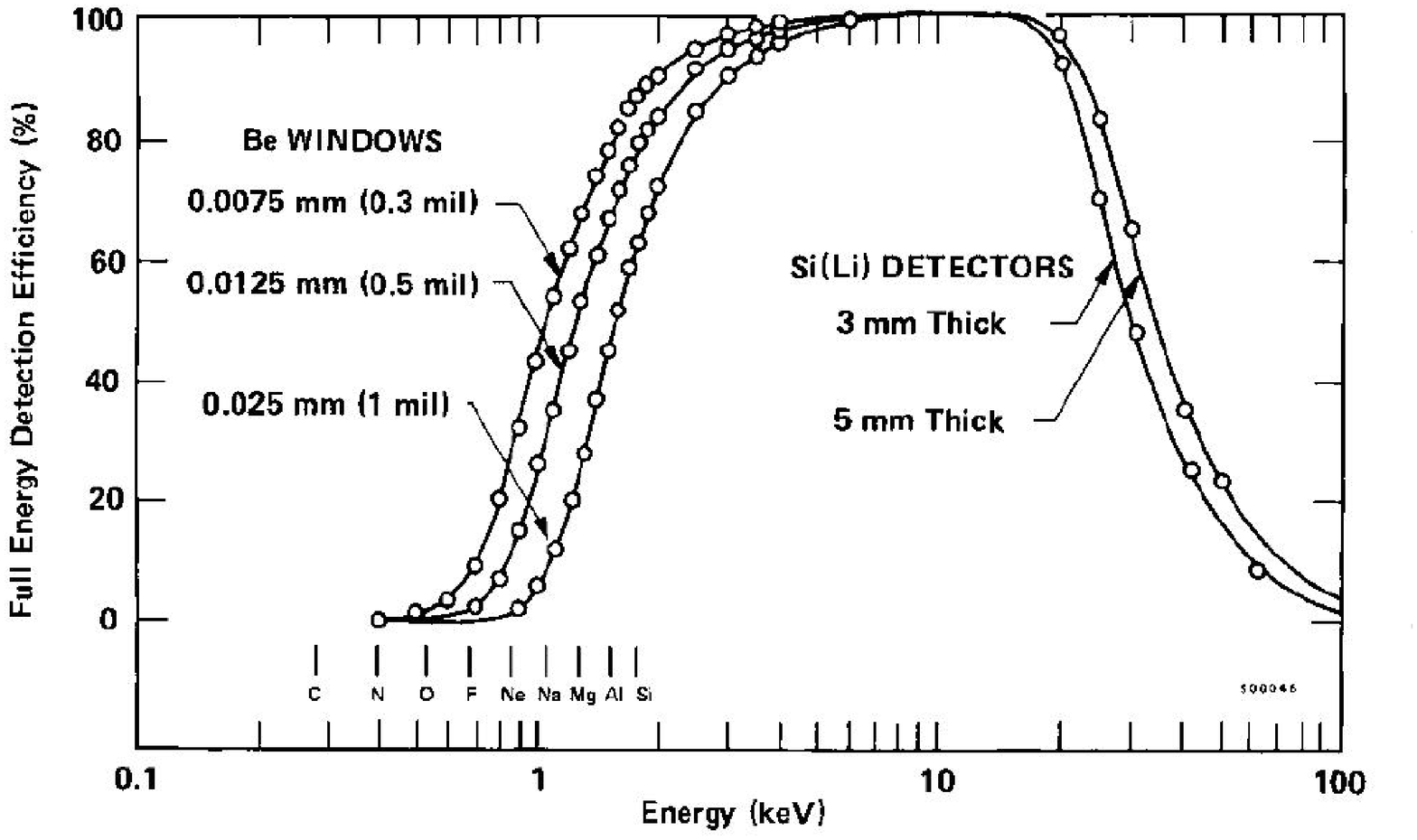}{Detection efficiency of ORTEC~Si(Li) detectors \cite{ortec}.}
%----------------------------------
Emitted x-rays were registered by an ORTEC single crystal Si(Li) detector placed perpendicular to the beam direction. The crystal of \(6\)~mm diameter and \(3\)~mm thickness, together with a \(7.5\)~$\mu$m Be-window, gave a detection efficiency in the energy range \(2\)-\(4\)~keV better than \(90\%\) (Fig.~\ref{fig:ortec}). The detector was energy calibrated with a standard $^{55}$Fe calibration source. Calibration was frequently repeated throughout the experiment in order to check the stability of the data acquisition system.

%----------------------------------
\munepsfig[1.15]{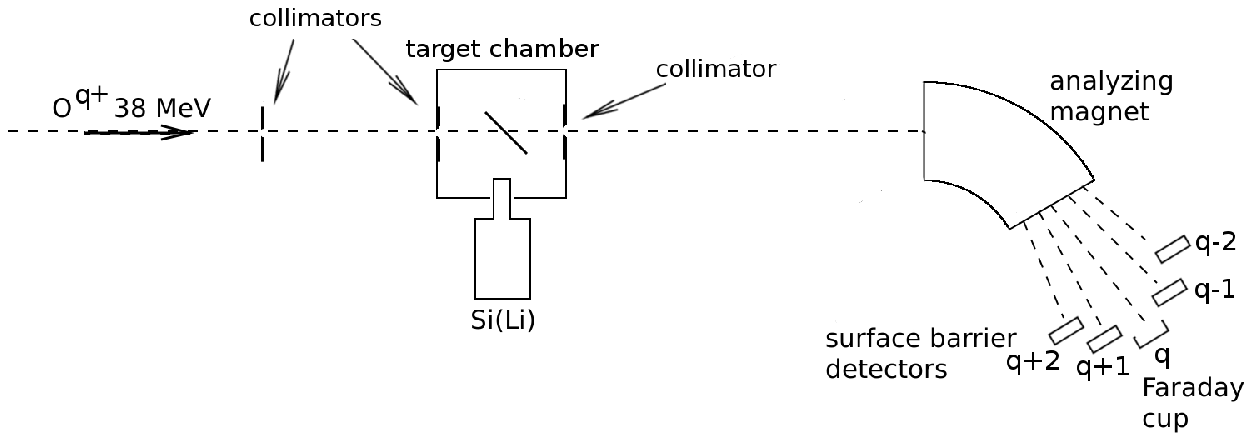}{Experimental setup.}
%----------------------------------

Along the beam direction, a set of two collimators was placed in front of the target chamber. The distance between collimators was about \(2\)~m. Collimator apertures of \(2\) and \(3\)~mm were to ensure a good beam collimation. An additional collimator between the target and magnet prevented scattered ions from entering the spectrometer and generating false coincidences (see Fig.~\ref{fig:setup}).

The target chamber was followed by a magnetic spectrometer. The magnetic field of the dipole magnet separated the final charge states of the ions and directed them towards four surface barrier detectors. The primary beam was registered by a Faraday cup. Surface barrier detectors counted ions with charge states equal to \(q-2\), \(q-1\), \(q+1\) and \(q+2\), where \(q\) is the charge state of the primary beam. Both the magnetic field of the spectrometer and the position of the surface barrier detectors were adjustable and created a versatile system, which could be used for various beam charge states and energies.
A schematic drawing of the experimental setup is presented in Fig.~\ref{fig:setup}.

%The total geometry factor for the experiment, 
%----------------------------------
%\muneqn{geom}{G=\dfrac{d\Delta\Omega}{4\pi}=4.2(1.4)\cdot10^{15},}
%----------------------------------
%where the uncertainty comes from the target thickness, which can vary by up to 10\% due to manufacturing processes \cite{tar}.

%----------------------------------
%\munepsfig[0.60]{}{Scheme of the experimental setup}.}
%----------------------------------
\section{Data acquisition system}
\label{section:time}
The data acquisition system was designed for registration of x-rays as well as particles with final charge states of \(q-2\), \(q-1\), \(q+1\) and \(q+2\) in a way that allowed for a software analysis of photon-particle coincidences. A schematic drawing of the electronics is presented in Fig.~\ref{fig:electronics_v}. For the clearness of presentation only a setup for one particle detector is drawn.

Signals from the particle detectors were first amplified by ORTEC~474 timing filter amplifiers (TFA), converted to a logic signal by constant fraction discriminators (ORTEC~463 or 473) and finally used as the STOP signals for time to amplitude converters (TAC, ORTEC~566).

A signal from the x-ray detector preamplifier was amplified and split (into two unipolar and one bipolar signals) by the Tennelec 244 spectroscopy amplifier. Additionally, a signal from the backside input of the amplifier was used (it is an unchanged signal from the front input) as an input for the TFA (ORTEC~474). The unipolar signals were processed by linear gate stretchers (LGS, ORTEC~542), while the bipolar signal was analyzed by a timing single channel analyzer and then converted to a logic signal by the delay generator (Phillips Sci 794), so that it could be used as a strobe signal for TAC, LGS modules and the ADC. The same signal was also used as a master trigger for the computer.
Additionally, the TFA signal was used as a gate for one of the LGSs. This resulted in registration of only x rays which gave the START signal for the TAC. The `non-gated' LGS served as a check of the x rays lost by the TFA. It was seen from the data that about 25$\%$ of the x-rays were not accepted by the TFA.

In Fig.~\ref{fig:electronics_v} a signal from the `gated' LGS is called `fast', while the one from the `non-gated' LGS is referred to as `slow'.
The x-ray timing signal was used as a START signal for TACs.
%As the system was originally prepared for two x-ray detectors, TAC START and strobe signals were generated as OR of both detectors. Unfortunately during the experiment one of the detectors had to be disconnected, due to high noise level, though all the electronics setup remained unchanged. 

All the TAC and LGS outputs were registered by the ORTEC~811 ADC and the data were written into SpecTCL \texttt{.evt} list mode files and converted afterward into \texttt{.ROOT} files for analysis purposes. The CERN-ROOT software was used for data analysis.

The construction of the data acquisition system allowed for registration of the time difference between photon and particle detection with a resolution of \(86(4)\)~ns (FWHM of the time peak, Fig.~\ref{fig:time}). This information was used for determination of the true coincidences between registered photons and ions. An example of a time spectrum registered during the experiment is presented in Fig.~\ref{fig:time}. The peak associated with true coincidences is clearly visible and the time window set for data analysis is marked with an arrow. The remaining parts of the time spectrum include random coincidences that were, after normalisation, subtracted from the coincidence spectrum.

%----------------------------------
\munepsfig[.75]{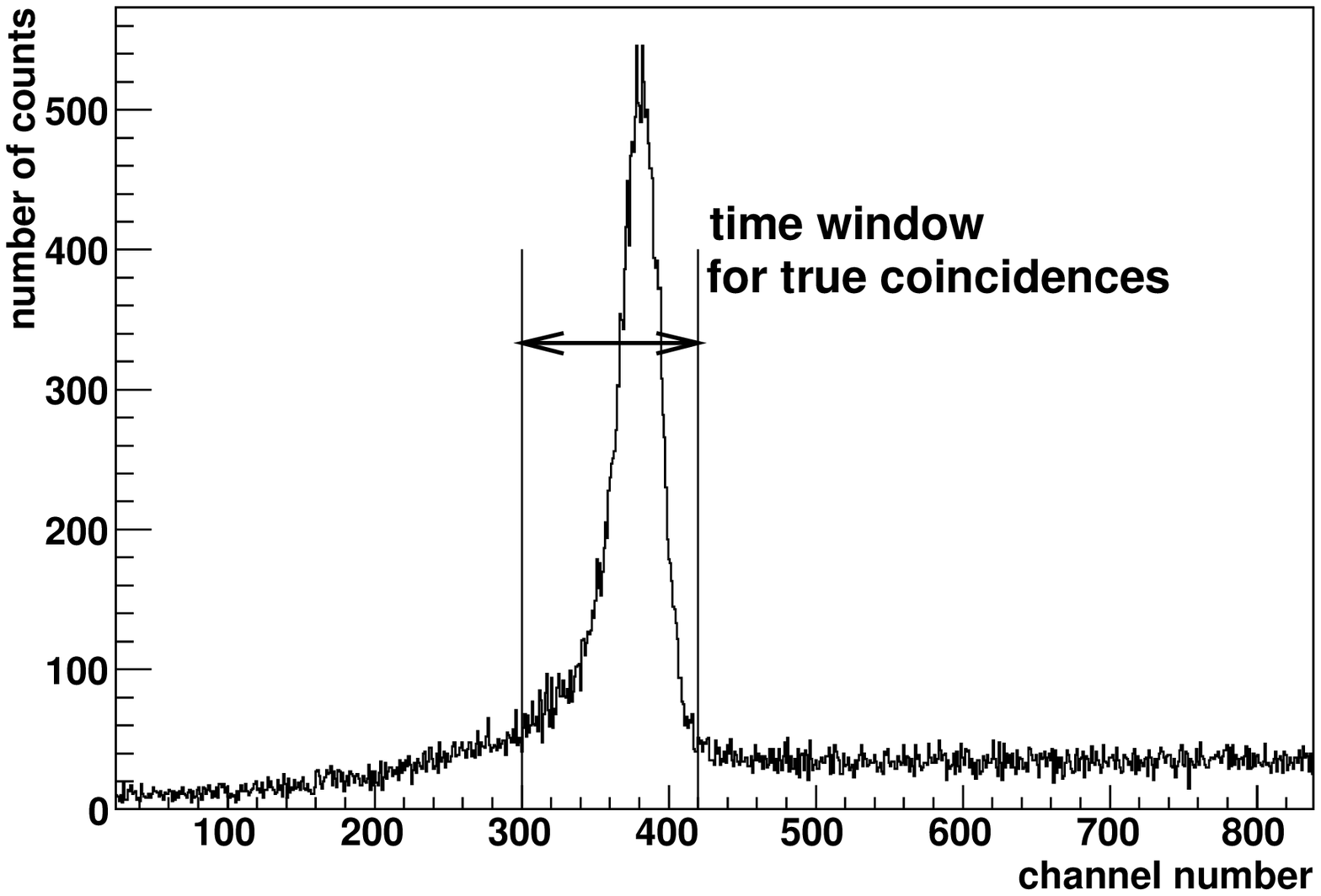}{Example of a time spectrum registered during the experiment. The arrow indicates the width of a time window for true coincidences (calibration $2$~ns/channel).}
%----------------------------------

%----------------------------------
\munepsfig[.855]{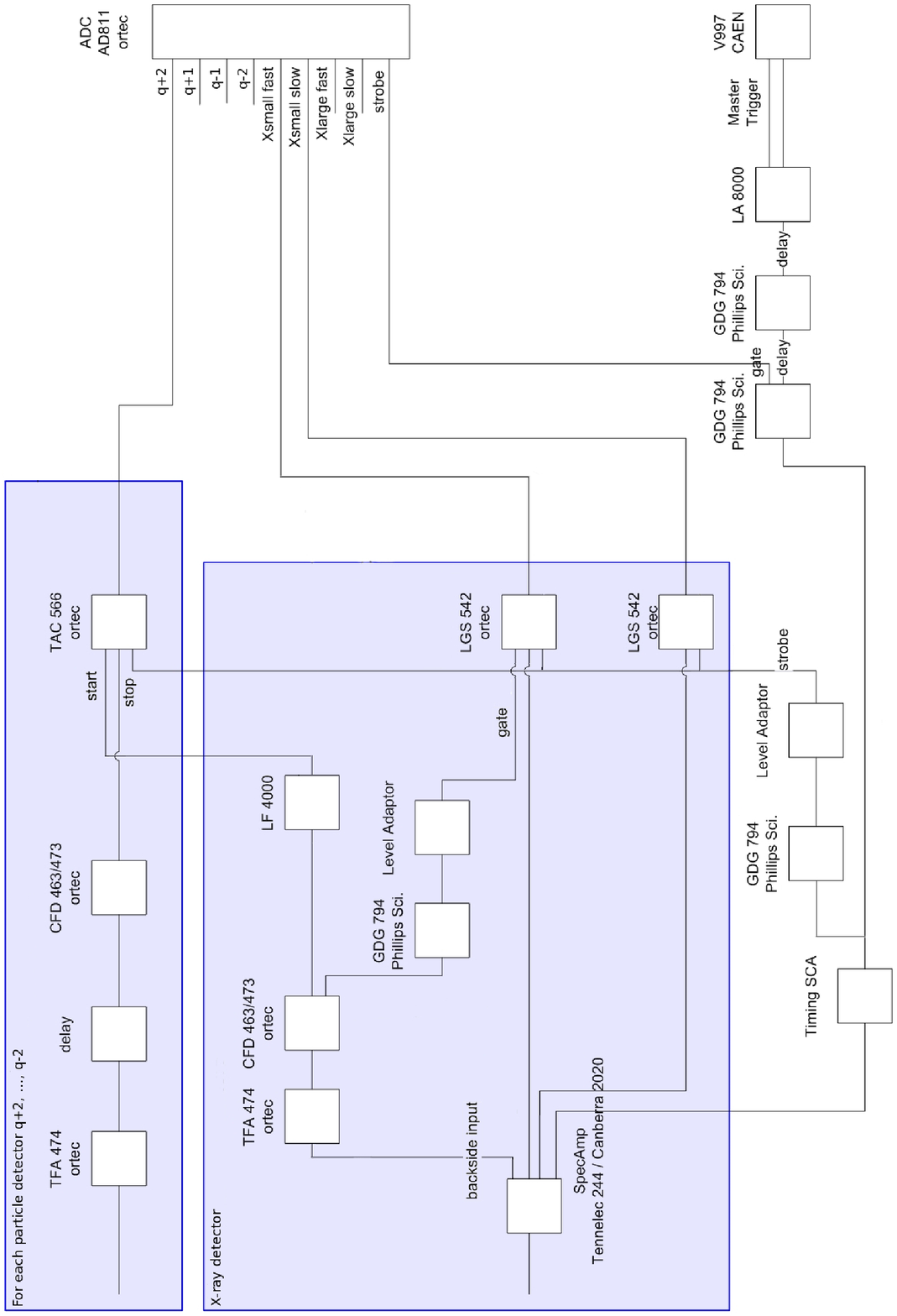}{Scheme of the data acquisition system.}
%----------------------------------

\chapter{Data analysis}
\label{chap:anal}
%\thispagestyle{fancy}
%\section{Overview of the obtained data}

The main goal of the experiment was observation of the x rays generated by bare oxygen ions (O$^{8+}$) impinging on carbon at an energy of \(38\)~MeV. This gives a projectile velocity of \(9.7\)~a.u., which corresponds to the adiabacity parameter \(\xi_O\) of \(0.82\).
As can be seen in Fig.~\ref{fig:3pion2}~(a), the single x-ray spectrum registered during O$^{8+}$~+~C collisions is dominated by the Al K-$\alpha$ line, which is produced by scattered ions hitting the aluminum target holder. In order to establish the shape of this line a run without the carbon foil was performed. The resulting spectrum, which contains only the Al K-$\alpha$ line, is represented by the dashed line in Fig.~\ref{fig:3pion2}~(a). In Fig.~\ref{fig:3pion2}~(b) the O$^{8+}$~+~C spectrum after subtraction of the Al K-$\alpha$ line is shown.
%-----------------------------------
\munepsfig[.8]{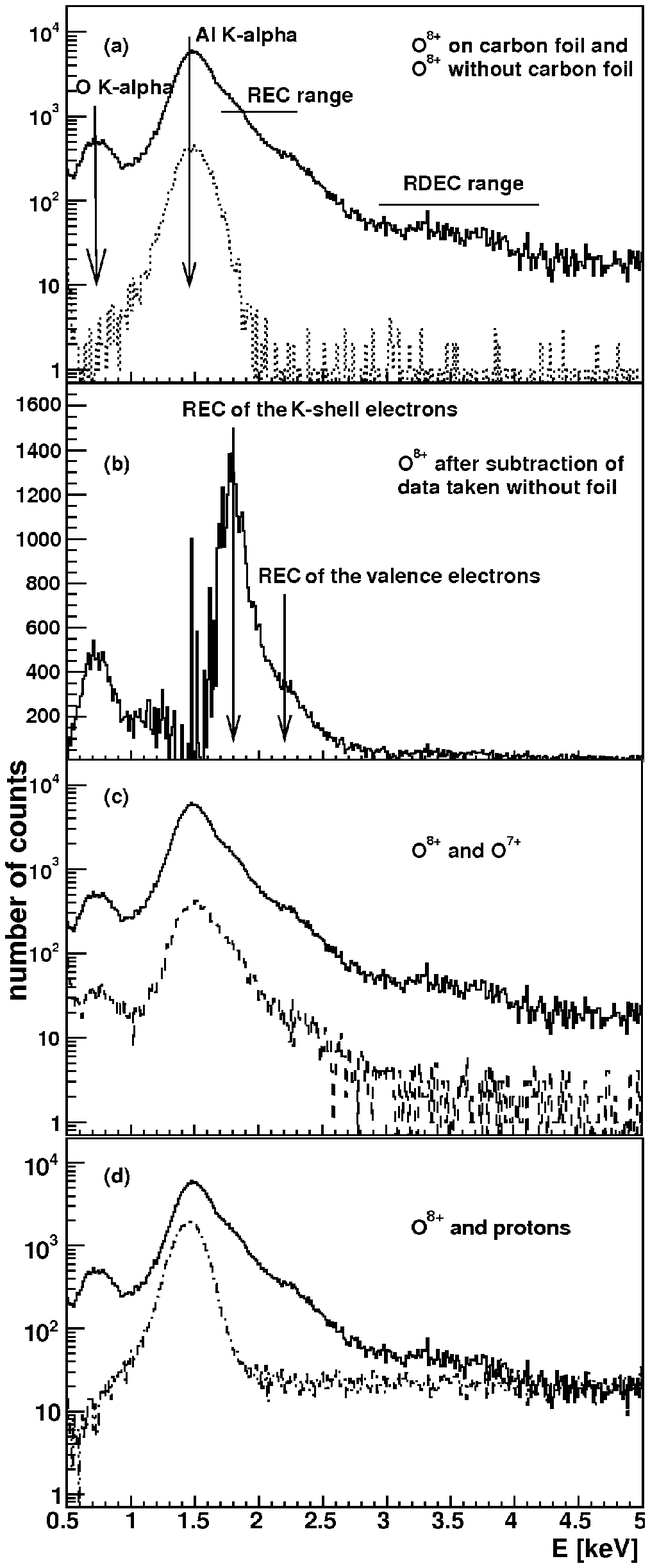}{Experimental single x-ray spectra. In all spectra: solid line -- $38$~MeV O$^{8+}$~+~C. (a) dashed line -- O$^{8+}$ data taken without the carbon foil, (b) O$^{8+}$ data after subtraction of the Al K-$\alpha$ line, (c) dotted line --  $38$~MeV O$^{7+}$~+~C, (d) dot-dashed line -- $2.375$~MeV protons on carbon.}
%-----------------------------------

A run with \(38\)~MeV O$^{7+}$ ions was made to check if the structure of the x-ray spectra in the RDEC region changes when one electron in the projectile K-shell is present. As the experiment was aimed at observation of RDEC mainly to the ground state (\(1s^2\)), presence of a \(1s\) electron in the H-like ion should block this process. Thus, a change in the RDEC range of the x-ray spectrum should be registered.
As shown in Fig.~\ref{fig:3pion2}~(c), the structure in the RDEC region is different for the O$^{7+}$ ions.

%-----------------------------------
\munepsfig[.6]{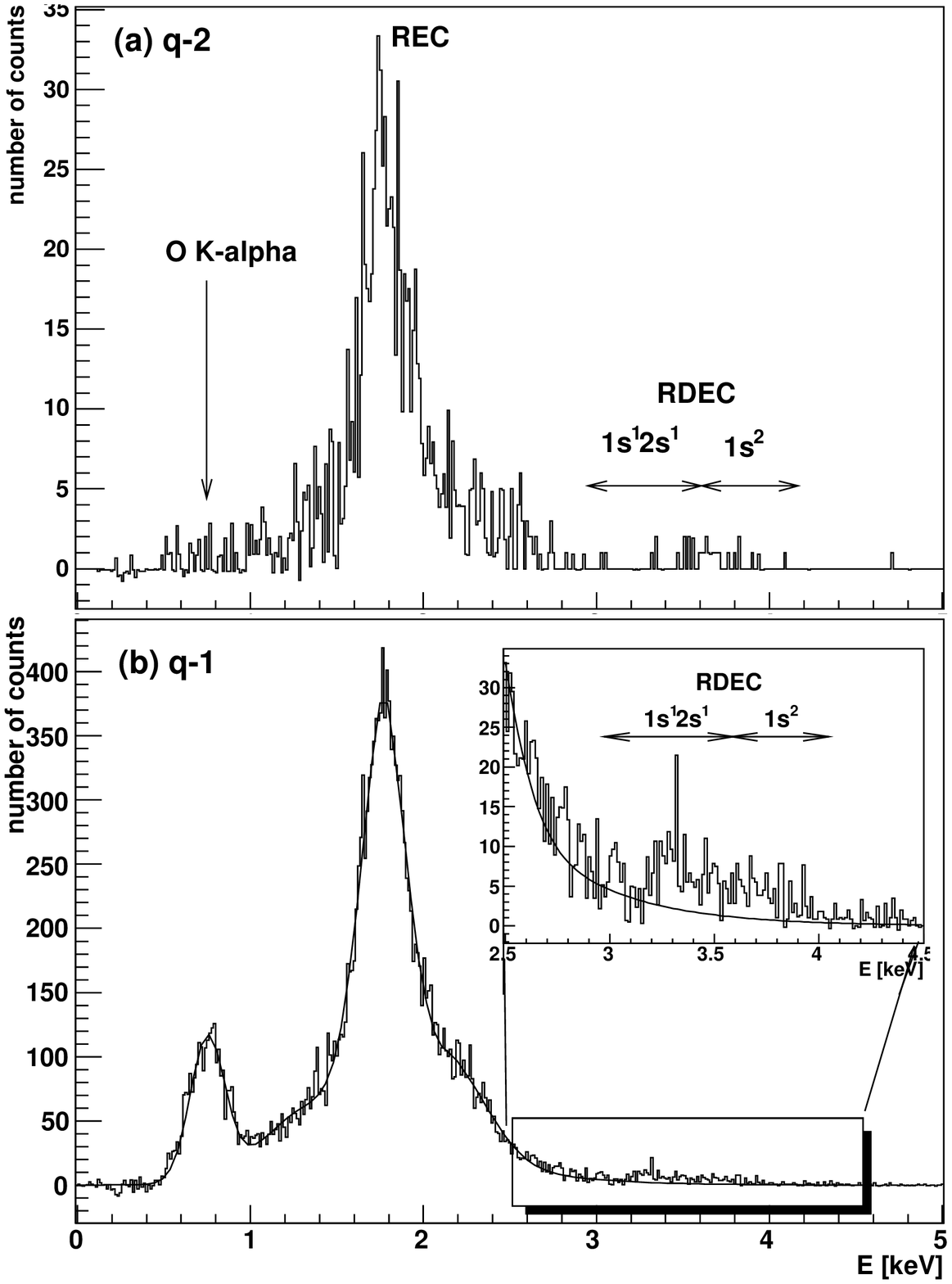}{X rays registered for O$^{8+}$~+~C collisions in coincidence with ions which captured (a) two electrons and (b) one electron. Solid line is the sum of the REC Compton profile and the Gaussian shape of the oxygen K-$\alpha$ line fitted to the spectrum.}
%-----------------------------------

As an additional test of the experimental conditions, PIXE analysis of the target foil was performed with \(2.375\)~MeV protons, i.e. for the same collision velocity as for O$^{8+}$ ions. During this process the impinging proton beam excites atomic states of the atoms within the material, which then deexcite emitting characteristic x rays. Intensities of the lines allow for estimation of the amount of the impurities in the material. Here, PIXE analysis was performed to check if any impurities that might produce x rays in the RDEC range are present in the foil. As can be seen in Fig.~\ref{fig:3pion2}~(d), no structure in the RDEC region was observed, nor was there evidence for REC around the photon energy of \(2\) keV.

Coincidence spectra obtained for bare O$^{8+}$ ions capturing one or two electrons are presented in Fig.~\ref{fig:coincidence}. These spectra were obtained using the coincidence condition on the appropriate time spectra as described in Section~\ref{section:time}. The contribution of random events was subtracted. In both spectra, a well separated structure in the RDEC region is evident. The REC counts associated with ions which captured two electrons come from multiple capture processes in the target with at least one radiative capture (REC). This could not be avoided due to a very high cross section for nonradiative electron capture, which is of the order of \(0.2\)~Mb, as estimated according to the commonly used scaling formula \cite{schla}. Another origin of these REC counts is the DREC process, which, due to a cross section of \(48\)~b (Eq.~\ref{eqn:mey_DREC}), is a source of about \(10\%\) of the total number of counts in the REC energy range.

The following subsections of this chapter will address the data analysis in detail.

\section{PIXE analysis of the target material}

To check if there is any statistically significant structure in the RDEC range of x-ray spectrum obtained for the proton beam, the method described in Appendix~\ref{app:A} was used. The value of $\alpha$ was set to \(0.05\), which means that the probability that any structure in the tested range of the spectrum is due to statistical fluctuations (\(H_0\) hypothesis) is equal to \(0.95\).
%-----------------------------------
\munepsfig[.6]{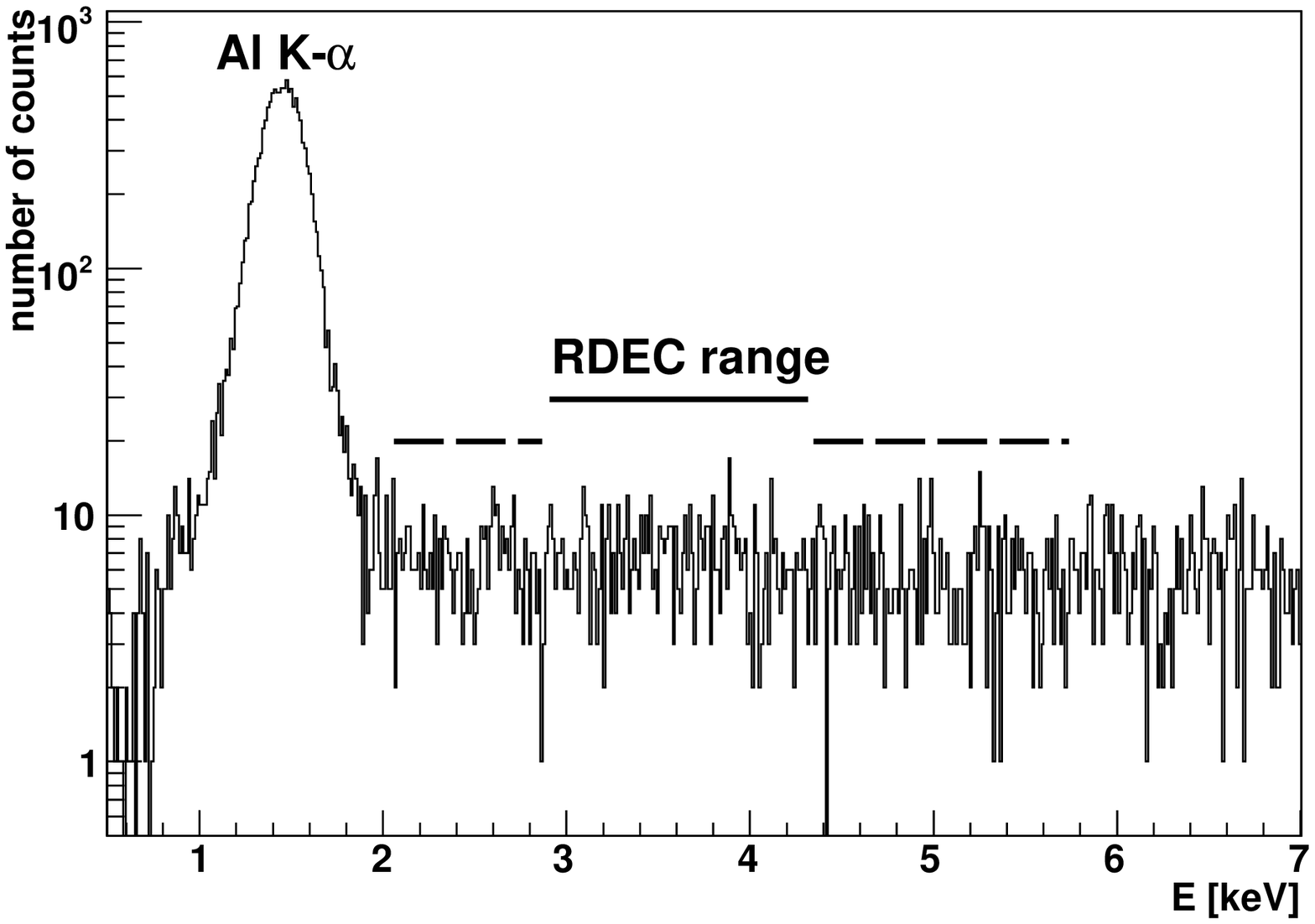}{Proton induced x-ray spectrum. Solid line: the RDEC range; dashed line: region considered during background estimation.}
%-----------------------------------
A very crucial step of this analysis was a proper estimation of the background in the vicinity of the RDEC range. As can be seen in Fig.~\ref{fig:protons} the background shape is not very smooth and a standard linear fit of the background, implementing the least squares method, was very sensitive to the choice of the fitting region. %thus the results of the linear fit strongly depended on the choice of the fitting region and varied even by \(30\%\). That is why the fitting method was not reliable in this case and another 
That is why the following method of estimation of the background parameters was used.
A region on both sides of the RDEC range (marked with dashed lines in Fig.~\ref{fig:protons}), without the Al K-$\alpha$ line, was integrated and the result was divided by the number of channels in the chosen range of the spectrum. This gave an average number, \(N_{ch}\), of the background counts per channel, \(N_{ch}=6.40(23)\). This number multiplied by the number of channels in the RDEC region of the x-ray spectrum gave the total number of the background counts of \(b=895(32)\). %The uncertainty is the result of the dispersion of the number of counts in the considered channels.
The total number, \(N\), of counts registered within the RDEC region of the spectrum is equal to \(903(30)\). 

The above values of \(N\) and \(b\) determined the value of the statistical variable \(T\) (Eq.~\ref{eqn:T}) to be equal \(0.372\). This gives \(1-\alpha_T=0.54\), which means that the \(H_0\) hypothesis cannot be rejected.
In other words, there is no evidence for any peak structure, that could be a result of any physical process, in the RDEC range of the proton induced spectrum. Thus, it can be assumed that there are no target impurities that could emit photons in the RDEC energy range.
Values of all the relevant parameters are given in Table~\ref{tab:h0_pixe}.

\muntab{@{\extracolsep{\fill}}ccccc}{h0_pixe}{Results of a $\chi^2$ test of the RDEC range of the proton induced spectrum.}{
\hline
$N_{ch}$ & $N$ & $b$ & $T$ & $1-\alpha_T$\\
\hline
6.40(23) & 903(30) & 895(32) & 0.372 & 0.54 \\
\hline
}

\section{Projectile K- and L-shell electron loss}

It could be already noticed that the events in the RDEC range are more frequent in the spectrum associated with the single charge exchange rather than with the double charge exchange channel (Fig.~\ref{fig:coincidence}). Similarly, most of the REC photons are observed in the singles spectrum and not in the single charge exchange channel (compare Figs~\ref{fig:3pion2}~(b) and \ref{fig:coincidence}~(b)). It is likely that after the observed capture process the ion may lose one of the captured electrons.
This is due to the fact that, even in a very thin carbon foil, the electron loss probability is very high for the weakly bound ionic systems that are formed during the collision. Electron loss cross sections obtained by interpolation of relevant data (see Chapter~\ref{chap:processes}) are given in Table~\ref{tab:ionis}. These cross sections are in fair agreement with the data of Shima et al. \cite{shim}, where the bare and H-like ions are indicated as the most populated charge states observed when a \(38\)~MeV oxygen beam traverses through a carbon foil. Together, they account for more than \(80\%\) of the final charge states, with \(50\%\) in the H-like (7+) state. For the system investigated, the electron loss cross sections are at least \(3\) orders of magnitude greater than the REC cross section. This means that there is a very high probability for an ion, after the capture process, to undergo another collision, which is most likely to be accompanied by the electron loss process.

For the system investigated, the cross section for the removal of the L-shell electron is about an order of magnitude greater than that for the K-shell electron (see Table~\ref{tab:ionis}). Thus, in case of double capture to the \(1s^12s^1\) state, the \(2s\) electron is promptly removed, while in case of capture to \(1s^2\) the final charge state of the ion is more likely to remain unchanged. Therefore, one would expect most of the photons originating from RDEC to the projectile excited state to show up in the single charge exchange channel, while the capture to the ground state will be less affected by the K-shell electron loss process. This can be observed in Fig.~\ref{fig:coincidence}, where the \(1s^12s^1\) peak is clearly visible in the \(q-1\) coincidence spectrum, while it is almost absent in the \(q-2\) channel, compared with the structure associated with the capture to the \(1s^2\) state which is still visible in the \(q-2\) channel.

\newcommand{\threelinebrace}{$\left\lbrace \begin{array}{c}  \\ \\ \\\end{array} \right.$}

%----------------------------------
\muntab{@{\extracolsep{\fill}}p{5cm}rx{2cm}x{6cm}}{ionis}{Electron loss cross sections for oxygen ions at $38$~MeV estimated from the data presented in \cite{bom,tan,hipp}.}{
\hline
				&&Process 				& Cross section [Mb]\tabularnewline
\hline
L-shell ionization	&			&O$^{5+}$ $\rightarrow$ O$^{6+}$ 	& 19.0 \tabularnewline
\tabularnewline%\hline
\multirow{3}{*}{K-shell ionization}& \multirow{3}{*}{\threelinebrace }		&O$^{6+}$ $\rightarrow$ O$^{7+}$ 	& 3.6 \tabularnewline
						&&O$^{7+}$ $\rightarrow$ O$^{8+}$ 	& 0.4 \tabularnewline
						&&O$^{6+}$ $\rightarrow$ O$^{8+}$ 	& 0.1 \tabularnewline
\hline
}
%----------------------------------

The cross section for double electron loss (O$^{6+}$ $\rightarrow$ O$^{8+}$), even though it is much smaller than that for single electron loss, was also considered. Due to this process, in some cases, both captured electrons could be removed from the ion and the RDEC events might be observed only in the singles spectrum. However, after subtraction of all the background contributions (Al K-$\alpha$, REC, bremsstrahlung) the remaining number of counts within the RDEC range of the singles spectra was consistent with the sum of all the counts from the $q-1$ and $q-2$ coincidence channels. This means that these two channels include all the registered events that could be associated with the RDEC process and the sum of the RDEC counts from both channels should be included in the cross section estimation.
In case of REC, in order not to lose any of the registered events, the total number of REC photons was obtained from the singles spectra.

%This is due to the fact that in a weakly bound ionic system (such as O ions) the ionization probability is very high in the ion velocity range under consideration. As a consequence the RDEC photons will be registered, to a large extent, in coincidence with a single charge exchange. Electron loss cross sections obtained by interpolation of relevant data  (see Chapter~\ref{chap:processes}) are given in Table~\ref{tab:ionis}. In the system used, the cross section for L-shell ionization is an order of magnitude greater than that for the K-shell. Therefore, one would expect most of the photons coming from the RDEC process to the projectile excited state (one of the electrons captured into the L-shell) to show up in the single capture coincidence channel. This can be observed in Fig.~\ref{fig:coincidence}, where the structure in the \(1s^12s^1\) region is clearly visible in the single coincidence spectrum, while it is almost absent in the double coincidence spectrum. 

\section{Background processes}
\label{section:bcgr}
For such a subtle effect as the RDEC process, estimation of the background shape in the RDEC range of the x-ray spectrum is of a great importance.
The most significant processes occurring during collisions of bare ions with solid target that can contribute to the observed x-ray spectrum, are:
\begin{itemize}
 \item
radiative electron capture to continuum (RECC),
\item
secondary electron bremsstrahlung (SEB),
\item
atomic bremsstrahlung (AB),
\item
nucleus-nucleus bremsstrahlung (NB),
\item
the high energy tail of the single REC profile caused by the momentum distribution of the target electrons (Compton profile).
\end{itemize}

Under the experimental conditions, the high energy limits for the RECC (\(T_r\)) and the SEB (\(T_m \)) processes are \(1.3\)~keV and \(5.2\)~keV, respectively (see Eqs~\ref{eqn:Tr} and \ref{eqn:Tm}). Therefore, RECC will not contribute to the RDEC background, as it appears in the low photon energy range. In addition, RECC will be suppressed by the absorption in the Be-window. The SEB process might be a significant background contribution and will be discussed further.

There are three bremsstrahlung contributions in the x-ray spectra within the RDEC photon energy range: SEB, AB and NB. Secondary electron bremsstrahlung and atomic bremsstrahlung can be estimated by simple scaling from the data obtained by Folkmann for protons of similar velocity to the oxygen ions used during the experiment. The x-ray spectrum for p~+~C collisions given in \cite{ishi1} can be easily transformed by means of appropriate scaling formulae (Eqs~\ref{eqn:scalingRECC}-\ref{eqn:scalingSEB}).
As can be seen in Fig.~\ref{fig:ishii_background}, this x-ray spectrum is dominated by electron bremsstrahlung (SEB~+~AB) for photon energies up to \(10\)~keV, which is far beyond the region of interest (\(2.8\)-\(4.2\)~keV). According to the formula given by Eqs~\ref{eqn:nnb}-\ref{eqn:x}, the cross section for the NB process is of the order of \(1\)~mb/keV~sr in the RDEC range, which is a few orders of magnitude smaller than the cross sections for any of the electron bremsstrahlung processes. Thus, the contribution from NB can be neglected in the discussed case.

%The differential cross sections for SEB and AB processes in the experiment were estimated based on the data presented in \cite{ishi1}. The total bremsstrahlung spectrum, shown in Fig.~\ref{fig:ishii_background} was scaled by \(Z^2\), according to scaling laws given by Eqs \ref{eqn:scalingAB} and \ref{eqn:scalingSEB}.
%-----------------------------------
\munepsfig[.735]{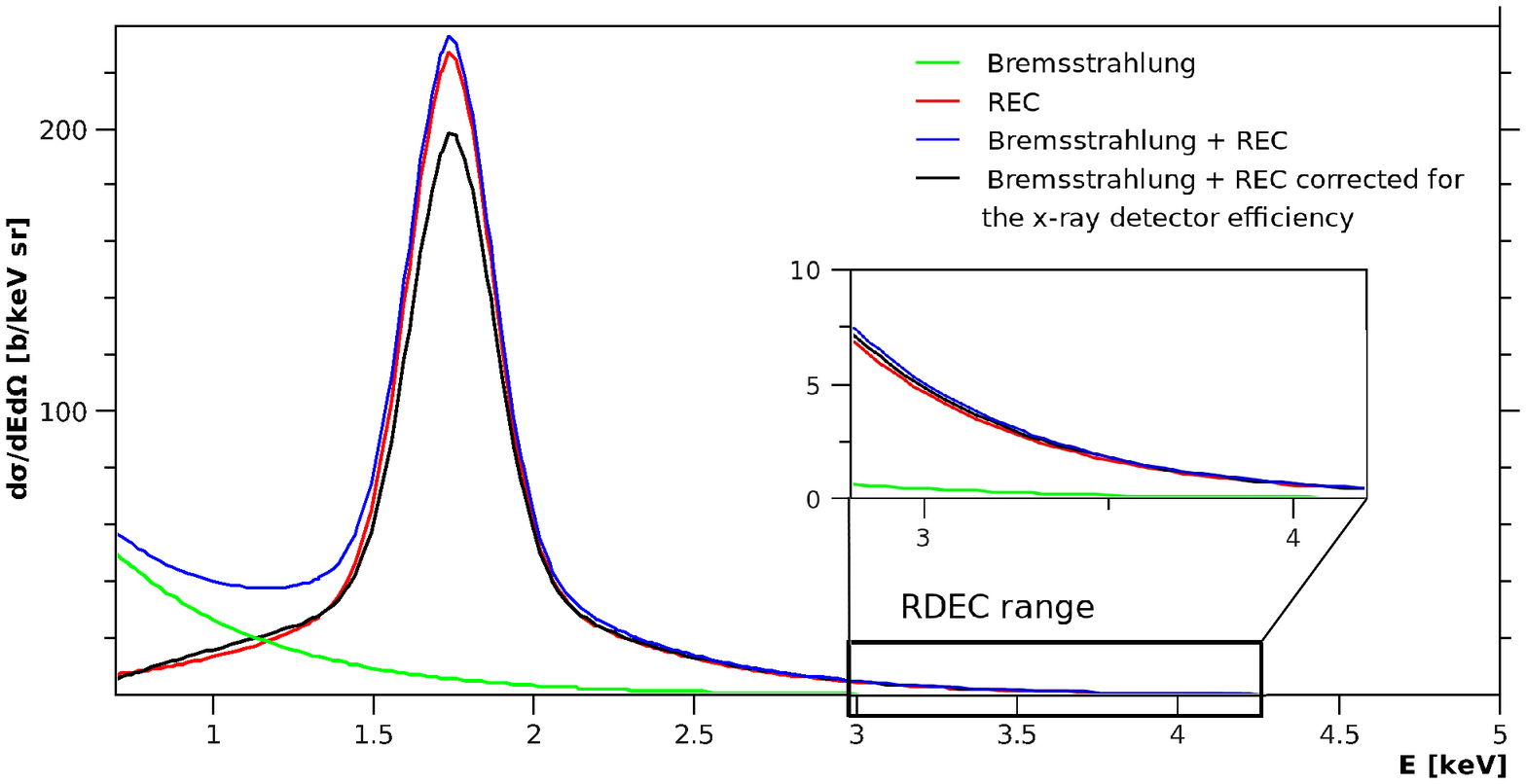}{Background structure in the single x-ray spectrum. The bremsstrahlung contribution includes all the relevant processes (SEB~+~AB~+~NB) discussed in the text. The spectrum is completely dominated by the REC structure.}
%-----------------------------------
Another contribution that should be taken into account is the high energy tail of REC. The observed REC structure originates not only from single radiative electron capture, but can also be associated with the noncorrelated double electron capture (DREC).
However, estimation based on Eq.~\ref{eqn:mey_DREC} shows that the DREC cross section is an order of magnitude smaller than that for a single REC process, as shown in Table~\ref{tab:proc}.

All these background contributions are shown in Fig.~\ref{fig:bremss+REC}. In the expected RDEC energy region (\(2.8\)-\(4.2\)~keV) the broad REC spectrum significantly exceeds the bremsstrahlung contribution (see inset in Fig.~\ref{fig:bremss+REC}). As a consequence, the RDEC line will be placed mainly on the REC tail.

%-----------------------------------
\muntab{@{\extracolsep{\fill}}l c}{proc}{Total cross sections for the background processes that were taken into account during data analysis.}
{\hline
Process 	& Cross section [kb] \\
\hline
Radiative electron capture (REC)	& 0.512  \\
Nonradiative electron capture (NREC)	& 240.0  \\
Double radiative electron capture (DREC)	& 0.048\\
\hline}
%-----------------------------------

The cross section for background contributions was also corrected for the detection efficiency of the Si(Li) detector. Both crystal efficiency and Be-window absorption were taken into account by using the data shown in Fig.~\ref{fig:ortec}. The shape of the background, after efficiency correction, is also shown in Fig.~\ref{fig:bremss+REC}. It can be noticed that the spectrum is completely dominated by the tail of the single REC line and the contribution of the bremsstrahlung is almost negligible. Additionally, the low energy photons (energy not exceeding \(0.5\)~keV) are completely absorbed in the beryllium window of the x-ray detector and the curve, representing the sum of all background contributions corrected for the detector efficiency, drops almost to zero at low photon energies.

\section{Pile-up of single REC photons and its contribution to the RDEC energy range of the spectrum}
\label{section:double}
As the energy of a RDEC photon is about twice as large as the energy of a single REC photon, it might be impossible to distinguish a real RDEC event from a situation when two REC photons are simultaneously registered by the x-ray detector (pile-up effect). This process may produce an additional background within the RDEC range of the x-ray spectrum.
Simultaneous detection of two REC photons may occur in three situations:
\begin{itemize}
 \item 
both photons emitted during the DREC process propagate in the same direction and are both registered by the x-ray detector,
\item
the ion undergoes multiple collisions during which at least twice the REC process occurrs and both photons are emitted towards the x-ray detector,
\item
if the beam intensity is high enough, there is a chance that two REC photons emitted by two sequential ions are simultaneously detected by the x-ray detector.
\end{itemize}

If the cross section for RDEC is of the order of \(0.1\)~b \cite{nef2} and the given geometry of the experiment is taken into account, one obtains the probability of observation of a RDEC photon during an ion-atom collision equal to \(5.3\cdot10^{-11}\). If the beam intensity is equal to \(I_{medium}=3\cdot10^{4}\)~ions/s, which is the mean value obtained during the experiment, the RDEC count rate will be of about \(1.6\cdot10^{-6}\)~events/s for the geometry applied (Table~\ref{tab:probab}). 

For a low-\(Z\) system, as the one used during the experiment, the cross section for the noncorrelated double electron capture (DREC) is rather high (see Table~\ref{tab:proc}) when compared to the RDEC cross section. Thus, simultaneous observation of both DREC photons might contribute to the count rate in the RDEC range. The probability that both DREC photons will be registered as one event is estimated to be comparable with the probability of registration of a true RDEC event (Table~\ref{tab:probab}). As the angular correlations of these two emitted photons have so far neither been measured nor calculated, it was assumed that the angular distribution of the two DREC photons is the same as for the REC photons (\(\sim \sin^2\theta\)) and the emission angles for the photons are independent of each other. This assumption may cause a significant overestimation of the corresponding probability given in Table~\ref{tab:probab}. It is more likely that the photons are emitted in opposite directions (momentum conservation) and the corresponding count rate is much lower than estimated above.

%-----------------------------------
\muntab{@{\extracolsep{\fill}}lcc}{probab}{Probabilities and count rates of the processes that might contribute to the x-ray spectrum in the RDEC range. For more information see text.}
{\hline
Process 		& Probability 		& Count rate [events/s] \\
\hline
RDEC 			& 5.3$\cdot$10$^{-11}$	& 1.6$\cdot$10$^{-6}$\\
DREC		 	& 6.0$\cdot$10$^{-11}$	& 1.8$\cdot$10$^{-6}$\\
double REC collisions 	& 5.2$\cdot$10$^{-14}$	& 1.6$\cdot$10$^{-9}$\\
sequential REC 		& -- 			& 2.2$\cdot$10$^{-10}$	\\
\hline
}
%-----------------------------------

For an  REC cross section of \(512\) b the probability that one ion undergoes two collisions in the target, with the REC process occurring in each of them, and that both photons are registered as one event, is about \(5.2\cdot10^{-14}\). This, together with the mean beam current value, results in a count rate of double REC photons as a single event on the order of \(10^{-9}\)~events/s, which is negligible when compared to the expected count rate of the RDEC photons (Table~\ref{tab:probab}).

%-----------------------------------
%\muntab{@{\extracolsep{\fill}}lccc}{ratio}{Ratio of the numbers of counts in the RDEC and REC regions in single and $q-1$ coincidence spectra as a function of the beam intensity.}
%{\hline
%Beam intensity [\(10^{3}\) ions/s]			& 19.7		& 29.5 		& 42.9 \\
%\hline
%\(N_{RDEC}/ N_{REC}\) in single spectra			&0.423(21)	&0.320(11)	&0.302(13) \\
%\(N_{RDEC}/ N_{REC}\) in \(q-1\) coincidence spectra 	&0.048(5) 	&0.040(3)	&0.047(4)  \\
%\hline}
%-----------------------------------
Another possibility of simultaneous detection of two REC photons is the observation from two incoming ions successively undergoing REC processes in the target. This random event may be registered only in the time window of the order of the charge collection time in the silicon detector. For a typical Si(Li) crystal the collection time does not exceed \(100\)~ns \cite{lip}. In case of an ORTEC silicon detector cooled with LN$_{2}$ it is on the order of \(10\)~ns \cite{ortec2}. Therefore, the corresponding count rate for the mean beam intensity value is on the order of \(10^{-10}\)~events/s and can be neglected.
This estimation can be tested by the analysis of the \( N_{RDEC} /  N_{REC}\) ratio as a function of beam intensity \(I\). The count rate of this random event scales with \(I^2\), while all the other count rates (Table~\ref{tab:probab}), including that for single REC, scale with \(I\). Thus, if only true (RDEC) events are registered, the number of counts in both REC and RDEC regions will be proportional to the beam intensity and the ratio \( N_{RDEC} /  N_{REC}\) should be independent of \(I\). If a pile-up effect, due to sequential REC collisions was observed in the RDEC region, \(N_{RDEC}\) would be proportional to the square of the beam intensity and the \( N_{RDEC} /  N_{REC}\) ratio would increase with the beam intensity.
%-----------------------------------
\munepsfig[.8]{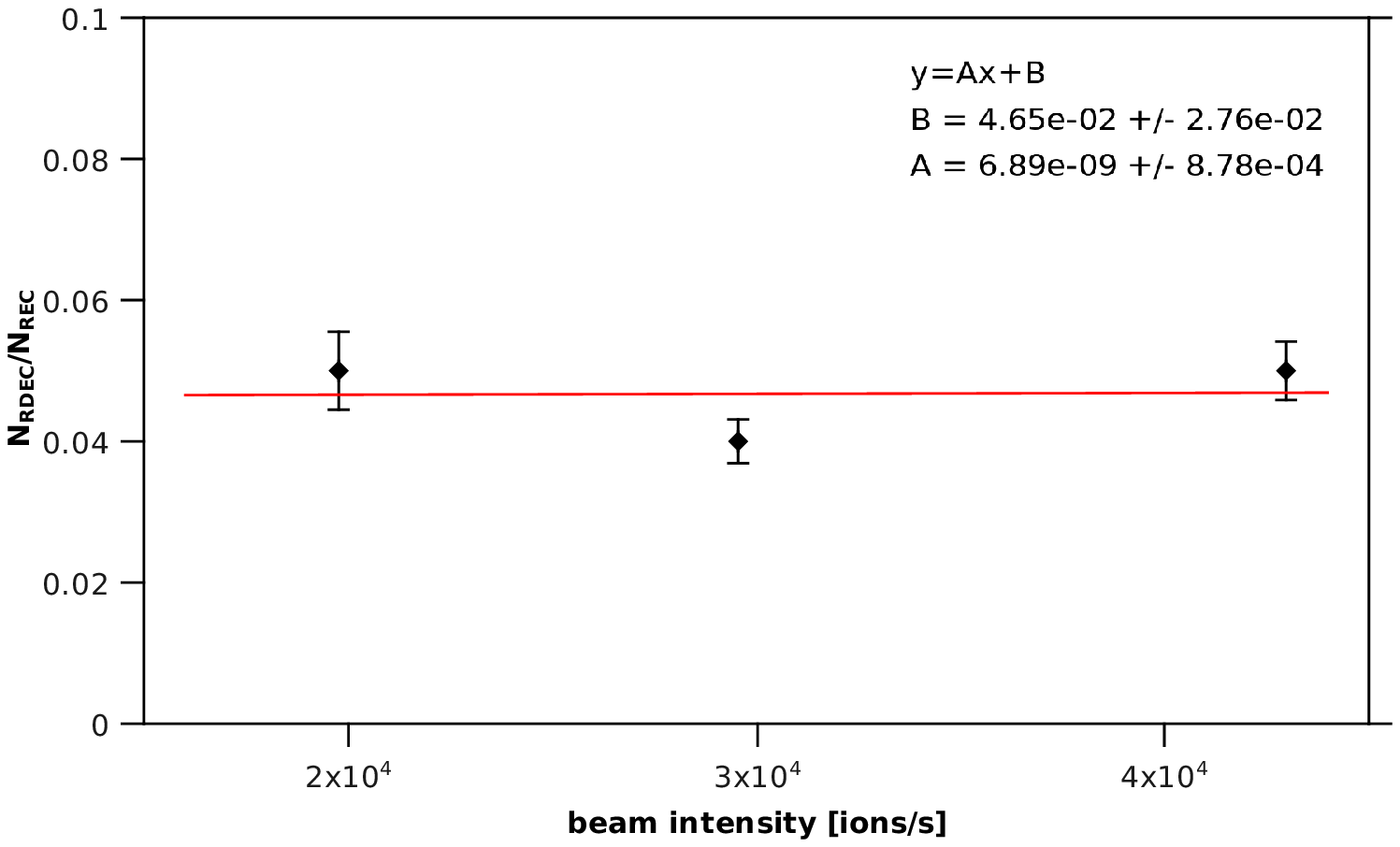}{$ N_{RDEC} /  N_{REC} $ ratio in the $q-1$ coincidence spectra as a function of beam intensity.}
%-----------------------------------
%However, the chance of such observation strongly depends on the beam intensity as only the photons that reach the x-ray detector during its charge collection time may be registered as a single event. The mean beam intensity registered \(I\) gives about \(30\) REC photons per second. This corresponds to \(0.1\) single REC photons per second registered by the detector. This gives the count rate of double photons equal to \(0.01\) events/s, which might seem extremely high. But one has to take into account the charge collection time of the silicon detector, as only photons registered during this time period might be seen by the spectroscopy amplifier as a single event. For a typical Si(Li) crystal the collection time does not exceed \(100\)~ns \cite{lip}. In case of ORTEC silicon detectors in LN$_{2}$ it is of the order of \(10\)~ns \cite{ortec2} which means that the chance that two sequential REC photons will be registered as one event is of the order of \(10^{-11}\), which is comparable to the probability of the RDEC event.
In order to separate the data collected with different beam intensities registered during data taking, data files were sorted according to the average beam intensity and separated into three groups:
\begin{itemize}
\item 
data taken with average beam current not exceeding \(2.5\cdot10^{4}\)~ions/s (average of all files: \(I_{low}=2.0\cdot10^{4}\)~ions/s) -- further referred to as low intensity beam;
\item
data taken with average beam current greater than \(4.0\cdot10^{4}\)~ions/s (average of all files: \(I_{high}=4.3\cdot10^{4}\)~ions/s) -- further referred to as high intensity beam;
\item
all the remaining data (average beam intensity: \(I_{medium}=3.0\cdot10^{4}\)~ions/s).
\end{itemize}
This gives a beam intensity ratio \(I_{high} / I_{low} = 2.15\), which, if the pile-up effect was observed in the RDEC region, would increase the RDEC to REC ratio by a factor of almost \(5\).

For the discussed beam intensities the \( N_{RDEC} /  N_{REC}\) ratio is shown in Fig.~\ref{fig:ratio} for the \(q-1\) coincidence channel. It can be seen in this figure that the ratio does not change within the uncertainty limits, which means that no pile-up events were registered in the RDEC region. These results do not exclude the contribution of the DREC process. However, the upper limit of this contribution is known and will be used in Chapter~\ref{chap:mc} for further discussion.

\section{Single spectra analysis}
In order to eliminate the Al K-$\alpha$ line from the O$^{8+}$~+~C data, single spectra obtained for bare ions without the carbon target were normalized and subtracted from the data taken with the foil.
As the shape of the Al K-$\alpha$ line, despite a good statistics, was not smooth enough, the line in the spectrum taken without the target foil could not be normalized to the one in the O$^{8+}$~+~C spectrum by simple comparison of the maxima of the lines. To establish the peak height of the Al K-$\alpha$ line in both spectra, the numbers of counts around the centroid of the Al K-$\alpha$ line in each x-ray spectrum were integrated and divided by the number of channels in the region of integration. This procedure prevented the contribution of the Al K-$\alpha$ line from being overestimated. Al K-$\alpha$ line normalized to the spectrum taken with the foil is shown in Fig.~\ref{fig:empty_and_foil}.

%----------------------------------
\muntab{@{\extracolsep{\fill}}l x{3cm} x{4.5cm} x{3cm}}{energies}{Calculated positions of the RDEC and REC peaks in the x-ray spectrum corresponding to different combinations of the initial and final states of the captured electrons. All values are given in~keV.}
{\hline
\multirow{2}{*}{Process} & \multicolumn{3}{c}{Captured target electrons}\tabularnewline
\cline{2-4}
%		& \multirow{2}{*}{valence}	& one K-shell		&  \multirow{2}{*}{K-shell}\\
%			&			& and one valence		&  \\
		&Valence	&One K-shell and one valence 	&K-shell\tabularnewline
\hline
REC		& 2.16		& --				& 1.88 \tabularnewline
RDEC \(1s^2\) 	& 4.18		& 3.91				& 3.64 \tabularnewline
RDEC \(1s^12s^1\) &3.58		& 3.31				& 3.04 \tabularnewline
\hline
}
%---------------------------------- 

After subtraction of the Al K-$\alpha$ line, the REC structure in the single spectrum was clearly resolved, as shown in Fig.~\ref{fig:single_rec}. %It can be noticed that the K-shell Compton profile is not sufficient to explain the shape of the observed line.
%-----------------------------------
\munepsfig[.6]{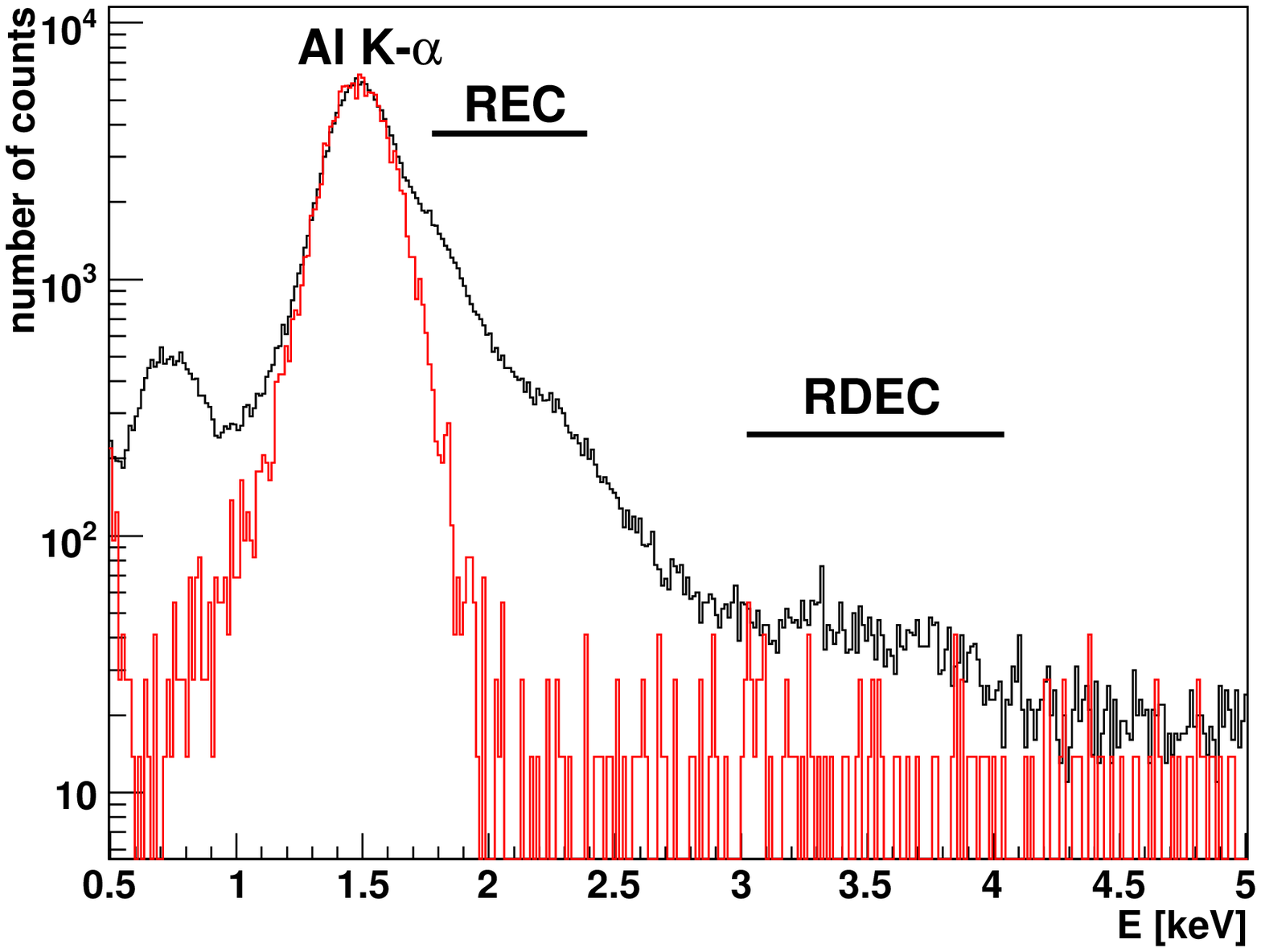}{O$^{8+}$ spectrum taken without the carbon foil (red line) normalized to the data taken with the foil (black line).}
%-----------------------------------
%-----------------------------------
\munepsfig[.6]{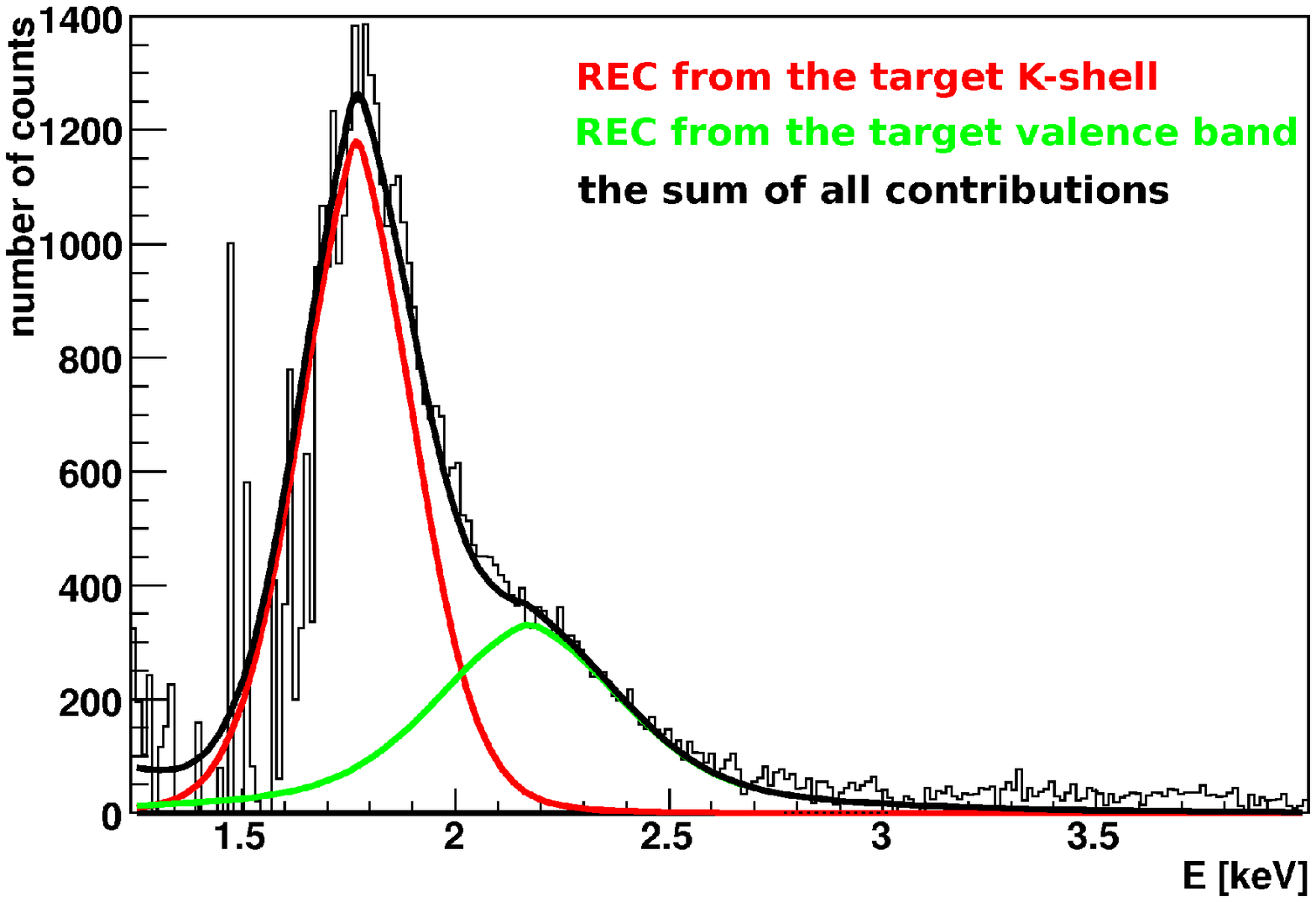}{Double structure of the REC line resolved after subtraction of the Al K-$\alpha$ line.}
%-----------------------------------
For the investigated low energy collisions the binding energy of the target electron cannot be neglected, as it can significantly contribute to the REC photon energy (see Eq.~\ref{eqn:Erec}). For the considered collision system the captured electron could come either from the target K-shell or from the target valence band. Thus, a double peak structure in the REC region was expected, with the peaks separated in the spectrum by the difference in the binding energy of the K-shell and valence electrons in the carbon foil, which is about \(280\)~eV \cite{booklet}. Expected positions of the corresponding REC peaks are given in Table~\ref{tab:energies}.
As can be seen in Fig.~\ref{fig:single_rec}, the obtained x-ray spectrum is well described by the sum of the target K-shell and valence band Compton profiles. 

The total number \(N_{REC}\) of the REC photons collected during the experiment was \(N_{REC}=39800(200)\). The uncertainty includes both the statistical error and the uncertainty generated during the subtraction of the background contribution.

%The main goal of the experiment was to look for the x-ray spectra in coincidence with particles which captured two electrons during collisions in the target. The single REC photons appearing in such coincidence spectrum are mainly due to multiple collisions in the solid target accompanied by at least one REC process. In the background analysis all combinations of double collisions, with radiative (REC) and nonradiative (NREC) electron capture, were taken into account.

\section{Coincidence spectra analysis}
\label{section:coincidence} 

%It can be seen in Fig.~\ref{fig:coincidence} that application of the coincidence conditions eliminated almost all the background counts in the region of interest. The only remaining background structure is the REC line. REC results in a very broad structure in the x-ray spectrum with a long tail extending towards high photon energies \cite{kleb}, which is clearly visible in Fig.~\ref{fig:coincidence}~(b).

%----------------------------------
\muntab{@{\extracolsep{\fill}}ccccccc}{h0_coinc}{Results of a $\chi^2$ test of the RDEC range of the coincidence spectra.}{
\hline
     & $N$ & $b$ & $T$ & $1-\alpha_T$ & $N_{RDEC}$ & $\beta$\\
\hline
$q-1$& 567(23) & 241(13) & 25000 & <0.0001 & 326(27) & <0.0001 \\
$q-2$& 31(6) & 0.035(3) & 420 & <0.0001 & 31(6) & <0.0001 \\
\hline
}
%----------------------------------

In order to establish the significance of the observed structure within the RDEC energy range of the coincidence spectra (Fig.~\ref{fig:coincidence}) the method described in Appendix~\ref{app:A} was applied. The first step was to check whether the observed structure was due to statistical fluctuations (\(H_0\) hypothesis). Similar to the PIXE analysis, \(\alpha=0.05\) was assumed. Then, the background contribution in each spectrum was obtained by integration of the REC Compton profile fit over the RDEC energy range (\(2.8\)-\(4.2\)~keV). This gave the total number of the background counts in the \(q-1\) and \(q-2\) spectra of \(b^{q-1}= 241(13)\) and \(b^{q-2}= 0.035(3)\). The total number of counts accumulated in the RDEC energy range during the experiment was \(N^{q-1}= 567(23)\) and \(N^{q-2}= 31(6)\) for the single and double charge exchange channels, respectively. These gave the value of the statistical variable (Eq.~\ref{eqn:T}) of \(T^{q-1} \approx 25000\) and \(T^{q-2} \approx 420\), which corresponds to the value of \(\alpha_T > 0.9999\) for both $q-1$ and $q-2$ channels. Thus, in both cases the hypothesis \(H_0\) has to be rejected, as the probability that the observed structure is due to statistical fluctuations is less than \(0.0001\).
This means that the observed structure is a result of a physical process (\(H_1\) hypothesis) and the number of counts associated with this process is:
\(N^{q-i}_{RDEC}=N^{q-i}-b^{q-i}\), where \(i=1\), \(2\), with the uncertainty given by \(\Delta N^{q-i}_{RDEC}=\sqrt{\left(\Delta  N^{q-i}\right) ^2+\left( \Delta b^{q-i}\right)^2 }\).
%-----------------------------------
\munepsfig[.96]{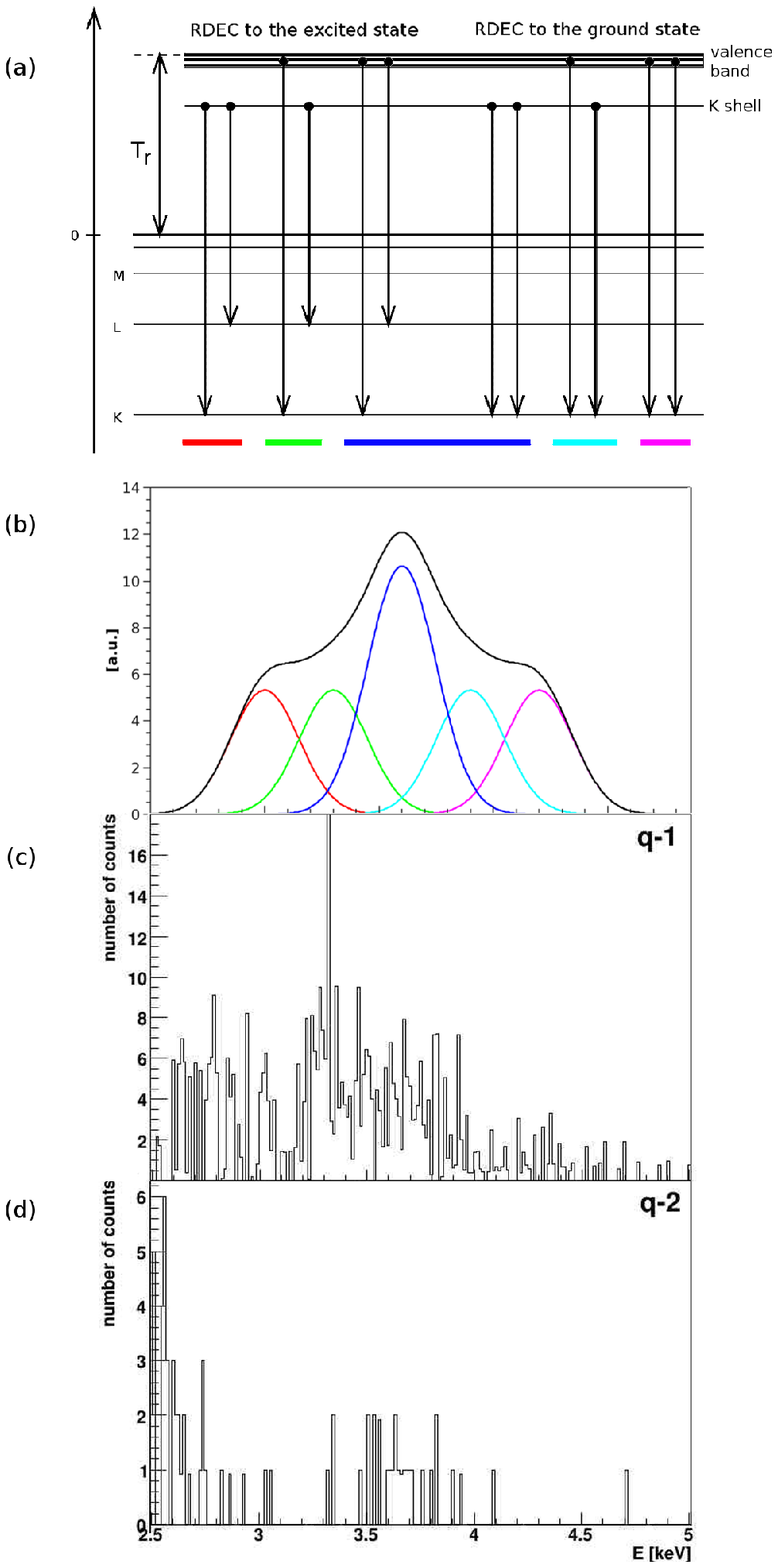}{Possible RDEC transitions (a) and the structure of the produced x-ray spectrum (b) when equal cross sections for all the partial processes are assumed. Black line -- the sum of all contributions. Additionally, corresponding RDEC spectra obtained experimentally in single (c) and double (d) charge exchange channels are presented.}
%-----------------------------------
This leads to numbers of the RDEC counts of \(N^{q-1}_{RDEC}=326(27)\) and \(N^{q-2}_{RDEC}=31(6)\) for the single and double charge exchange channels, respectively.
Additionally, as given in Eq.~\ref{eqn:beta2}, the type two error (\(\beta\)) can be estimated. In the case discussed here, the value of \(\beta\) is less than \(0.0001\) which is beyond the statistical significance.
Values of all the parameters for both single and double charge exchange channels are given in Table~\ref{tab:h0_coinc}.

Subtraction of the Compton profile based on Ref.~\cite{big} from the coincidence spectra (Fig.~\ref{fig:coincidence}) revealed a complex structure of the RDEC line. The resulting spectra are presented in Figs~\ref{fig:rdec_excited}~(c) and (d). The observed structure comprises at least two maxima which can be assigned to the RDEC process. It is not only a result of capture to the ground (\(1s^2\)) and excited (\(1s^12s^1\)) states of the projectile (see Fig.~\ref{fig:rdec_excited}~(b)), but can also be attributed to the capture of either K-shell or valence target electrons. Combinations of the initial and final electron states accessible for the process and the resulting RDEC peaks positions are given in Table~\ref{tab:energies}. The possible transitions are shown in Fig.~\ref{fig:rdec_excited}~(a) with the positions of the lines representing their contributions to the RDEC x-ray spectrum (Fig.~\ref{fig:rdec_excited}~(b)). For this presentation, probabilities of all the possible RDEC transitions were assumed to be equal. This spectrum should be compared with the data obtained experimentally in the single and double charge exchange channels after background subtraction, shown in Figs~\ref{fig:rdec_excited}~(c) and (d). The latter fgure suggests that the transitions from the target valence band to the projectile K-shell are negligible.

%This suggests that the capture from the valence band is not observed in the spectrum. This observation might suggest that correlated capture of two electrons is possible only when the velocities of the captured electrons are comparable. This condition is more likely to be fulfilled in case of the K-shell electrons, thus, the RDEC from the target K-shell should be more probable. This observation is in contradiction to the theory \cite{mikh2}, which predicted a significant contribution to the RDEC process from the capture of the target valence electrons.

In order to make at least an estimate of the contributions of all the possible transitions, the sum of lines representing all the possible processes was fitted to the sum of the x-ray spectra registered in the \(q-1\) and \(q-2\) coincidence channels. The result is presented in Fig.~\ref{fig:ratio_fit} and the fitting parameters are given in Table~\ref{tab:rdec_fit}. It can be noticed that in case of transitions which included at least one target valence electron the corresponding areas below the fitted curve are close to zero. Based on this observation, it can be deduced that the transitions from the valence band to the \(1s^12s^1\) state, which overlap with the K-shell $\rightarrow$ \(1s^2\) transition, are also negligible.
%-----------------------------------
\munepsfig[.65]{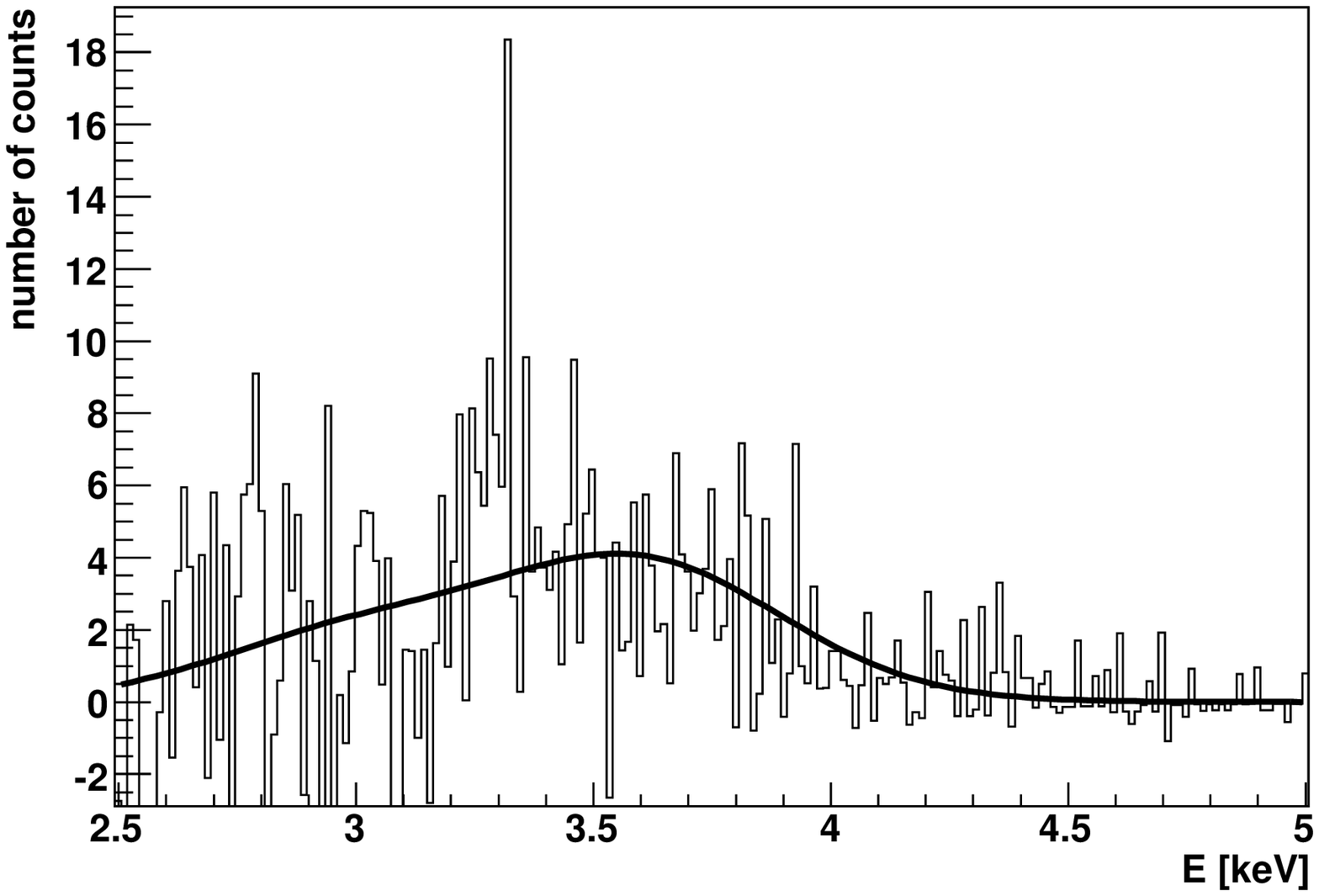}{The sum of spectra registered in single and double charge exchange channels with a fit of all possible combinations of the RDEC transitions. Fitting parameters are given in Table~\ref{tab:rdec_fit}.}
%-----------------------------------

Thus, transitions from the valence band were practically not present in the RDEC spectrum. Moreover, the ratio, \(R_A\), of the areas under the curves representing 
transitions from the target K-shell to the excited and ground projectile states, \(R_A=A_{1s^12s^2}/A_{1s^2}\), is equal to \(0.500(68)\) This value reflects the ratio of the RDEC cross sections for capture to the excited and ground states \(\sigma_{RDEC}^{1s^12s^2}/\sigma_{RDEC}^{1s^2}\). The theoretical value of this ratio can be estimated from Fig.~\ref{fig:sigma} for the adiabacity parameter of \(\xi_O=0.82\), which corresponds to the experimental conditions. The theoretical value of this cross section ratio is equal \(0.7\), which is close to the experimental result.

%---------------------------------- 
\muntab{@{\extracolsep{\fill}}lcc}{rdec_fit}{Areas ($A$) of the shapes of the RDEC contributions fitted to the sum of $q-1$ and $q-2$ spectra. The FWHM of all lines was set to $0.3$~keV which is the width of the carbon Compton profile.}
{\hline
Transition							&Peak position [keV]		&Area ($A$)\tabularnewline
\hline
2~$\times$~K-shell $\rightarrow$ \(1s^12s^1\)				&3.04			&118(11)\tabularnewline
1~$\times$~K-shell~+~1$~\times$~valence $\rightarrow$ \(1s^12s^1\)			&3.31			&0.0(1.9)\tabularnewline
2~$\times$~valence $\rightarrow$ \(1s^12s^1\) or 2~$\times$~K-shell $\rightarrow$ \(1s^2\)	&3.62		&237(11)\tabularnewline
1~$\times$~K-shell~+~1~$\times$~valence $\rightarrow$ \(1s^2\)				&3.91			&0.0(2.5)\tabularnewline
2~$\times$~valence $\rightarrow$ \(1s^2\)					&4.18			&2.5(9.9)\tabularnewline
\hline
}
%---------------------------------- 

%These contributions cannot be separated in the x-ray spectrum as in case of O$^{8+}$+C collisionsthe difference in the binding energies of K and L shell of the projectile is two times greater than the binding energy of the target K-shell electron: \(E_B^K-E_B^L \approx 2 E_{Bt}\) (see Eq.~\ref{eqn:Erdec}). 

\chapter{The RDEC cross section}
\label{chap:cross}
\section{Experimental value of the RDEC cross section} 
\label{sect:cross_section}
As it was discussed in Section~\ref{section:coincidence}, RDEC photons appeared both in single and double charge exchange channels. The numbers of counts in the coincidence spectra, \(N^{q-1}_{RDEC}=326(27)\) and \(N^{q-2}_{RDEC}=31(6)\), in single and double charge exchange channel, respectively, gave the total number of RDEC counts \(N_{RDEC}=357(28)\). The total number of the REC counts collected during the experiment in the single x-ray spectrum was \(N_{REC}=39800(200)\).
The ratio \( N_{RDEC} /  N_{REC}\) of the total RDEC to REC numbers of counts obtained during the experiment was equal to \(0.0092(6)\). One can assume that:
%----------------------------------
\muneqn{cross_ratio}{
\dfrac{N_{RDEC}}{N_{REC}}=\dfrac{ \dfrac{d\sigma _{RDEC}}{d\Omega}\Big{|}_{\theta=90^\circ}}{\dfrac{d\sigma _{REC}}{d\Omega}\Big{|}_{\theta=90^\circ}},
}
%----------------------------------
which, for the angular differential cross section for the REC process obtained from Eq.~\ref{eqn:diff}, gives \((d\sigma _{RDEC}/d\Omega){|}_{\theta=90^\circ} = 0.71(5)\)~b/sr. When the angular distribution of the RDEC photons is assumed to be the same as for the REC photons (\(\sim \sin^2{\theta}\)), one obtains the total RDEC cross section value \(\sigma_{RDEC}=5.9(4)\)~b.

The registered REC photons can be emitted not only during the REC process but can also originate from the DREC process. Thus, the probability \(P_{REC}\) of observation of a single REC photon is given by:
%----------------------------------
\muneqn{REC_prob}{
P_{REC}=\left( \dfrac{d\sigma _{REC}}{d\Omega}\Big{|}_{\theta=90^\circ} + 2 \dfrac{d\sigma _{DREC}}{d\Omega}\Big{|}_{\theta=90^\circ} \right) \Delta \Omega d,
}
%----------------------------------
where \(d\) is the target thickness.
In case of the DREC process the cross section has to be multiplied by two, as either of the two emitted photons can be registered by the x-ray detector. Similarly, there is a certain probability that both REC photons emitted during theDREC process are registered as a single event (see discussion of this problem in Section~\ref{section:double}). Then, the probability \(P_{RDEC}\) of registration of a photon in the RDEC energy range can be expressed as:
%----------------------------------
\muneqn{RDEC_prob}{
P_{RDEC}=\left( \dfrac{d\sigma _{RDEC}}{d\Omega}\Big{|}_{\theta=90^\circ} + \dfrac{d^2\sigma _{DREC}}{d\Omega_1 d\Omega_2}\Big{|}_{\theta_1, \theta_2=90^\circ} \Delta \Omega \right) \Delta \Omega d,
}
%----------------------------------
where the indices \(1\) and \(2\) indicate each of the emitted DREC photons. In this case:
%----------------------------------
\muneqn{_RDECcross_ratio}{
\dfrac{N_{RDEC}}{N_{REC}}=\dfrac{P_{RDEC}}{P_{REC}}
}
%----------------------------------
and the RDEC differential cross section is estimated to be \((d\sigma _{RDEC}/d\Omega){|}_{\theta=90^\circ} = 0.66(39)\)~b/sr. When the angular distribution of the RDEC photons is again assumed to be the same as for REC photons (\(\sim\sin^2\theta\)), the total RDEC cross section is equal to \(\sigma_{RDEC}=5.5(3.2)\)~b. The uncertainty comes mainly from the uncertainty of the target thickness given by the maker of the foil \cite{tar}.
Both the obtained \(\sigma_{RDEC}\) values are comparable and both are significantly greater than the total cross section based on the theoretical expectations (\(\sigma_{RDEC}=0.15\)~b \cite{nef2}).

When, according to the observations described in Section~\ref{section:coincidence}, the contribution of the transitions from the target K-shell to the projectile excited state are taken into account, the cross section for RDEC to the ground state, \(\sigma_{RDEC}^{1s^2}\), and for the capture to the excited state, \(\sigma_{RDEC}^{1s^12s^1}\), can be calculated from:
%----------------------------------
\muneqn{RDEC_ratio6}{
\begin{cases}
\dfrac{\sigma_{RDEC}^{1s^12s^1}}{\sigma_{RDEC}^{1s^2}}=0.7,\\
\sigma_{RDEC}^{1s^12s^1}+\sigma_{RDEC}^{1s^2}=\sigma_{RDEC},
\end{cases}
}
%----------------------------------
which finally leads to the values of the RDEC cross sections \(\sigma_{RDEC}^{1s^2}\) and \(\sigma_{RDEC}^{1s^12s^1}\) of \(3.2(1.9)\)~b and \(2.3(1.3)\)~b, for the capture to the ground and excited states, respectively.
The \(\sigma_{RDEC}^{1s^2}\) value is still a factor of \(25\) greater than the theoretical one. This leads to the ratio, \(R_{exp}\), of the RDEC to REC cross sections of \(R_{exp}=\sigma^{1s^2}_{RDEC}/\sigma_{REC}=7.4(3.7)\cdot 10^{-3}\) which is much greater than the theoretical value based on Nefiodov calculations \cite{nef2} and the Stobbe formula for the REC cross section, \(R_{Nef} = 2.9 \cdot 10^{-4}\).

\section{Estimation of the $\sigma_{RDEC}/\sigma_{REC}$ ratio in the nonrelativistic approach}

Simple calculations based on the principle of detailed balance can be used to estimate the cross sections ratio \(R_{nrel}=\sigma^{1s^2}_{RDEC}/\sigma_{REC}\). Eq.~\ref{eqn:ratio2} expresses this ratio by means of single and double photoionization cross sections.
For a given projectile atomic number \(Z\), the ratio \(\sigma_{DPI} / \sigma_{PI}\) is almost constant and, for nonrelativistic photon energies, can be expressed as \cite{amu2}:
%----------------------------------
\muneqn{dph_ph}{
\dfrac{\sigma_{DPI}}{\sigma_{PI}} \approx \dfrac{0.0932}{Z^2}.
}
%----------------------------------
As the photoionization cross section scales with the photon energy \cite{fano}:
%----------------------------------
\muneqn{ph}{
\sigma_{PI} \sim \left( \dfrac{1}{\hbar \omega}\right) ^5,
}
%----------------------------------
the ratio \(\sigma_{PI}(2 \hbar \omega) / \sigma_{PI}(\hbar \omega) = 1/32\) and the cross sections ratio can be written as:
%----------------------------------
\muneqn{ratio_final}{
R_{nrel}=\dfrac{\sigma^{1s^2}_{RDEC}}{\sigma_{REC}}=0.00291 F \dfrac{Z_t-1}{Z^2}.
}
%----------------------------------

In case of the discussed experiment \(Z_t=6\), \(Z=8\) and the above formula gives, for \(F=1\), \(R_{nrel} = 2.3 \cdot 10^{-4}\), which is close to the value obtained by Nefiodov. This suggests that the \(R_{Nef}\) value calculated by Nefiodov is not very far from the nonrelativistic estimation of the RDEC cross section based on the cross section for double photoionization and the principle of detailed balance.

\section{RDEC cross section based on the Yakhontov approach}
\label{section:yakh}
In two papers \cite{yak1,yak2} Yakhontov calculated the value of the \(\sigma^{1s^2}_{RDEC}/\sigma_{REC}\) ratio for the \(11.4\)~MeV/u Ar$^{18+}$~+~C collisions. For a given value of the adiabacity parameter \(\xi\), the value of this ratio scales with the projectile atomic number as \(Z^{-5}\). In case of the argon and oxygen experiments, the value of the adiabacity parameter was comparable (\(\xi_{Ar}=0.85\) and \(\xi_{O}=0.82\)). Thus, it was possible to use this scaling for estimation of the value of \(R_{Yakh}\) for the oxygen experiment from the value calculated by Yakhontov for the argon experiment (\(R^{Ar}_{Yakh}\)). Since for these experiments:
%----------------------------------
\muneqn{yakh_r}{
\dfrac{R_{Yakh}}{R^{Ar}_{Yakh}}=\dfrac{Z_O^{-5}}{Z_{Ar}^{-5}}=57.66,
}
%----------------------------------
one obtains the value of the RDEC to REC cross section ratio for oxygen of \(R_{Yakh}=2.1\cdot10^{-4}\) which is consistent with both nonrelativistic estimations, i.e. Nefiodov calculations and the estimation based on the principle of detailed balance.
\vspace{1cm}

%----------------------------------
\muntab{@{\extracolsep{\fill}}lcccc}{rdec_summary}{Comparison of the experimental values of the RDEC cross section and the $R=\sigma^{1s^2}_{RDEC}/\sigma_{REC}$ ratio with the ones obtained from various theoretical approaches .}{
\hline
 &\multirow{2}{*}{Experiment} & Nefiodov  & Non-relativistic  & Yakhontov  \\
 &				& calculations & approach & calculations \\
\hline
$\sigma^{1s^2}_{RDEC}$ [b]&3.2(1.9)& 0.15& 0.14	& 0.14\\
$R=\sigma^{1s^2}_{RDEC}/\sigma_{REC}$ & 7.4(3.7)$\cdot$10$^{-3}$ & 2.9$\cdot$10$^{-4}$ & 2.3$\cdot$10$^{-4}$ & 2.1$\cdot$10$^{-4}$ \\
\hline
}
%----------------------------------

The estimations of the RDEC cross section and the \(\sigma^{1s^2}_{RDEC}/\sigma_{REC}\) ratio presented in this chapter suggest that the experimentally obtained value of the RDEC cross section is much greater than theoretical estimations. Consequently, the value of the \(\sigma^{1s^2}_{RDEC}\) cross section is much greater than the theoretical ones, estimated from the \(R=\sigma^{1s^2}_{RDEC}/\sigma_{REC}\) ratio, as summarised in Table~\ref{tab:rdec_summary}. However, according to Nefiodov \cite{nef2}, the system chosen for the experiment does not fully meet the theoretical assumptions. Since \(Z \sim Z_T\), captured electrons cannot be treated as quasifree and theoretical calculations given in \cite{mikh1,mikh2,nef} might provide an underestimated cross section value. As the same assumptions for the captured electrons are applied for the other theoretical approaches to the RDEC cross section that are given in this chapter, these calculations may also give underestimated values of the RDEC cross section.

\chapter{Monte Carlo simulations of the x-ray spectra}
\label{chap:mc}

It was shown in Section \ref{sect:cross_section} that the experimental value of the RDEC cross section is significantly greater than the one predicted by Nefiodov \cite{nef2}. Thus, a Monte Carlo simulation was performed in order to check the structure of the x-ray spectrum that could be obtained, assuming the value of the RDEC cross section predicted by Nefiodov was correct. 

A \texttt{C++} code written by \'{S}wiat \cite{sw1} was used. This is a simulation of x-ray spectra observed during ion-atom collisions. The x-ray spectra that can be simulated are due to radiative processes such as REC, DREC and RDEC as well as characteristic x-ray radiation. This code was thoroughly tested by its author for many collision systems \cite{sw2} and was accepted as a reliable tool for simulations of x-ray spectra resulting from atomic collision processes.
The simulation can reproduce almost any experimental conditions regarding beam, target and detector properties. It allows for implementation of any geometry of the x-ray detector. The parameters that have to be given in the input file are: beam diameter, target thickness, detector distance from the target center, observation angle with respect to the beam direction, detector shape and dimensions. Detection efficiency is provided as a list of data points which are interpolated for the given photon energy. A detailed description of the input parameters can be found in \cite{sw2}. Here, the detector parameters were provided according to information presented in Chapter~\ref{chap:exp} and the detection efficiency was based on data given by ORTEC \cite{ortec} (see Fig.~\ref{fig:ortec}).
%-----------------------------------
\munepsfig[.6]{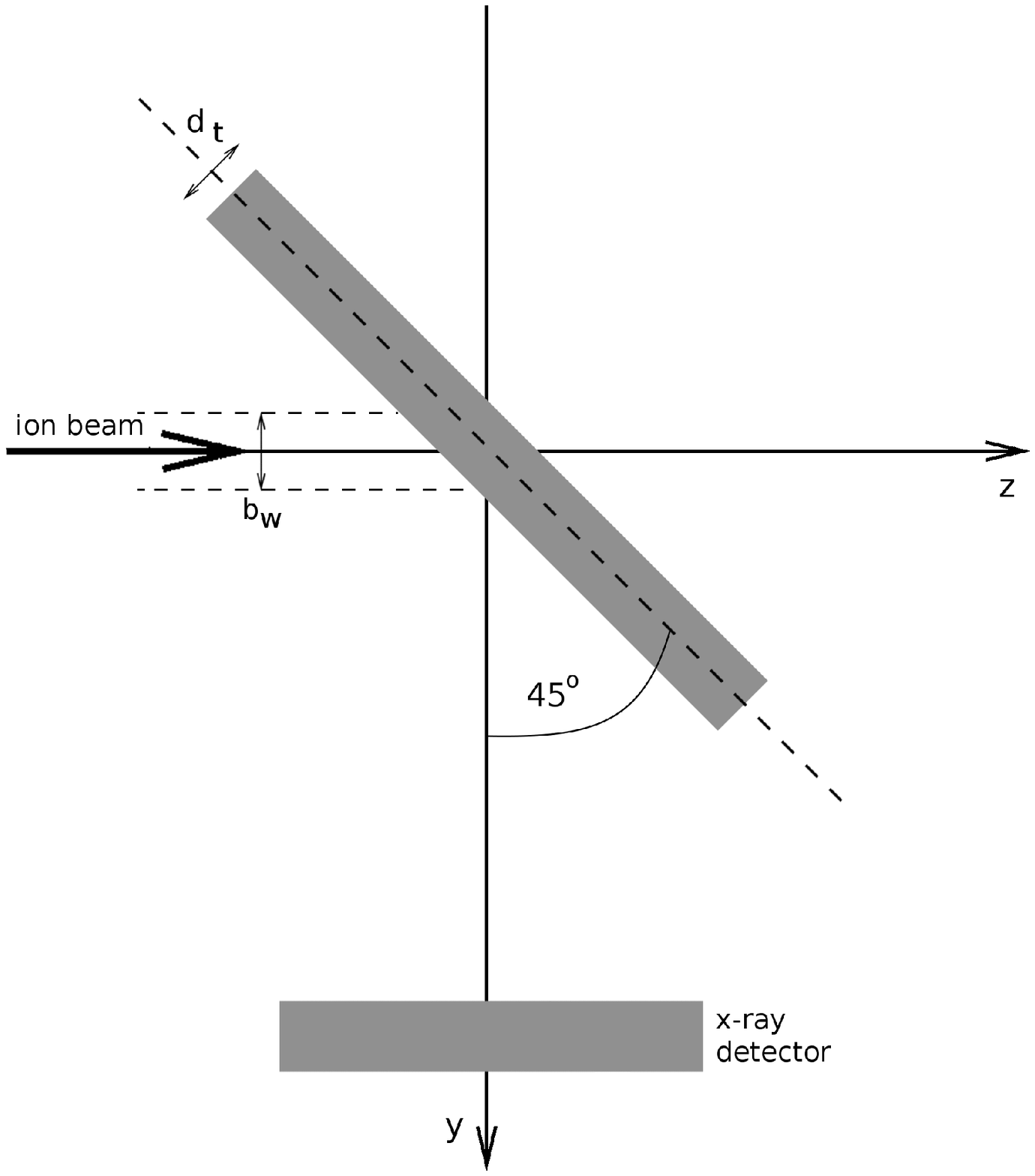}{Geometry of the experimental setup implemented in the Monte Carlo simulation, $b_{w}$ -- the beam diameter, $d_t$ -- target thickness in mm. The $x$-axis is perpendicular to the picture plane.}
%-----------------------------------
The code was originally designed to reproduce the conditions of gas-jet target experiments. Thus, by default, a Gaussian distribution of gas particles was used to describe the target density. Here, the code was adopted to the experimental conditions discussed in this thesis, a thin solid target positioned at \(45^\circ\) to the beam direction. The coordinate system used for simulation is shown in Fig.~\ref{fig:monte} for \(x=0\). It was assumed that the beam propagates in the \(z\) direction, while the center of the x-ray detector is positioned at \(x=0\) and at positive values of \(y\) (compare Figs.~\ref{fig:setup} and \ref{fig:monte}). The primary method for determination of the collision coordinates \((x,y,z)\) within the target (see Chapter~3 of \cite{sw2}) was replaced by:
%----------------------------------
\muneqn{mc_x}{
\begin{cases}
x=b_w(Rand-0.5),\\
y=b_w(Rand-0.5),\\
z=d_t(Rand-0.5)+y \tan{\left(\dfrac{\pi}{4}\right)},
\end{cases}
}
%----------------------------------
where \(b_w\) is the beam diameter and \(d_t\) is the target thickness, both given in mm. \(Rand \in \left[0;1 \right]\) is provided by a random number generator. The beam transverse cross section was assumed to be a circle, thus the \((x,y)\) coordinates were limited by:
%----------------------------------
\muneqn{mc_bw}{
x^2 + y^2 \leq \left(\dfrac{b_w}{2}\right)^2.
}
%-----------------------------------

The REC and DREC processes were implemented with the cross sections of \(512\)~b and \(48\)~b, respectively, as given in Table~\ref{tab:proc}. As shown in the previous chapter, the RDEC structure observed in x-ray spectra consisted mainly of two lines. In order to simulate this double structure of the RDEC line (i.e. capture to the ground (\(1s^2\)) and excited (\(1s^12s^1\)) state), the input data included two kinds of the RDEC processes: capture to the ground state with the cross section of \(\sigma^{1s^2}_{RDEC}\) and capture to the excited state with cross section of \(\sigma^{1s^12s^1}_{RDEC}=0.7\sigma^{1s^2}_{RDEC}\) estimated according to \cite{nef}. As the photon angular distributions for neither RDEC nor DREC is known, they were assumed to be the same as for the REC photons, that is \(\sim\sin^2\theta\). Additionally, the DREC photons were considered to be emitted independently of each other, as discussed in Section \ref{section:double}.
%Background processes were not included in the simulation, as the code allows only for simulation of a plain background, which is insufficient for a proper description of bremsstrahlung contributions..

%-----------------------------------
\munepsfig[.5]{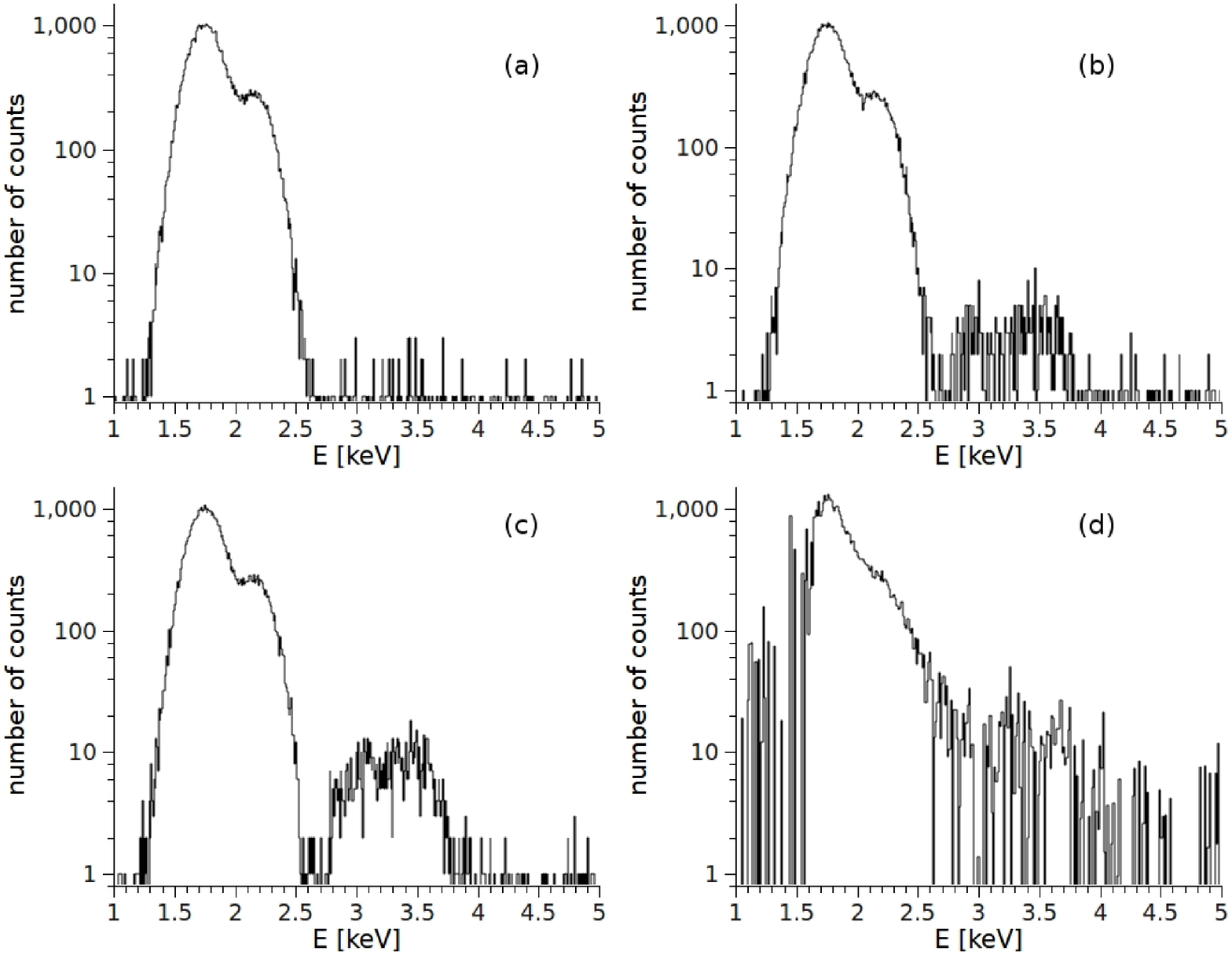}{Monte Carlo simulated x-ray spectra: (a) no RDEC included, (b) the RDEC cross section as given by Nefiodov, (c) $\sigma^{1s^2}_{RDEC}=3$~b and $\sigma^{1s^12s^1}_{RDEC}=2.1$~b -- cross sections values for which MC simulation gives the results closest to the experimental data. (d) Experimentally obtained singles spectrum.}
%-----------------------------------
%-----------------------------------
\munepsfig[.5]{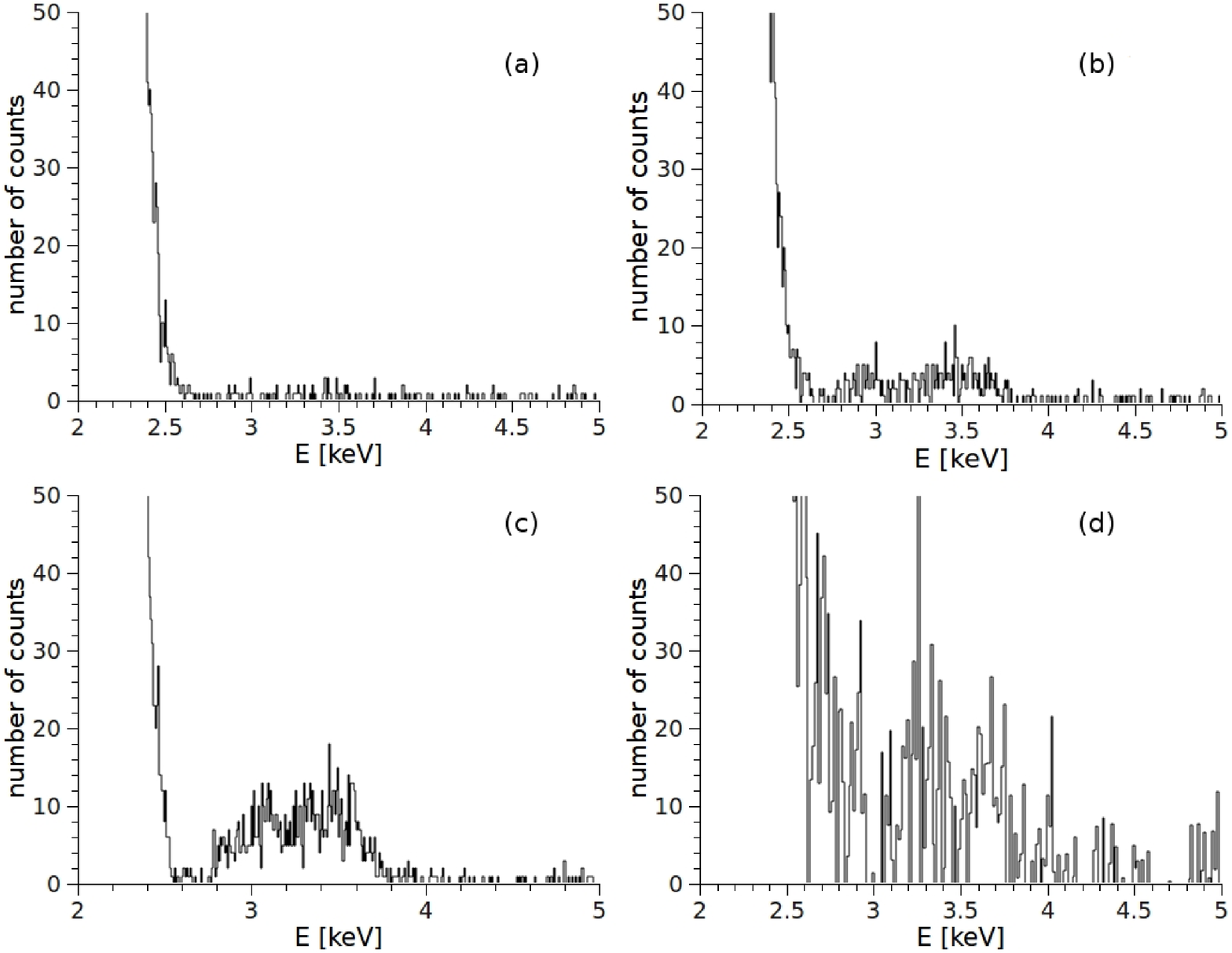}{The RDEC range of the x-ray spectra. Results of simulations: (a) no RDEC included, (b) the RDEC cross section as given by Nefiodov, (c) $\sigma^{1s^2}_{RDEC}=3$~b and $\sigma^{1s^12s^1}_{RDEC}=2.1$~b -- cross sections values for which MC simulation gives the results closest to the experimental data. (d) Experimentally obtained single spectrum}
%-----------------------------------

During the first step of the simulation, RDEC was not included in the input data. The obtained spectrum is shown in Fig.~\ref{fig:panel}~(a). It can be noticed that only a few counts showed up in the RDEC energy range. These counts are due to simultaneous detection of both the DREC photons. Comparison with the results in Fig.~\ref{fig:panel}~(d) shows clearly that DREC itself cannot explain the experimental results. Moreover, the result of the simulation shows that the contribution of a pile-up effect of the DREC photons is not significant as mentioned in Section~\ref{section:double}.

%----------------------------------
\muntab{@{\extracolsep{\fill}}lccccc}{mc_ratio}{Ratios of the numbers of counts in the RDEC and REC regions obtained during Monte Carlo simulations compared with the experimental value.}{
\hline
				&&RDEC cross section [b]		&\multicolumn{2}{c}{Number of counts}&\multirow{2}{*}{RDEC/REC ratio}\\
\cline{3-5}
           			&&$\sigma^{1s^2}_{RDEC}+\sigma^{1s^12s^1}_{RDEC}$&REC range 	&RDEC range  	&\\
\hline

\multirow{3}{*}{simulation}	&\multirow{3}{*}{\threelinebrace }& 0 & 40100(180) & 71(8) &0.0018(2)\\
				&& 0.15~+~0.105 & 39900(180) & 225(8) &0.0056(5)\\
				&& 3~+~2.1 & 39700(180) & 449(20) &0.0113(3)\\
\\
experiment			&&3.2~+~2.3		& 39800(200) & 357(28) & 0.0092(6) \\

\hline
}
%----------------------------------

The second step of the simulations was implementation of the RDEC process with the cross sections \(\sigma^{1s^2}_{RDEC}=0.15\)~b \cite{nef2} and \(\sigma^{1s^12s^1}_{RDEC}=0.7\sigma^{1s^2}_{RDEC}=0.105\)~b \cite{nef} for the capture to the ground and the excited state, respectively. Again, the results showed that this was not sufficient to explain the experimental data (compare Figs~\ref{fig:panel}~(b) and (d)).

During the third step of the simulation the RDEC cross section was increased until the resulting spectrum was comparable with the experimental one (Fig.~\ref{fig:panel}~(d)). The best results, shown in Fig.~\ref{fig:panel}~(c), were obtained for \(\sigma^{1s^2}_{RDEC}=3\)~b and \(\sigma^{1s^12s^1}_{RDEC}=0.7\sigma^{1s^2}_{RDEC}=2.1\)~b. However, it can be noticed that the RDEC structure within the experimentally obtained spectrum (Fig.~\ref{fig:panel}~(d)) is placed on the REC tail, which seems to be much broader than the simulated one. This may be due to the fact that for the simulation the Compton profile of diamond was used as given by Reed \cite{reed}, which might differ from the one for the amorphous carbon foil that was used during the experiment. However, the main purpose of the simulation was to investigate the ratio of the numbers of counts related to the RDEC and REC processes, thus the Compton profile here is of a minor importance.

In Fig.~\ref{fig:panel_rdec} only the RDEC range of all the spectra was shown on the linear scale, for better visualization of the double structure of the RDEC line. 
As a test of the result of the simulations, the ratio of the numbers of counts in the RDEC and REC range was calculated. The obtained values are shown in Table~\ref{tab:mc_ratio}. It can be seen that the results of the third step of the simulation are in the best agreement with the experimental data. This again shows that the cross section value calculated by Nefiodov, even when capture to the excited state is included, is insufficient to explain the experimental results.

\chapter{Conclusions}
\label{chap:conclusions}

In this dissertation an experiment dedicated to the radiative double electron capture process (RDEC) was presented. The experiment was carried out at Western Michigan University using the \(6\)~MV tandem Van de Graaff accelerator. The choice of the collision system, O$^{8+}$~+~C at 38~MeV, was based on the recent theoretical calculations of the RDEC cross section \cite{mikh1,mikh2,nef}, which pointed to mid-\(Z\) ions and low collision energy as the best systems for observation of RDEC. The theoretical approach also suggested an enhancement of the RDEC cross section in such systems due to capture of electrons from the target valence band. Moreover, it was pointed out that the capture to the excited state of the projectile may be a significant contribution to the process.

The results allowed for the first experimental verification of RDEC and provided a test of the main theoretical predictions \cite{mikh1,mikh2,nef}. The obtained x-ray spectra revealed a complex structure of the RDEC line. However, the capture from the target valence band, which, according to the theory, was supposed to significantly contribute to the RDEC process, was not confirmed. The observed structure allowed for identification of capture to the projectile ground (\(1s^2\)) and excited (\(1s^12s^2\)) states. The ratio of the counts which could be associated with these two processes gave a ratio of the RDEC cross sections for the capture to the excited and the ground states \({\sigma_{RDEC}^{1s^12s^2}}/{\sigma_{RDEC}^{1s^2}}=0.500(68)\), which is close, considering the data statistics, to the theoretical value of \(0.7\).

The latter value, together with the observed ratio of the numbers of counts in the RDEC and REC range of the x-ray spectra, \({N_{RDEC}}/{N_{REC}}=0.0092(6)\), allowed for estimation of the RDEC cross sections:
\begin{itemize}
 \item \(\sigma_{RDEC}^{1s^2}=3.2(1.9)\)~b for the capture to the ground state,
\item  \(\sigma_{RDEC}^{1s^12s^1}=2.3(1.3)\)~b for the capture to the excited state.
\end{itemize}
The main results of the experiment have already been published \cite{sim}. The obtained value of the cross section is a factor of about \(25\) greater than the theoretical value provided by Nefiodov \cite{nef2}. Consequently, the ratio \(R={\sigma^{1s^2}_{RDEC}}/{\sigma_{REC}}=0.0074(37)\) is also significantly greater than estimated from the theory, as shown in Table \ref{tab:summary}.

%-----------------------------------
\muntab{@{\extracolsep{\fill}}lcccccc}{summary}{Summary of the results of the theoretical calculations and the experiments dedicated to the RDEC process.}
{\hline
\multirow{2}{*}{\(Z\)} &\multirow{2}{*}{\(E\) [MeV/u] }  & \multirow{2}{*}{\(\xi\) } & \multirow{2}{*}{ \(Z_t\) } & \multicolumn{3}{c}{$\sigma$$_{RDEC}^{1s^2}$ [mb]}\\
\cline{5-7}
 	& 	& 	& 	& Ref. \cite{mikh1} 	& Ref. \cite{yak2} 	& experiment\\
\hline
18 	& 11.4 	& 0.840 & 6 	& 3.2 			& 1.85 			& $\leq$5.2 \cite{war} \\
92 	& 297 	& 0.841	& 18 	& 2.5$\cdot$10$^{-2}$ 	& 5000 			& $\leq$10 \cite{bed} \\
8	&2.375	&0.820	&6	& 1.5$\cdot$10$^{2}$~$^{(a)}$ 	& 1.4$\cdot$10$^{2}$~$^{(b)}$		& 3.2(1.9)$\cdot$10$^{3}$~$^{(c)}$\\
\hline
\multicolumn{7}{l}{$^{(a)}$ provided by Nefiodov \cite{nef2}}\\
\multicolumn{7}{l}{$^{(b)}$ estimated from the $R_{Yakh}={\sigma^{1s^2}_{RDEC}}/{\sigma_{REC}}$ ratio}\\
\multicolumn{7}{l}{$^{(c)}$ this experiment}}
%-----------------------------------

The results of the so far conducted experiments dedicated to the RDEC process are given in Table~\ref{tab:summary} . The discrepancies between the theories and experiments, as well as the differences between various theoretical approaches, show that further investigation of the RDEC process is necessary. A similar experiment chould be performed, with the same experimental setup available at WMU, but with an additional absorbing material in front of the x-ray detector. This will exclude all the pile-up effects and contributions of the DREC process, which formed a crucial problem of the background analysis in the presented experiment. Moreover, the angular distribution and correlations of the DREC photons should be measured as DREC interferes with other radiative capture processes.

As the role of the capture from the valence band of the solid target was not verified, application of a gas jet target with a light gas (He) would be useful, due to a significant reduction of background and a much smaller probability of multiple collisions. Additionally, a system with a greater difference between projectile and target atomic number (for example Ca$^{20+}$, Ar$^{18+}$ or Xe$^{54+}$ on He) would allow for better separation of photons originating from the capture to the ground and excited state of the projectile.
All these conditions can be satisfied during an experiment on the ESR gas jet target at GSI.

\renewcommand{\theequation}{A.\arabic{equation}}
\setcounter{equation}{0}
\appendix

\chapter{Statistical analysis of the observed signal}
\label{app:A}

In order to check if the observed structure within a given region of the spectrum is a result of a physical process or just a statistical fluctuation, a method suggested in \cite{eadie} was applied. Here, a brief description of the approach is presented.

Alternatively, one can calculate the probability of observation of a statistical fluctuation within the range of interest. This method assumes that the background distribution is known.

A null hypothesis can be defined as:

\begin{center}
\textit{\(H_0\): there is no physical effect within the \(AB\) range.}
\end{center}

It is assumed that the background shape can be described by a function \(b({\bf x},{\bf \Theta})\), which depends on the observed variable \({\bf x}\) and unknown parameters \({\bf \Theta}\).
%-----------------------------------
\munepsfig[.7]{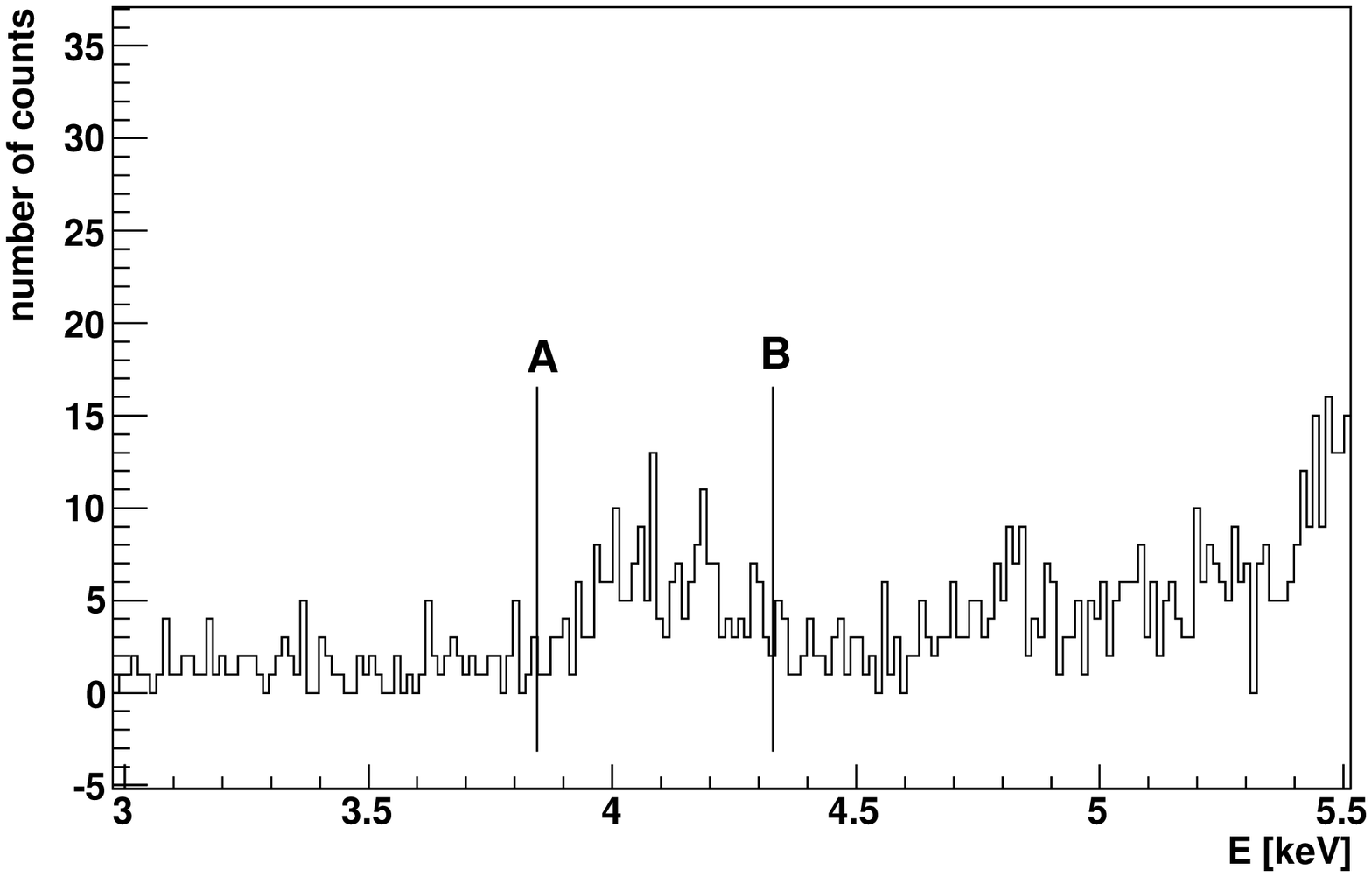}{Example of an experimentally obtained spectrum with a structure within the \textit{AB} range.}
%-----------------------------------
In this case the number of the background counts in the \(AB\) region (Fig. \ref{fig:app_ex}) is given by:
%-----------------------------------
\muneqn{bcgr}{ 
\hat{b}_{AB}=\int_{A}^{B}b({\bf x},\hat{{\bf \Theta}})d{\bf x}.
}
%-----------------------------------

As \(\hat{b}_{AB}\) is a function of estimators \(\hat{{\bf \Theta}}\), the variance can be obtained by a substitution of variables:
%-----------------------------------
\muneqn{sigma}{ 
\hat{\sigma}_{AB}^2={\bf D}^T\tilde{V}{\bf D},
}
%-----------------------------------
where \({\bf D}\) denotes a vector of derivatives:
%-----------------------------------
\muneqn{Di}{
D_i=\dfrac{\partial \hat{b}_{AB}}{\partial \Theta_i}\Big{|}_{\hat{{\bf \Theta}_i}} = \int_{A}^{B}\dfrac{\partial}{\partial \Theta_i}b({\bf x},\hat{{\bf \Theta}})dx
}
%-----------------------------------
If \(N_{AB}\) denotes the number of events in the \(AB\) range, the optimal test statistic, which checks if \(N_{AB}\) significantly differs from \(\hat{b}_{AB}\), is:
%-----------------------------------
\muneqn{T}{
T=\dfrac{(N_{AB}-\hat{b}_{AB})^2}{V(N_{AB}-\hat{b}_{AB})},
}
%-----------------------------------
where \(V(N_{AB}-\hat{b}_{AB})\) is a variance. If the \(H_0\) hypothesis is true, then:
%-----------------------------------
\muneqn{En}{
E(N_{AB})=b_{AB}
}
\muneqn{Vn}{
V(N_{AB})=b_{AB}
}
%-----------------------------------
where \(E(N_{AB})\) is the estimator of the number of counts in the AB region and the estimator of \(b_{AB}\) is \(\hat{b}_{AB}\). Thus:
%-----------------------------------
\muneqn{Vn-b}{
V(N_{AB}-\hat{b}_{AB})\approx\hat{b}_{AB}+\hat{\sigma}_{AB}^2 -2cov(N_{AB},\hat{b}_{AB}).
}
%-----------------------------------
As during the estimation of \(\Theta\) parameters the \(AB\) range was excluded, \(N_{AB}\) and \(\hat{b}_{AB}\) are uncorrelated:
%-----------------------------------
\muneqn{Vn_b2}{
V(N_{AB}-\hat{b}_{AB})\approx\hat{b}_{AB}+\hat{\sigma}_{AB}^2
}
%-----------------------------------
and
%-----------------------------------
\muneqn{T2}{
T=\dfrac{(N_{AB}-\hat{b}_{AB})^2}{\sigma_{AB}^2+\hat{b}_{AB}}.
}
%-----------------------------------
If the number of counts \(N_{AB}\) is large, it has a normal distribution around \(\hat{b}_{AB}\) and the statistic \(T\) behaves as a \(\chi^2\) distribution for \(DoF=1\) (degrees of freedom).

For the obtained value of the statistical variable \(T\) the value of \(\alpha_T\) can be estimated based on Table \ref{tab:chi}. 
If \((1-\alpha_T) > \alpha\), where \(\alpha\) is the probability of the type one error, the hypothesis can be accepted. The deviation is small enough for chance alone to account for it. 
If \((1-\alpha_T) < \alpha\), the hypothesis should be rejected, as there may be some factor other than chance operating for the deviation to be so great. 

%-----------------------------------
\muntab{@{\extracolsep{\fill}}lcccccccccc}{chi}{Quantiles of the $\chi^2$ distribution for $DoF=1$ \cite{kamys}.}
{
\hline
$\alpha$ &0.005	&0.010	&0.020	&0.025	&0.050	&0.100	&0.250	&0.300	&0.500\\ 
$\chi^{2}$ &4.0E-5	&0.0002	&0.0006	&0.0010	&0.0039	&0.0642	&0.1015	&0.1485	&0.4549\\
\hline
$\alpha$ &0.700	&0.750	&0.800	&0.900	&0.950	&0.975	&0.990	&0.995	&0.999\\
$\chi^{2}$ &1.0742	&1.3233	&1.6424	&2.7055	&3.8415	&5.4119	&6.6349	&7.8794	&10.8270\\
\hline
}
%-----------------------------------

If, based on statistics \ref{eqn:T2}, \(H_0\) was rejected, estimation of the signal \(s=N_{AB}-\hat{b}_{AB}\) can still be used, but with a different estimator variance. The problem one is about to solve is now a test of hypothesis \(H_1\) against hypothesis \(H_0\).

\begin{center}
\textit{\(H_1\): there is a physical signal \(s\) and background \(\hat{b}_{AB}\) in the \(AB\) range.}
\end{center}

In this case the variance will be given by:
%-----------------------------------
\muneqn{Vn_b3}{
V(N_{AB}-\hat{b}_{AB})\approx N_{AB}+\hat{\sigma}_{AB}^2,
}
%-----------------------------------
where, for the same reason as in Eq. \ref{eqn:Vn_b2}, covariance is not included. The risk of making a type two error is equal to:
%-----------------------------------
\muneqn{beta}{
\beta = P( d \leq \lambda_\alpha | H_1 ),
}
%-----------------------------------
where \(d=\sqrt{T}\) and \(\lambda_\alpha\) is defined by \(\Phi(\lambda_\alpha)=\alpha\), \(\Phi\) being a cumulative normal distribution function, given by:
%-----------------------------------
\muneqn{phi}{
\Phi(x)=\dfrac{1}{\sqrt{2 \pi}} \int_{-\infty}^x e^{-t^2/2}dt.
}
%-----------------------------------

For the \(H_1\) hypothesis the mean value and variance are given by:
%-----------------------------------
\muneqn{EVn}{
E(N_{AB})=V(N_{AB})=b_{AB}+s,
}
%-----------------------------------
while \(d\) has a normal distribution \({\bf N}(\mu,\sigma^2)\), where:
%-----------------------------------
\muneqn{muN}{
\mu=\dfrac{s}{\sqrt{\hat{b}_{AB}+\hat{\sigma}_{AB}^2}},
}
\muneqn{sigmaN}{
\hat{\sigma}^2=\dfrac{\hat{b}_{AB}+s+\hat{\sigma}_{AB}^2}{\hat{b}_{AB}+\hat{\sigma}_{AB}^2}.
}
%-----------------------------------
Thus, finally:
%-----------------------------------
\muneqn{beta2}{
\beta=\Phi\left[ \dfrac{\lambda_\alpha \sqrt{\hat{b}_{AB}+\hat{\sigma}_{AB}^2}-s}{\sqrt{\hat{b}_{AB}+s+\hat{\sigma}_{AB}^2}}  \right].
}
%-----------------------------------

\addcontentsline{toc}{chapter}{List of References}
\addtocontents{bbl}{\protect\thispagestyle{fancy}}
\bibliography{ref}

\end{document}